\documentclass[12pt,oneside,letter]{article}
\usepackage[left=1in, right=1in, top=1in, bottom=1in]{geometry}

\usepackage{iftex}
\ifPDFTeX
	\usepackage[T1]{fontenc}
	\usepackage{lmodern}
\fi

\usepackage{amsfonts}
\usepackage{float}
\usepackage{placeins}
\usepackage{amsmath}
\usepackage{amssymb}
\usepackage{amsthm}
\usepackage[toc]{appendix}
\usepackage{changepage}
\usepackage{color, colortbl}
\usepackage{xcolor}
\usepackage[bottom]{footmisc}
\usepackage{graphicx}
\usepackage{array}
\usepackage{graphics}
\usepackage{epstopdf}
\usepackage{epsfig}
\usepackage{tabularx}
\usepackage{xr}
\usepackage{enumitem}
\usepackage[font=large, labelsep=period, labelfont=bf]{caption}

\usepackage{epstopdf} 
\usepackage{setspace} 

\usepackage[hyperfootnotes=true]{hyperref}
\usepackage{longtable}
\usepackage{lscape}
\usepackage{multirow} 
\usepackage[round]{natbib}
\usepackage{newfloat}
\usepackage{pdflscape}
\usepackage{pdfpages}
\usepackage{rotating}
\usepackage[detect-all]{siunitx}
\usepackage{subcaption}
\usepackage[autostyle]{csquotes}
\usepackage{siunitx,booktabs, makecell}
\usepackage{etoolbox}
\usepackage{epigraph}
\usepackage{soul}
\usepackage{titlesec}
\usepackage[USenglish]{babel}

\usepackage{tikz}
\usepackage{pgfplots}
\usepackage{pgfplotstable}
\pgfplotsset{compat=1.17}
\usepgfplotslibrary{groupplots}
\usetikzlibrary{pgfplots.dateplot,fillbetween}

\newcolumntype{d}[1]{D{.}{.}{#1}}

\definecolor{blue}{rgb}{0.00,0.07,1.00}
\definecolor{red}{rgb}{1.00,0.00,0.00}
\definecolor{black}{rgb}{0.00,0.00,0.00}
\definecolor{LightCyan}{rgb}{0.88,1,1}
\definecolor{White}{rgb}{1.00,1.00,1.00}
\definecolor{Gray}{gray}{0.95}
\definecolor{Oran}{rgb}{0.97,1.00,0.67}
\definecolor{darkred}{rgb}{0.6,0.0,0.0}
\definecolor{orange}{rgb}{1,0.5,0}
\definecolor{w}{rgb}{1.00,1.00,1.00}
\definecolor{b}{rgb}{0.00,0.00,0.00}

\def\sym#1{\ifmmode^{#1}\else\(^{#1}\)\fi}

\newcommand*{\mytab}[1]{\hyperref[{#1}]{Table~\ref*{#1}}}
\newcommand*{\myfig}[1]{\hyperref[{#1}]{Figure~\ref*{#1}}}
\newcommand*{\myprp}[1]{\hyperref[{#1}]{Prompt~\ref*{#1}}}
\newcommand*{\myexmpl}[1]{\hyperref[{#1}]{Example~\ref*{#1}}}
\newcommand*{\mysec}[1]{\hyperref[{#1}]{Section~\ref*{#1}}}
\newcommand*{\myeq}[1]{\hyperref[{#1}]{Equation~(\ref*{#1})}}
\newcommand*{\myeqs}[1]{\hyperref[{#1}]{(\ref*{#1})}}
\newcommand*{\mypred}[1]{\hyperref[{#1}]{Prediction~\ref*{#1}}}
\newcommand*{\mylemma}[1]{\hyperref[{#1}]{Lemma~\ref*{#1}}}
\newcommand*{\myprop}[1]{\hyperref[{#1}]{Proposition 1}}
\newcommand*{\myass}[1]{\hyperref[{#1}]{Assumption~\ref*{#1}}}

\newcommand*{\mytabIA}[1]{\hyperref[{#1}]{Internet Appendix Table~\ref*{#1}}}
\newcommand*{\myfigIA}[1]{\hyperref[{#1}]{Internet Appendix Figure~\ref*{#1}}}

\newcommand*{\mytabsIA}[1]{\hyperref[{#1}]{Internet Appendix Tables~\ref*{#1}}}
\newcommand*{\mytabxIA}[1]{\hyperref[{#1}]{\ref*{#1}}}

\newcommand*{\mysecIA}[1]{\hyperref[{#1}]{Internet Appendix~\ref*{#1}}}
\newcommand{\mysecIAheader}[1]{\hyperref[{#1}]{Internet Appendix}}

\titlespacing\section{0pt}{-10pt plus 0pt minus 0pt}{-5pt plus 0pt minus 0pt}
\titlespacing\subsection{0pt}{-10pt plus 0pt minus 0pt}{-10pt plus 0pt minus 0pt}
\titlespacing\subsubsection{0pt}{-10pt plus 0pt minus 0pt}{-10pt plus 0pt minus 0pt}

\DeclareFloatingEnvironment[
fileext=loa,    
listname={List of Prompts}, 
name=Prompt,    
placement=ht,   
within=none,    
]{prompt}

\DeclareFloatingEnvironment[
fileext=loa,    
listname={List of Examples}, 
name=Example,    
placement=ht,   
within=none,    
]{examples}

\hypersetup{
	colorlinks=true,
	raiselinks=true,
	breaklinks=false,
	linkcolor=blue,
	anchorcolor=blue,
	citecolor=blue,
	pdftitle={AI and Exchange Rate Predictability},
	pdfauthor={Amin Izadyar},
	pdfsubject={Foreign Exchange, Return Predictability, Large Language Models},
	pdfkeywords={Foreign Exchange, Return Predictability, Large Language Models,
		ChatGPT, Artificial Intelligence, Exchange Rate Disconnect, Taylor Rule},
	pdfcreator={pdfLaTeX},
	pdflang={en-US}
}
\begin{document}
	\thispagestyle{empty}
	
	\title{{\LARGE \textbf{AI and Exchange Rate Predictability}}\vspace{0.5cm}}
	
	\date{}
	
	\thispagestyle{empty}
	\author{%
		\begin{tabular}{c}
			\textbf{Amin Izadyar} \\
			Imperial College Business School, Imperial College London \\
			\texttt{Email:a.izadyar23@imperial.ac.uk} \\
			\large\today
		\end{tabular}
	}
	\maketitle
	
	\thispagestyle{empty}
	\newpage
	
	\vspace{4cm}
	
	\begin{abstract}
		\noindent 
		I revisit the exchange rate disconnect puzzle, first documented by \citet{Meese/Rogoff:1983}, using generative artificial intelligence (AI) to forecast currency returns based on economic fundamentals. Using ChatGPT and DeepSeek, I analyze a comprehensive dataset of economic data releases for major currency pairs and measure the fundamental strength of each currency. These AI-powered fundamentals exhibit significant cross-sectional predictive power. A simple trading strategy that goes long currencies with strong fundamentals and short currencies with weak fundamentals generates a Sharpe ratio exceeding 0.7 per annum. The excess returns of this strategy remain significant after controlling for traditional currency factors. To mitigate concerns of look-ahead bias, I run multiple exercises to ensure that predictability stems from AI reasoning rather than memorization. Finally, I explore the potential sources of predictability and find evidence that the Taylor rule framework, generally used by central banks to set interest rates, is a key mechanism connecting exchange rates to economic fundamentals.
		
	\end{abstract}
	
	\bigskip
	\noindent\textit{Keywords:} Foreign Exchange, Return Predictability, Large Language Models, ChatGPT, Artificial Intelligence \\
	\medskip
	\noindent\textit{JEL Classification}: C53, F31, F37, G12, G15.

	\thispagestyle{empty}
	\newpage
	
	\setcounter{page}{1}
	

		
	\section{Introduction} \label{sec:introduction} 
	The ability of economic fundamentals to forecast exchange rates remains elusive, since models based on fundamentals are often outperformed by a simple random walk, a phenomenon known as the “exchange rate disconnect” puzzle \citep*[e.g.,][]{Meese/Rogoff:1983}.  Although the recent literature has identified a few economic variables that appear to have predictive power, the answer to this empirical puzzle remains unresolved \citep*[e.g.,][]{MARK:1995, ENGEL/WEST:2005, ROSSI:2013}.\\
	Against this backdrop, the emergence of artificial intelligence (AI) offers new opportunities to re-examine this puzzle. AI's recent advancements have enabled it to solve problems once considered too complex or data-intensive for traditional methods. Motivated by these developments, I leverage AI’s reasoning power and proficiency in handling large datasets to study a comprehensive dataset of economic data releases, covering over 500 indicators from 1996 to 2024 for major economies. This study aims to find a link between exchange rates and economic fundamentals, thus enhancing our understanding of price discovery in the largest and deepest financial market in the world. To preview my results, I find evidence that AI-derived fundamentals can predict future exchange rate returns and the most important predictors are \textit{Inflation data}, \textit{Employment data}, and \textit{Broad economic activity indicators}.\\
	Using large language models (LLMs) like GPT-4o and DeepSeek-V3, I analyze a large dataset comprising realized values, previous figures, and consensus forecasts of key economic data releases, such as GDP reports, employment statistics, inflation indices, and central bank decisions. I interact with the AI model using a structured prompt and ask it to generate a concise analysis and a directional signal indicating whether the data release implies the currency would STRENGTHEN, WEAKEN, or has an INSIGNIFICANT OR UNCERTAIN impact. Notably, the input provided to the model includes only the realized, previous, and forecast values, along with the name of the currency associated with the data release, excluding any information about the time or date of the release. Using AI's output, I then construct a simple measure, called the \textit{AIFX index}, that captures the net fundamental strength of each currency. Specifically, for each currency, the \textit{AIFX index} is defined as the difference between the number of positive (directional signal as “STRENGTHEN”) and negative (“WEAKEN”) signals, divided by the total number of signals, over a given lookback window.\\
	The \textit{AIFX index} exhibits significant cross-sectional predictive power. A simple trading strategy, called the \textit{AIFX strategy}, that goes long currencies with a strong \textit{AIFX index} and short currencies with a weak \textit{AIFX index} produces an annualized Sharpe ratio larger than 0.7. Moreover, after controlling for traditional currency factors like dollar, dollar carry, carry, momentum, and value, I uncover a statistically significant alpha that accounts for 74\% of the \textit{AIFX strategy}’s average return. I further validate the result through a panel regression exercise, showing that the \textit{AIFX index} effectively predicts next month exchange rate returns. Taken together, these findings suggest that AI can help uncover previously underexplored sources of return predictability in the FX market, thus shedding light on the role of economic fundamentals, as advocated by the theoretical literature. \newline
	I conduct multiple robustness checks to ensure the reliability of the baseline results. First, while the main analysis uses GPT-4o to interpret data releases, I replicate the entire exercise using DeepSeek V3 to test the sensitivity of the core findings to the choice of AI model. The results remain consistent, with both models exhibiting very similar performance. Second, I construct an alternative measure, the \textit{Weighted AIFX index}, which assigns a weight to each directional signal based on its estimated level of importance. I use this weighted index as the signal in a cross-sectional trading strategy and find that the core results remain robust. Third, I also implement a time-series trading strategy as an alternative to the cross-sectional strategy used in the baseline specification. The time-series strategy also generates economically meaningful Sharpe ratios, and its performance remains significant after controlling for benchmark currency factors. \newline
	A major concern when using LLMs for prediction tasks is look-ahead bias, which occurs when a model is trained using information not available at the time of the prediction. As a result, the model’s performance may look better than it would be in real-time. To mitigate this concern, I implement four different exercises. In the first one, I investigate whether the AI model may implicitly “remember” the timing of a data release. Specifically, I use the same data that was fed to the AI model to generate the directional signals, but this time I ask it to indicate the year (not the exact date) when the data was released. If the AI model can recall the timing of the release, we should expect it to correctly identify the release year in a high proportion of cases. However, the distribution of years guessed by the AI model differs markedly from the true distribution of data releases in the dataset. I show that only 5.6\% of the model’s guesses are correct on average within each year. In the second exercise, I exploit the fact that the knowledge cut-off date for GPT-4o is October 2023, while for GPT-3.5 it is September 2021. This two-year gap provides an opportunity to examine whether the relative performance of the two models differs significantly. Specifically, I compare their performance during the period from 1996 to 2021, covered by both models' training sets, to the period from 2021 to 2023, which only GPT-4o was trained on. If look-ahead bias were present, we would expect a sharp decline in the relative performance of GPT-3.5 after its training period ends, compared to GPT-4o. To test this, I conduct a difference-in-differences analysis and find that the relative performance of the two models does not differ significantly across the two periods. In the third exercise, I test for the possibility that the AI model may have a memory of the overall relationship between exchange rate returns and certain macro-variables. For example, if the model was trained during a period when inflation and currency returns were positively correlated, it might predict higher exchange rates in response to rising inflation, based on memory and not reasoning. Therefore, I investigate whether the AI model has a memory of the realized correlation of macro variables with next month currency returns over the sample period, but find no evidence indicating so. In the fourth exercise, I construct a portfolio based on what the AI model can remember about monthly currency returns during the sample period, referred to as the \textit{pure hindsight portfolio}, and use it as a control factor. I find that the return of the \textit{AIFX strategy} is orthogonal to the return of the \textit{pure hindsight portfolio}. Overall, these findings collectively suggest that the AI model’s performance is unlikely to be driven by look-ahead bias, and should reflect genuine reasoning based on the information available at the time of prediction. \newline
	After establishing the predictive power of the AI-derived variables, I investigate the underlying sources of this predictability. This step is crucial from an economic standpoint, as it sheds light on the possible mechanisms through which fundamentals influence exchange rate movements. First, I find that \textit{Inflation data}, \textit{Employment data}, and \textit{Broad economic activity indicators} are the most important categories for forecasting exchange rates. These variables are closely linked to the monetary policy framework proposed by \citet{TAYLOR:1993}, suggesting that central banks' policy responses to economic conditions play a pivotal role in exchange rate determination. This interpretation is supported by prior empirical studies such as \citet{CLARIDA/WALDMAN:2007}, \citet{MOLODTSOVA/PAPELL:2009}, and \citet{ENGLE/WU:2024}, which document the predictive power of Taylor-rule fundamentals for exchange rates. Second, I find that the predictive signal is largely driven by positive news (news implying currency appreciation) rather than negative news (implying depreciation). Further analysis reveals that negative news tends to trigger a stronger immediate market reaction than positive news, consistent with the findings of \citet{ANDERSEN/ETAL:2003}. As a result, negative news may leave less room for delayed exchange rate adjustments, thereby reducing its predictive content at longer horizons. A plausible interpretation is that this asymmetry in market reaction is partly driven by the way central banks implement monetary policy. In particular, several studies have documented that monetary authorities tend to respond more aggressively to negative output gaps than to positive ones, leading to a general bias toward lower interest rates \citep{DOLADO/ETAL:2004, BRUGGEMANN/RIEDEL:2011, HOFMANN/BOGDANOVA:2012, KOMLAN:2013}. The political economy of monetary policy also reinforces this asymmetric tendency. Rate hikes can be politically unpopular as they may slow the economy or increase borrowing costs. This asymmetric policy stance can influence investors' expectations, prompting stronger immediate reaction to negative news and contributing to the asymmetric predictive power documented in this study. Taken together, the empirical findings point to the importance of the Taylor rule and monetary policy in explaining exchange rate movements. Nonetheless, alternative explanations cannot be definitively ruled out.\newline
	This research contributes to two strands of literature. The first involves the well-known “exchange rate disconnect” puzzle, first observed by \citet{Meese/Rogoff:1983}. Their findings, seen as shocking at the time, prompted a large literature that re-examined the robustness of the results \citep{MARK:1995, KILIAN:1999, CHEUNG/ETAL:2005, MOLODTSOVA/PAPELL:2009}. However, the early empirical studies were inconclusive in addressing the puzzle. A notable contribution in this context is \citet{ENGEL/WEST:2005}, who offer a potential resolution. They demonstrate analytically that exchange rates can be consistent with present value asset pricing models and follow a process arbitrarily close to a random walk if certain conditions are met. Following this, \citet{ENGEL/ETAL:2007} present a defense of exchange rate models by arguing that a random walk model is a tough benchmark to beat and propose alternative methods for evaluating the performance of exchange rate models. In addition, recent empirical studies suggest a connection between currency returns and countries’ external imbalances \citep{GOURINCHAS/REY:2007, DELLACORTE/ETAL:2012, DELLACORTE/ETAL:2016}, sovereign risk \citep{AUGUSTIN/ETAL:2020, DELLACORTE/ETAL:2022, DELLACORTE/ETAL:2023}, the output gap \citep{COLACITO/ETAL:2020}, macroeconomic uncertainty \citep{BERG/MARK:2018, DELLA/KRECETOVS:2024}, and unemployment \citep{NUCERA:2017}. Notably, the pattern of predictability documented in this paper is consistent with the findings of \citet{DAHLQUIST/HASSELTOFT:2020}, who examine how past trends in key macroeconomic indicators, referred to as economic momentum, can predict currency returns. This paper advances the existing literature by using a novel AI-powered methodology to analyze an expanded set of economic indicators. This innovative approach helps uncover new predictability patterns and provides new insights into exchange rate movements.\newline Second, this project adds to the body of literature on novel research methods in financial economics that leverage generative AI. For example, \citet{EISFELDT/SCHUBERT:2024} conduct a comprehensive survey of how this emerging technology can decrease the time and costs associated with traditional research designs in finance while enabling novel analytical approaches. Emerging applications include generating data embeddings \citep{GABIAX/ETAL:2024, KIM:2024}, text classification \citep{CHANGAND/ETAL:2024, KROCKENBERGER/ETAL:2024}, retrieval-augmented generation \citep{BARTIK/ETAL:2024, CHEN/WANG:2024}, simulating agent behavior \citep{HORTON:2023, FEDYK/ETAL:2024, Hewitt:2024}, and hypothesis generation \citep{SIETAL:2024, LUDWIG/MULLAINATHAN:2024}. Specifically, the prompting technique employed in this study is most similar to the approaches used in the following papers. \citet{Bybee:2023} uses AI to generate economic expectations from historical news data spanning 120 years. In addition, \citet{LOPEZLIRA/TANG:2023} and \citet{CHEN/ETAL:2023} explore the ability of generative AI, specifically ChatGPT, to predict stock price movements based on sentiments extracted from business news headlines. In contrast to these papers, which focus on sentiment extraction from textual data, this study does not aim to extract signals from text. Instead, it relies on structured, numerical data, and the AI model is prompted to generate analysis based on the numerical values of economic data releases. Overall, this paper contributes to the existing literature by demonstrating AI's capability to analyze large volumes of structured data in the context of currency markets. In addition, it introduces new techniques to address look-ahead bias.\newline
	The remainder of the paper is organized as follows; \mysec{sec:data} presents the data; \mysec{sec:Analysing Data Releases} outlines the construction of the AI-powered variables; \mysec{sec:Performance} evaluates the predictive power of these variables; \mysec{sec:Look-ahead Bias; Reasoning or Memorization} investigates the issue of look-ahead bias; \mysec{sec:Mechanism} explores the underlying mechanisms driving the predictability; and \mysec{sec:Conclusions} concludes. A separate \mysecIAheader{appendix} provides additional results not included in the main body of this paper.
	\section{Data}   \label{sec:data} 
	I focus on G-10 currencies that include the United States dollar (USD), Euro (EUR), Japanese yen (JPY), British pound sterling (GBP), Swiss franc (CHF), Canadian dollar (CAD), Australian dollar (AUD), New Zealand dollar (NZD), Swedish krona (SEK), and Norwegian krone (NOK). I limit my focus to these currencies because of the long history of economic data releases available. To measure the economic fundamentals of each currency, I have collected the economic calendar data from Investing.com. The economic calendar aggregates key economic data releases, such as GDP reports, employment statistics, inflation readings, central bank decisions, and other economic indicators (544 unique indicators), across multiple countries. It provides, for each data release, the realized value, the previous figure, and the consensus forecast. The dataset spans from January 1996 to October 2024 and comprises a total of 174,820 data points. \mytab{tab:data:headline frequency table} displays the number of observations collected for each currency. In addition, I have collected end-of-day (London time) exchange rates and one-month forward rates from Bloomberg. Notably, there are nine exchange rates in the cross-section, and all rates are defined as the amount of U.S. dollars (USD) required to purchase one unit of foreign currency (FCU). \mytab{tab:fx returns summary stats} reports summary statistics for exchange rate returns.
	\begin{center}
		\textsc{\mytab{tab:data:headline frequency table} and \mytab{tab:fx returns summary stats} about here} \vspace{0cm}
	\end{center} 
	\section{Analysing Data Releases}   \label{sec:Analysing Data Releases} 
		\subsection{AI as a financial Analyst}   \label{sec:ChatGPT as a financial Analyst} 
		For each data release in the economic calendar data, I feed structured prompts, as in \myprp{prompt:VariableConstruction:financial analyst prompt}, to GPT-4o using APIs. The prompt contains the title, realized value, previous figure, and the consensus forecast of the data release and instructs the AI model to generate a concise analysis and a directional signal indicating whether the data release implies the currency would STRENGTHEN, WEAKEN, or has an INSIGNIFICANT OR UNCERTAIN impact. In \myprp{prompt:VariableConstruction:financial analyst prompt}, \{currency\} will correspond to the currency associated with the data release. Notably, I exclude any information about the time or date of the release.
		\begin{prompt}[H] 
			\centering
			\fbox{ 
				\parbox{0.97\textwidth}{ 
					\textit{\textbf{Prompt:} Forget all previous instructions. You are a financial analyst. You will be given a news headline. Please analyze the news's effect on \{currency\} and provide a brief response. Then indicate whether this headline implies \{currency\} will STRENGTHEN, WEAKEN, or have an INSIGNIFICANT OR UNCERTAIN effect. Generate the output in this format: \{(ANALYSIS: short analysis discussing the channel), (DIRECTION: one of STRENGTHEN, WEAKEN, INSIGNIFICANT OR UNCERTAIN)\}}
				}
			}
			\caption{}
			\label{prompt:VariableConstruction:financial analyst prompt}
		\end{prompt}
		\vspace{-20pt}As illustrated in \myexmpl{examples:VariableConstruction:first example}, \myexmpl{examples:VariableConstruction:second example}, and \myexmpl{examples:VariableConstruction:third example}, the output explains the mechanism of how the release could affect the associated currency and indicates the potential direction of change.
		\begin{examples}[H] 
			\centering
			\fbox{ 
				\parbox{0.97\textwidth}{ 
					\begin{itemize}[left=0pt]
						\setlength{\itemsep}{-5pt} 
						\item Currency: AUD
						\item \textbf{Input}: \textit{News headline:} Trade Balance (monthly), \textit{Actual:} {4.502B}, \textit{Forecast:} {6.100B}, \textit{Previous:} {7.180B}
						\item \textbf{Output}: \textit{Analysis}: The trade balance figure came in lower than both the forecast and the previous month, indicating a smaller than expected surplus. This could suggest that exports have decreased or imports have increased more than anticipated, potentially signaling weaker economic activity. A lower trade balance can negatively impact the currency as it implies that fewer Australian dollars are demanded in international markets to purchase Australian goods and services., \textit{Direction}: WEAKEN.
					\end{itemize}
				}
			}
			\caption{}
			\label{examples:VariableConstruction:first example}
		\end{examples}
		\vspace{-20pt}
		\begin{examples}[H] 
			\centering
			\fbox{ 
				\parbox{0.97\textwidth}{ 
					\begin{itemize}[left=0pt]
						\setlength{\itemsep}{-5pt} 
						\item Currency: CAD
						\item \textbf{Input}: \textit{News headline:} GDP (MoM), \textit{Actual:} {0}, \textit{Forecast:} {0.001}, \textit{Previous:} {0.006}
						\item \textbf{Output}: \textit{Analysis}: The Canadian GDP month-on-month growth rate is recorded at 0\%, which is below the forecast of 0.1\% and significantly lower than the previous month's 0.6\%. This stagnation signals a slowdown in economic growth and may dampen investor sentiment towards the Canadian dollar. As GDP is a key indicator of economic health, a lower than expected performance can lead to expectations of a more accommodative monetary policy from the Bank of Canada, potentially putting downward pressure on the CAD., \textit{Direction}: WEAKEN.
					\end{itemize}
				}
			}
			\caption{}
			\label{examples:VariableConstruction:second example}
		\end{examples}
		\vspace{-20pt}
		\begin{examples}[H] 
			\centering
			\fbox{ 
				\parbox{0.97\textwidth}{ 
					\begin{itemize}[left=0pt]
						\setlength{\itemsep}{-5pt} 
						\item Currency: GBP
						\item \textbf{Input}: \textit{News headline:} CPI (MoM), \textit{Actual:} {0.004}, \textit{Forecast:} {0.003}, \textit{Previous:} {0.002}
						\item \textbf{Output}: \textit{Analysis}: The Consumer Price Index (CPI) measures the change in the price level of a basket of goods and services. A higher-than-expected CPI indicates increased inflationary pressures. The actual CPI of 0.004 surpasses both the forecast of 0.003 and the previous value of 0.002. This suggests stronger inflationary trends, potentially leading to expectations of tighter monetary policy by the Bank of England. Higher interest rates generally lead to currency appreciation as they attract foreign investment seeking higher returns., \textit{Direction}: STRENGTHEN.
					\end{itemize}
				}
			}
			\caption{}
			\label{examples:VariableConstruction:third example}
		\end{examples}
		 \vspace{-20pt}To clarify the terminology, throughout this paper, I refer to data releases with the direction labeled as STRENGTHEN in the output as positive news, those labeled as WEAKEN as negative news, and those labeled as INSIGNIFICANT OR UNCERTAIN as neutral news. \mytab{tab:data:headline frequency table} presents a detailed breakdown of the count and percentage of positive, negative, and neutral news for each currency during the sample period, providing insights into the distribution of outputs from the AI model.
		\subsection{Variable Construction}			\label{sec:Variable Construction} 
		I follow a simple and intuitive approach to construct three AI-powered variables. Suppose we are at time \(t\) and let \(\tau\) denote the lookback period. Therefore, the time interval \(L=(t-\tau,t]\) would be the lookback window at time \(t\). Based on this, I construct the following variables for currency \(c\):
		\begin{align}
			\text{Strength}_{c,t,\tau} &= \frac{\text{Number of \textbf{positive} news related to currency $c$ in $L$}}{\text{Total number of news related to currency $c$ in $L$}} 
			\label{eq:VariableConstruction:definition of pos ratio}\\[8pt]
			\text{Weakness}_{c,t,\tau} &= \frac{\text{Number of \textbf{negative} news related to currency $c$ in $L$}}{\text{Total number of news related to currency $c$ in $L$}} 
			\label{eq:VariableConstruction:definition of neg ratio}
		\end{align}
		\begin{equation}
			\text{AIFX}_{c,t,\tau} = \text{Strength}_{c,t,\tau} - \text{Weakness}_{c,t,\tau}
			\label{eq:VariableConstruction:definition of diff ratio}
		\end{equation}
		The \textit{Strength ratio} captures the proportion of positive news, while the \textit{Weakness ratio} measures the proportion of negative news. The \textit{AIFX index} is defined as the net balance between positive and negative news, providing a single composite metric of implied currency strength derived from AI-classified data.
		\section{Performance Evaluation}				 			\label{sec:Performance}
		In this section, I analyze the predictive power of the AI-derived variables. I begin by constructing cross-sectional trading strategies that use the \textit{AIFX index} as the signal, evaluated across a range of lookback periods. The performance of these strategies is then assessed relative to common currency factors. To formally test the statistical significance of the \textit{AIFX index} in predicting future returns, I estimate panel regressions. To ensure robustness, the entire analysis is replicated using DeepSeek-V3, a leading alternative to the baseline model GPT-4o. Next, I introduce an alternative specification of the AI-powered variables by weighting each data release according to its estimated economic importance. Finally, I implement a time-series strategy based on the same signal and further decompose the predictive component to isolate the role of U.S. dollar fundamentals.
		
		\subsection{Cross-sectional Trading Strategy}  \label{sec:Cross Sectional Trading Strategy: Diff Ratio as the signal}
		I assess the predictive power of the \textit{AIFX index} (as defined in \myeq{eq:VariableConstruction:definition of diff ratio}), using stylized cross-sectional trading strategies. Notably, to evaluate the sensitivity of performance to the length of lookback window, I consider lookback periods of 1 to 60 months. Specifically, for each choice of lookback period, currencies are sorted by their \textit{AIFX index} at the end of each month and I take long positions in the top two currencies with the highest \textit{AIFX index} and short positions in the bottom two with the lowest. I refer to this strategy as the \textit{AIFX strategy}. \myfig{fig:Performance:sharpe ratio cs strategy diff} presents the annualized Sharpe ratios of the strategy across different lookback periods.
		\begin{center}
			\textsc{\myfig{fig:Performance:sharpe ratio cs strategy diff} about here} 
		\end{center}
		 The \textit{AIFX strategy} consistently yields positive Sharpe ratios across all lookback periods, with economically significant performance in most cases. Predictive power appears particularly strong for lookback periods of 36 to 60 months. Additionally, \mytab{tab:Performance:cs AI strategy diff performance statistics} reports the performance statistics of the \textit{AIFX strategy}, including the mean, standard deviation, skewness, excess kurtosis and first-order autocorrelation of returns. \myfig{fig:Performance:cs strategy performance over time 48 months diff positive} illustrates the dollar value of an initial \$1 investment in the \textit{AIFX strategy}\footnote{To save space, \myfig{fig:Performance:cs strategy performance over time 48 months diff positive} also displays the cumulative return of a strategy that uses \textit{Strength ratio} (defined in \myeq{eq:VariableConstruction:definition of pos ratio}) as the signal. This strategy will be discussed in \mysec{sec:CS Trading Strategy: Positive and Negative Ratio as Signal}.}. A visual inspection of the figure indicates a general upward trend in performance, with gains distributed relatively evenly throughout the sample period. \myfig{fig:portfolio composition 4 months CS diff ratio} shows the portfolio composition over time. The strategy exhibits moderate turnover, implying that transaction costs are unlikely to significantly erode returns.
		\begin{center}
			 \textsc{\mytab{tab:Performance:cs AI strategy diff performance statistics}, \myfig{fig:Performance:cs strategy performance over time 48 months diff positive} and \myfig{fig:portfolio composition 4 months CS diff ratio} about here} 
		\end{center}
		To further examine the performance, I run contemporaneous regressions based on:
		\begin{align}
			RX_{t,\tau} &= \alpha_{\tau} + \beta_{1,\tau} \text{Dollar}_{t} + \beta_{2,\tau} \text{Dollar 	Carry}_{t} + \beta_{3,\tau} \text{Carry}_{t} + \notag\\ &\quad  \beta_{4,\tau} \text{Momentum}_{t}  +  \beta_{5,\tau} \text{Value}_{t}  + \epsilon_{t,\tau}
			\label{eq:Performance:regression for performance over fx strategies CS Strategy Diff}
		\end{align}
		where \( RX_{t,\tau} \) denotes the monthly excess return of the \textit{AIFX strategy} with lookback period \( \tau \); \textit{Dollar} is a long-only portfolio that takes equal-weighted long positions in all currencies; \textit{Dollar Carry} is a directional strategy that goes long (short) all currencies when the average forward discount is positive (negative); \textit{Carry}, \textit{Momentum}, and \textit{Value} are cross-sectional strategies that rank currencies by their forward discount, previous month's return, and five-year return, respectively. The construction of currency factors is further detailed in \mysecIA{Appendix:Definition of FX factors} and their performance statistics are reported in \mytabIA{tab:Performance:fx strategy performance statistics}. I report the regression results of \myeq{eq:Performance:regression for performance over fx strategies CS Strategy Diff} in \mytab{tab:Performance:cs performance over fx strategies}.
		\begin{center}
			\textsc{\mytab{tab:Performance:cs performance over fx strategies} about here}  
		\end{center}
		 The findings indicate that the \textit{AIFX strategy}’s returns are not fully explained by the common currency factors, and this conclusion holds across different lookback periods. For example, with a 48-month lookback period, the strategy yields a statistically significant alpha that accounts for 74\% of the average return of the strategy, suggesting that only about one-quarter of the return is subsumed by traditional currency factors. Notably, for lookback periods of 54 and 60 months, the alpha becomes only marginally significant, as the Value factor gains more explanatory power. Overall, the analysis presented in this section provides strong evidence that the \textit{AIFX index} possesses significant predictive power. These findings suggest that AI can help uncover previously underexplored sources of return predictability in the FX market, contributing to a renewed connection between exchange rates and underlying economic fundamentals.

		\subsection{Panel Regression}	\label{sec:Panel Regression}
		To assess the statistical significance of the predictive power of the \textit{AIFX index}, I estimate the following panel regression model for each choice of lookback period:
		\begin{equation}
			R_{c,t+1} = \alpha_{t,\tau} + \beta_{\tau} \text{AIFX}_{c,t,\tau} + \epsilon_{c,t,\tau}
			\label{eq:Performance:regression return on diff ratio cs}
		\end{equation}
		In \myeq{eq:Performance:regression return on diff ratio cs}, \(R_{c,t+1}\) represents the monthly excess return of currency \(c\) at time \(t+1\), and \(\text{AIFX}_{c,t,\tau}\) denotes the \textit{AIFX index} for currency \(c\) at time \(t\) when the lookback period is set equal to \(\tau\). Observations for monthly returns are non-overlapping, and both \(R_{c,t+1}\) and \(\text{AIFX}_{c,t,\tau}\) are measured at the end of calendar months. Accordingly, this regression examines whether the \textit{AIFX index} of a currency at the end of a given month can predict the currency's return in the subsequent month. Time fixed effects (\(\alpha_t\)) are included to simulate a cross-sectional setting where the focus is not on average returns, but rather on the cross-sectional differences in currency returns. The primary coefficient of interest is \(\beta\), which captures the predictive relationship between the \textit{AIFX index} and future returns. Consistent with the methodology in \mysec{sec:Cross Sectional Trading Strategy: Diff Ratio as the signal}, I consider a range of lookback periods to investigate how the predictive power of the \textit{AIFX index} varies with the length of lookback window. \myfig{fig:Performance:t-stat for regression of return on diff ratio CS} presents the \(t\)-statistics associated with \(\beta\) from \myeq{eq:Performance:regression return on diff ratio cs} across different lookback periods. The results indicate that the predictive relationship is statistically significant for most lookback periods. In addition, the t-statistic profile in \myfig{fig:Performance:t-stat for regression of return on diff ratio CS} resembles the Sharpe-ratio pattern in \myfig{fig:Performance:sharpe ratio cs strategy diff}. Furthermore, \myfigIA{fig:Performance:t-stat for regression of spot return on diff ratio CS} displays the regression results when spot returns, rather than excess returns, are used as the dependent variable. While the coefficient of \textit{AIFX} is still statistically significant across a range of lookback periods, its predictive power is somewhat weaker than in the excess-return specification. The findings of this section provide additional evidence that the \textit{AIFX index} possesses predictive ability for future exchange rate returns.
		
		\begin{center}
			\textsc{\myfig{fig:Performance:t-stat for regression of return on diff ratio CS} about here}  
		\end{center}
	
		\subsection{GPT-4o vs DeepSeek-V3}
		The core analysis of this paper makes use of GPT-4o. As a robustness check, I also consider DeepSeek-V3, a key competitor to GPT-4o. \myfig{fig:Performance:GPT-4o vs DeepSeek-V3} compares the Sharpe ratios of the \textit{AIFX strategy} constructed using GPT-4o and DeepSeek-V3. The results reveal similar performance, with both exercises delivering comparable Sharpe ratios across different lookback periods and exhibiting a consistent pattern. These findings provide supporting evidence that the baseline results are robust to the choice of alternative AI models. 
		\begin{center}
			\textsc{\myfig{fig:Performance:GPT-4o vs DeepSeek-V3} about here}  
		\end{center}
		
		\subsection{Alternative ways of constructing the AIFX index}
		I now examine whether the results are robust to an alternative method for constructing the \textit{AIFX index}. In the core analysis, to construct the variables, I only consider the count of positive, negative, or neutral news items. While this method is straightforward and easy to interpret, it overlooks the heterogeneity in the economic significance of different data releases. To address this limitation, I adopt an alternative approach that incorporates the relative importance of each data release. Using \myprp{prompt:Performance:WeightedDiffRatio}, I further ask GPT-4o to assign a numerical score from 1 to 100 indicating the magnitude of the expected impact of the data release on the exchange rate. \myexmpl{examples:Performance:Weighting example I} and \myexmpl{examples:Performance:Weighting example II} illustrate this methodology in more detail. The resulting score, scaled by 100, is then used as a weight for each data release in the construction of the variables.
		\begin{prompt}[H] 
			\centering
			\fbox{ 
				\parbox{0.97\textwidth}{ 
					\textit{\textbf{Prompt:} Forget all previous instructions. On a scale of 1 to 100, how big is the potential impact of the data release on the currency? Provide a short explanation of your reasoning. Generate the output in this format: \{(ANALYSIS: short analysis discussing your reasoning), (Score: a number between 1 and 100)\}}
				}
			}
			\caption{}
			\label{prompt:Performance:WeightedDiffRatio}
		\end{prompt} 
		\vspace{-20pt}\begin{examples}[H] 
			\centering
			\fbox{ 
				\parbox{0.97\textwidth}{ 
					\begin{itemize}[left=0pt]
						\setlength{\itemsep}{-5pt} 
						\item Currency: CAD
						\item \textbf{Input}: \textit{Data release headline: Core CPI (YoY)}
						\item \textbf{Output}: \textit{Analysis}: The Core Consumer Price Index (CPI) is a critical indicator of inflation, excluding volatile items like food and energy. A change in the Core CPI can signal shifts in purchasing power and inflation trends, influencing the Bank of Canada's monetary policy decisions, such as interest rate adjustments. Such decisions, in turn, can significantly impact the Canadian dollar's strength. If the Core CPI shows unexpected inflation, it could lead to speculation on rate hikes, strengthening CAD, or if it suggests low inflation, it may lead to rate cuts, weakening CAD., \textit{Score}: 85.
					\end{itemize}
				}
			}
			\caption{}
			\label{examples:Performance:Weighting example I}
		\end{examples}
		\vspace{-20pt}\begin{examples}[H] 
			\centering
			\fbox{ 
				\parbox{0.97\textwidth}{ 
					\begin{itemize}[left=0pt]
						\setlength{\itemsep}{-5pt} 
						\item Currency: GBP
						\item \textbf{Input}: \textit{Data release headline: U.K. Construction Output (YoY)}
						\item \textbf{Output}: \textit{Analysis}: The construction sector is a significant component of the U.K. economy, but it is not the largest. A year-on-year increase or decrease in construction output can provide insights into economic health and potential future economic conditions, which could impact investor confidence in the GBP. However, it is not as impactful as major economic indicators like GDP, inflation, or employment data. Thus, while it holds relevance, its direct influence on the GBP is moderate.), \textit{Score}: 45.
					\end{itemize}
				}
			}
			\caption{}
			\label{examples:Performance:Weighting example II}
		\end{examples}
		\vspace{-20pt}Similar to \mysec{sec:Variable Construction}, I construct three AI-powered variables for currency \(c\) at time \(t\) and for a lookback period of \(\tau\):
		\begin{align*}
			\text{Weighted Strength}_{c,t,\tau} 
			&= \frac{\text{Weighted sum of positive news related to currency $c$ in $L$}}{\text{Weighted sum of all news related to currency $c$ in $L$}} 
			\\[10pt]
			\text{Weighted Weakness}_{c,t,\tau} 
			&= \frac{\text{Weighted sum of negative news related to currency $c$ in $L$}}{\text{Weighted sum of all news related to currency $c$ in $L$}}
		\end{align*}
		\vspace*{-20pt}
		\begin{equation}
			\text{Weighted AIFX}_{c,t,\tau} = \text{Weighted Strength}_{c,t,\tau} - \text{Weighted Weakness}_{c,t,\tau}
			\label{eq:Performance:definition of weighted diff ratio}
		\end{equation}
		\myfig{fig:Performance:diff ratio weighted squared and unweighted} compares the performance of the strategies based on the \textit{AIFX index} and the \textit{Weighted AIFX index} across a range of lookback periods. The figure shows that incorporating weights enhances the Sharpe ratios for nearly all lookback windows. This finding suggests that the baseline strategy can be further improved by refining the specification of the input variables. Moreover, the results demonstrate that the predictive performance of the strategy is robust to alternative methods of constructing the AI-powered variables.
		\begin{center}
			\textsc{\myfig{fig:Performance:diff ratio weighted squared and unweighted} about here}  
		\end{center}
		
		\subsection{Time-Series Strategies}				\label{Time-Series Strategies}		
		So far, I have worked with cross-sectional trading strategies based on the \textit{AIFX index}. Here, I consider time-series strategies based on the same signal to examine an alternative portfolio formation approach. Let \(\text{AIFX}_{c,t,\tau}\) be the indicator for currency \(c\) and \(\text{AIFX}_{US,t,\tau}\) be the corresponding value for the U.S. dollar. I then define the following variable:
		\begin{equation}
			\text{Diff AIFX}_{c,t,\tau} = \text{AIFX}_{c,t,\tau} - \text{AIFX}_{US,t,\tau}
			\label{eq:Performance:definition of diff currency usd ratio}
		\end{equation}
		For each choice of lookback period \(\tau\), at  the end of each month \(t\), I take a long position in currency \(c\) if \(\text{Diff AIFX}_{c,t,\tau}\) is positive and a short position if negative. Thus, portfolio weights are either \(+1\) or \(-1\), depending on the sign of the signal.\footnote{To maintain comparability in volatility with the cross-sectional strategy, I scale these sign-based weights by \(N\), the number of currencies included in the portfolio.} Unlike the cross-sectional strategy, which is dollar neutral, the time-series strategy may carry exposure to the dollar, either positive or negative. Portfolios are then rebalanced every month. As in \mysec{sec:Cross Sectional Trading Strategy: Diff Ratio as the signal}, I evaluate performance across lookback windows ranging from 1 to 60 months. \myfigIA{fig:Performance:sharpe ratio TS strategy diff currency usd} reports the Sharpe ratios of the time-series strategy that uses \textit{Diff AIFX index} as the trading signal. While the strategy exhibits somewhat lower Sharpe ratios compared to its cross-sectional counterpart, performance remains economically meaningful across a wide range of lookback periods. Notably, the predictive power of the signal is stronger for longer lookback windows, particularly those between 45 and 60 months. To further investigate the nature of time-series predictability, I decompose \textit{Diff AIFX index} into its components and examine \(\text{AIFX}_{c}\) and \(\text{AIFX}_{USD}\) separately. Specifically, I follow the same time-series portfolio formation approach, but in one case use \(\text{AIFX}_{c}\) as the trading signal, and in the other, use \(\text{AIFX}_{USD}\). \myfigIA{fig:Performance:sharpe ratio TS strategy diff currency diff usd} presents the Sharpe ratios of the two strategies. The results show that the strategy based on \(\text{AIFX}_{USD}\) delivers economically significant Sharpe ratios, generally in the range of 0.4 to 0.5, across a wide spectrum of lookback periods. In contrast, the strategy based on \(\text{AIFX}_{c}\) does not yield Sharpe ratios significantly different from zero; moreover, the performance fluctuates in sign across different lookback periods. These findings suggest that the time-series predictability is primarily driven by \(\text{AIFX}_{USD}\). This may reflect either the dominant role of U.S.-related economic fundamentals relative to domestic fundamentals in forecasting currency returns, or simply the greater availability of data related to U.S. fundamentals. In addition, \mysecIA{Appendix: Time-Series Strategy; More Results} provides further insights by reporting the strategy’s cumulative returns, key performance statistics, and the results of regressions of the strategy’s returns on benchmark currency factors. To conclude, the findings of this section provide further evidence that the \textit{AIFX index} contains valuable predictive information, and that its predictive power is robust across both cross-sectional and time-series settings.
		\section{Look-ahead Bias: Reasoning or Memorization?} \label{sec:Look-ahead Bias; Reasoning or Memorization}
		A major concern when using LLMs for forecasting tasks is look-ahead bias, which occurs when a model is trained using information not available at the time of prediction. This can make the model’s performance appear better than it would be in real-time. To address this concern, I design and implement four different tests to determine whether the model’s performance reflects genuine reasoning or merely the recall of memorized information. First, I investigate whether the model might implicitly “remember” the timing of a data release. Second, I exploit the difference in knowledge cut-off dates between GPT-4o and GPT-3.5 to test for performance divergence in the post-training period of GPT-3.5. Third, I assess whether the model has memorized the correlation between macroeconomic variables and future currency returns. Finally, I construct a portfolio based entirely on what the AI model remembers of currency returns during the sample period and use it as a benchmark to control for predictive signals that are contaminated by look-ahead bias.
		\subsection{Guess the Year} 	\label{sec:Guess the Year}
		In the core analysis, I do not provide any information about the timing or date of the release as part of the input. Nevertheless, there remains the possibility that the AI model could infer the release date from the provided inputs. To test for this possibility, I conduct an experiment using the same data that was originally fed into the AI model via \myprp{prompt:VariableConstruction:financial analyst prompt}, but instead of asking for an economic analysis, I ask the model to identify the year (not the exact date) in which the data release occurred. This exercise is implemented using \myprp{prompt:LookAheadBias:guess the year for lookahead bias analysis}.
		 \begin{prompt}[H] 
		 	\centering
		 	\fbox{ 
		 		\parbox{0.97\textwidth}{ 
		 			\textit{\textbf{Prompt:} Forget all previous instructions. You are a financial analyst. You will be given a news headline related to \{currency\}. The news headline was published sometime between 1996 and 2024 (inclusive). Based on the information available and your memory, indicate the year this headline was published. Generate the output in this format: \{(YEAR: a 4-digit number indicating the year)\}}
		 		}
		 	}
		 	\caption{}
		 	\label{prompt:LookAheadBias:guess the year for lookahead bias analysis}
		 \end{prompt}
		 \vspace{-20pt}If the AI model is indeed able to recall the timing of the release, we would expect it to correctly identify the release year in a high proportion of cases. \myfig{fig:LookAheadBias:economic events per year} displays the number of data releases published each year in the dataset, representing the true distribution of data releases over time. In contrast, \myfig{fig:LookAheadBias:how many guess each year} presents the distribution of GPT-4o’s year-level guesses. It is evident that the model’s guessed distribution diverges significantly from the actual distribution of data releases.
		 \begin{center}
		 	\textsc{\myfig{fig:LookAheadBias:economic events per year} and \myfig{fig:LookAheadBias:how many guess each year} about here}  
		 \end{center}
		 To provide a more nuanced view of the model’s classification performance at the year level, for each year \(y\), I calculate three standard evaluation metrics commonly used in the machine learning literature:
		 \begin{align}
		 	\text{Precision}_y 
		 	&= \frac{\text{Correct guesses for year } y}{\text{Total instances guessed as year } y},\\[6pt]
		 	\text{Recall}_y 
		 	&= \frac{\text{Correct guesses for year } y}{\text{Total actual data releases in year } y}
		 \end{align}
		 \vspace{-25pt}
		 \begin{equation}
		 	\text{F1}_y 
		 	= \frac{2}{\frac{1}{\text{Precision}_y} + \frac{1}{\text{Recall}_y}}.
		 	\label{eq:Lookahead:definition of f1 score}
		 \end{equation}
		 Precision\(_y\) measures how accurate the model is when it predicts year \(y\). That is, among all instances the model guessed as year \(y\), how many were actually correct. Recall\(_y\) measures how well the model identifies instances from year \(y\). It captures the proportion of actual data releases in year \(y\) that the model correctly guessed. F1\(_y\) is the harmonic mean of precision and recall for year \(y\), providing a balanced measure that accounts for both false positives and false negatives. \mytabIA{tab:LookaheadBias:guess_year_detailed_table} reports a breakdown of the model’s classification performance across years. On average, the model achieves a precision of 5.68\%, a recall of 4.17\%, and an F1 score of 2.36\%, which are all very low. Further analysis, reported in \myfigIA{fig:LookAheadBias:f1 score annual return}, investigates whether better classification performance (higher F1 scores) coincides with stronger strategy performance and finds no evidence indicating so. To conclude, the results of this section undermine the hypothesis that the AI model’s predictive power stems from memorization by showing that its signal is not driven by implicit knowledge of historical release dates.
		\subsection{Cut-off Test: GPT-4o vs GPT-3.5} \label{sec:LookAheadBias:KnowledgeCutoff}
		The knowledge cut-off date for GPT-4o is October 2023, whereas for GPT-3.5, it is September 2021. This two-year gap offers a unique opportunity to examine whether the performance of the two models diverges significantly based on their access to post-2021 data. To investigate this, I compare the performance of GPT-4o and GPT-3.5 over two distinct subperiods: (i) 1996–2021, a period fully covered by both models’ training data, and (ii) 2021–2023, a period included only in GPT-4o’s training set. If look-ahead bias is present, then GPT-3.5’s performance should drop sharply after 2021, while GPT-4o maintains its predictive accuracy. To implement this test, I feed both GPT-3.5 and GPT-4o two separate prompts: \myprp{prompt:LookAheadBias:chatgpt3.5vs4/19962021} and \myprp{prompt:LookAheadBias:chatgpt3.5vs4/20212023}. These prompts are identical in structure to \myprp{prompt:VariableConstruction:financial analyst prompt}, but they explicitly indicate the time period in which the data release occurred, either between January 1996 and September 2021 or between September 2021 and October 2023.
		\begin{prompt}[H] 
			\centering
			\fbox{ 
				\parbox{0.97\textwidth}{ 
					\textit{\textbf{Prompt:} Forget all previous instructions. You are a financial analyst. You will be given \textbf{a news headline which was published sometime between January 1996 and September 2021}. Please analyze the news's effect on \{currency\} and provide a brief response. Then indicate whether this headline implies \{currency\} will STRENGTHEN, WEAKEN, or have an INSIGNIFICANT OR UNCERTAIN effect. Generate the output in this format: \{(ANALYSIS: short analysis discussing the channel), (DIRECTION: one of STRENGTHEN, WEAKEN, INSIGNIFICANT OR UNCERTAIN)\}.}
				}
			}
			\caption{}
			\label{prompt:LookAheadBias:chatgpt3.5vs4/19962021}
		\end{prompt}
		\vspace{-20pt}\begin{prompt}[H] 
			\centering
			\fbox{ 
				\parbox{0.97\textwidth}{ 
					\textit{\textbf{Prompt:} Forget all previous instructions. You are a financial analyst. You will be given \textbf{a news headline which was published sometime between September 2021 and October 2023}. Please analyze the news's effect on \{currency\} and provide a brief response. Then indicate whether this headline implies \{currency\} will STRENGTHEN, WEAKEN, or have an INSIGNIFICANT OR UNCERTAIN effect. Generate the output in this format: \{(ANALYSIS: short analysis discussing the channel), (DIRECTION: one of STRENGTHEN, WEAKEN, INSIGNIFICANT OR UNCERTAIN)\}}
				}
			}
			\caption{}
			\label{prompt:LookAheadBias:chatgpt3.5vs4/20212023}
		\end{prompt}
		\vspace{-20pt}Having obtained the outputs from both AI models across the two subperiods, I construct the \textit{AIFX index}, as defined in \mysec{sec:Variable Construction}, for each currency on the last day of each calendar month using a one-month lookback window. In the first exercise, I compute the monthly correlation between the \textit{AIFX index} values generated by GPT-3.5 and GPT-4o over the two periods. \mytab{tab:LookAheadBias:correlation diff ratio before after chatgpt3.5vs4} shows that the correlation between the two models' outputs remains high and largely stable across both periods, indicating no substantial shift in model behavior after 2021. To further assess potential divergence in outputs, I implement a difference-in-differences analysis as specified in \myeq{eq:LookAheadBias:diff-in-diff gpt3.5vs4}. In this setup, \(T_i\) is a treatment indicator equal to 1 for GPT-4o and 0 for GPT-3.5, while \(\text{After}_t\) is a time indicator equal to 1 for the period from September 2021 to October 2023 and 0 for the period from January 1996 to September 2021. The interaction term \(T_i \times \text{After}_t\) captures the differential change in outputs between the two models after GPT-3.5’s training cut-off.
		\begin{equation}
			\text{AIFX}_{i,t} = \beta_0 + \beta_1 T_i + \beta_2 \text{After}_t + \beta_3 T_i \text{After}_t + \epsilon_{it}
			\label{eq:LookAheadBias:diff-in-diff gpt3.5vs4}
		\end{equation}
		If look-ahead bias were at work, we would expect the coefficient \(\beta_3\) to be statistically significant. However, as shown in \mytab{tab:LookAheadBias:diff in diff gpt3.5vs4}, the interaction term is statistically insignificant across specifications. These results suggest that GPT-3.5’s performance remains comparable to GPT-4o even after its training window ends. That said, the relatively short length of the two-year gap may limit the statistical power to detect small differences. Taken together, the evidence from this analysis further supports the conclusion that the AI model's predictive power is not driven by look-ahead bias.
		\begin{center}
			\textsc{\mytab{tab:LookAheadBias:correlation diff ratio before after chatgpt3.5vs4} and \mytab{tab:LookAheadBias:diff in diff gpt3.5vs4} about here} \vspace{0cm}
		\end{center} 
		\subsection{Correlations Between Macro Variables and Currency Returns}		\label{sec:Correlation Estimation}
		Another potential source of look-ahead bias may arise if the AI model has memorized general patterns between exchange rate returns and macroeconomic variables, during its training period. For example, suppose the model was trained on data in which inflation and currency returns were positively correlated. In that case, it might predict higher exchange rates in response to rising inflation, not through active reasoning, but by recalling historical associations. Therefore, I examine whether the AI model has a memory of the realized correlations between key macroeconomic variables and future currency returns over the period from January 1996 to October 2023. This period spans the full sample used in this study and ends at the knowledge cut-off date of GPT-4o. I focus on two macro variables: the monthly consumer price index (CPI) and unemployment rate. Using \myprp{prompt:LookAheadBias:remember the correlaton macro var currency returns}, I ask the AI model to report the correlation between each macro variable and next-month currency returns for each currency\footnote{For this exercise, I only consider currencies with available monthly CPI and unemployment rate data.}. I run this prompt 1,000 times for each currency–macro variable pair to obtain a distribution of the model’s estimated correlations.
		\begin{prompt}[H] 
			\centering
			\fbox{ 
				\parbox{0.97\textwidth}{ 
					\textit{\textbf{Prompt:} Forget all previous instructions. You are a financial analyst. Please indicate what is the correlation between \{country\}'s \{macro\_variable\} and next month \{currency\}USD returns over the period from January 1996 to October 2023. Do not include any extra explanations in the output. Generate the output in this format: \{(CORRELATION: number indicating the correlation)\}.}
				}
			}
			\caption{}
			\label{prompt:LookAheadBias:remember the correlaton macro var currency returns}
		\end{prompt}
		\vspace{-20pt}\myfig{fig:LookAheadBias:GBP CPI(MOM)} and \myfig{fig:LookAheadBias:GBP monthly unemployment rate} display the resulting distributions for GBP, alongside the true realized correlation values computed over the same period. If the AI model had a strong memory of the historical co-movement between macro variables and currency returns, we would expect its estimates to be narrowly concentrated and closely aligned with the true realized value. However, the two figures reveal substantial variation across the model’s estimates, and the means of the distributions are not aligned with the true realized values. Similar patterns are observed across other currencies, although figures are not reported to save space. These findings indicate that GPT-4o does not have a strong memory of the general correlation between macroeconomic variables and future exchange rate returns, further mitigating concerns of look-ahead bias.
		\begin{center}
			\textsc{\myfig{fig:LookAheadBias:GBP CPI(MOM)} and \myfig{fig:LookAheadBias:GBP monthly unemployment rate} about here}  
		\end{center}

		\subsection{Pure Hindsight Portfolio}	\label{sec:Pure Hindsight Portfolio}
		In this subsection, I construct a portfolio based solely on what the AI model may remember about historical currency returns. The aim is to isolate any performance that might arise purely from memorization, rather than reasoning. Specifically, I use \myprp{prompt:LookAheadBias:remember the direction for lookahead bias analysis} to directly ask GPT-4o whether each currency strengthened, weakened, or remained unchanged in a given month of the sample period.
		\begin{prompt}[H] 
			\centering
			\fbox{ 
				\parbox{0.97\textwidth}{ 
					\textit{\textbf{Prompt:} Forget all previous instructions. You are a financial analyst. Please indicate whether the currency \{currency\} has STRENGTHENED, WEAKENED, or was UNCHANGED over the period from \{start\_date\} to \{end\_date\}. Do not include any extra explanations in the output. Generate the output in this format: \{(DIRECTION: one of STRENGTHENED, WEAKENED, UNCHANGED)\}.}
				}
			}
			\caption{}
			\label{prompt:LookAheadBias:remember the direction for lookahead bias analysis}
		\end{prompt} 
		\vspace{-20pt}The output of this prompt reflects the model’s recollection of directional return movements, independent of any economic analysis or reasoning. Using these outputs, I form a trading strategy, referred to as the \textit{pure hindsight portfolio}, by taking long positions in currencies the model indicates will strengthen and short positions in those it indicates will weaken. The portfolio is rebalanced at the end of each calendar month, same as the \textit{AIFX strategy}. This portfolio exhibits highly positively skewed returns (skewness of 2.52), extremely fat tails (excess kurtosis of 18.61), and delivers a Sharpe ratio of 0.91\footnote{The performance statistics are for the period from January 1996 to October 2023}. \myfigIA{fig:LookAheadBias:pure hindsight performance over time} displays the cumulative return of the portfolio over time. Notably, the cumulative return is relatively flat prior to 2008, a period that is less represented in the AI model’s training data, as well as after the knowledge cut-off date in 2023. In contrast, the portfolio performs strongly between 2008 and 2023, with its best performance occurring during the 2008 financial crisis, suggesting that the AI model is better able to recall return directions from this well-covered period. The \textit{pure hindsight portfolio} could serve as a good control for predictive signals and the associated strategy returns that are contaminated by look-ahead bias. Therefore, I regress the excess returns of the \textit{AIFX strategy} on the excess returns of the \textit{pure hindsight portfolio}, as shown in \myeq{eq:LookAheadBias:regression ai strategy on pure hindsight strategy}:
		\begin{align}
			RX_{t,\tau}^{\text{AIFX strategy}} &= \alpha_{\tau} + \beta_{\tau} RX^\text{Pure hindsight portfolio}_{t}  + \epsilon_{t,\tau}
			\label{eq:LookAheadBias:regression ai strategy on pure hindsight strategy}
		\end{align}
		If the performance of the \textit{AIFX strategy} were driven by memorized knowledge, we would expect to observe significantly positive betas, or statistically insignificant alphas. In addition, I report the \textit{Information Ratio}, defined in \myeq{eq:LookAheadBias:definition of information ratio}, which measures the Sharpe ratio of the component of strategy returns orthogonal to the benchmark:
		\begin{align}
			\text{Information Ratio} = \frac{\alpha}{\text{Standard error of }\epsilon_t}
			\label{eq:LookAheadBias:definition of information ratio}
		\end{align}
		The results, reported in \mytab{tab:LookaheadBias:pure hindsight portfolio regression}, show that the intercept (\(\alpha\)) is highly significant across all lookback periods, while the beta (\(\beta\)) is negative and statistically significant. Moreover, the reported information ratios are economically meaningful, ranging from 0.75 to 0.85. These findings show that the \textit{AIFX strategy}’s returns are not subsumed by those of the \textit{pure hindsight portfolio}. In fact, the negative beta suggests that the \textit{AIFX strategy} may operate in opposition to memorized return signals. Overall, this analysis provides further evidence that the predictive signals identified by the AI model are not driven by look-ahead bias.
		\begin{center}
			\textsc{\mytab{tab:LookaheadBias:pure hindsight portfolio regression} about here} \vspace{0cm}
		\end{center} 
		\section{Mechanisms Underlying Predictability} 				\label{sec:Mechanism}
		In this section, I examine the underlying economic mechanisms. Understanding these mechanisms is essential, as it provides insight into how and why economic fundamentals influence exchange rate movements. First, I study which category of fundamentals have more predictive power. Second, I explore the asymmetric predictive power of positive and negative news. Finally, I propose and discuss a potential mechanism.
		\subsection{Predictive Power of Different Categories of Fundamentals}		\label{sec:Impact of Different Categories of Fundamentals}
		In this section, I examine which categories of economic fundamentals contribute most to exchange rate predictability. The dataset used in this study includes a broad array of economic indicators, comprising 544 unique data series. Up to this point, the AI-powered variables used to predict exchange rates have aggregated information from all these indicators. A natural next step is to assess the heterogeneous impact of economic fundamentals on currency forecasting. To do this, I classify the data releases into eight distinct categories. This categorization process can be viewed as a form of dimensionality reduction. The methodology used to assign indicators to categories is detailed in \mysecIA{Appendix:Categories prodecure}. In addition, \mytab{tab:Categories:example headlines by category} presents the eight categories, along with representative examples of data releases included in each category and the number of observations in the dataset for each group. To assess the importance of each category, I follow a leave-one-out approach. Specifically, I re-estimate the baseline cross-sectional strategy (\textit{AIFX strategy}) multiple times, each time excluding one of the eight categories from the construction of the \textit{AIFX index}. I then compute the percentage drop in the Sharpe ratio relative to the \textit{AIFX strategy}. The intuition behind this procedure is that a larger drop in Sharpe ratio indicates a greater contribution of that category to predictive performance\footnote{An alternative approach would be to construct the strategy using only one category at a time. However, this method would lead to significant variation in the number of observations across categories, making it difficult to draw meaningful comparisons. By contrast, the leave-one-out method ensures a more consistent sample size across all exercises, thereby improving the comparability of the results.}. This analysis is conducted for lookback periods ranging from 36 to 60 months, where the predictive power of the strategy is strongest. I then compute the average percentage drop in Sharpe ratio across these lookback periods. The results are presented in \myfig{fig:Categories:margianl contribution to sharpe ratio categores cs strategy diff}. The figure reveals that \textit{Inflation data}, \textit{Employment data}, and \textit{Broad economic activity indicators} are the top three categories associated with the largest drops in Sharpe ratio when excluded. Additionally, I find that the marginal contributions of other categories are negative, in other words, excluding them leads to better performance. Notably, \myfig{fig:Categories:CS strategy sharpe ratio all categories vs top three} compares the performance of the \textit{AIFX strategy}, constructed using all eight categories, with an alternative strategy that uses only the top three categories: ‘Inflation data’, ‘Employment data’, and ‘Broad economic activity indicators’. The figure shows a clear improvement in performance when the strategy is restricted to these three categories. Specifically, the Sharpe ratios increase by approximately 40\% on average, and in some instances, more than double. The results provided in this section indicate the key role of ‘Inflation data’, ‘Employment data’, and ‘Broad economic activity indicators’ in driving exchange rate predictability.
		\begin{center}
			\textsc{\mytab{tab:Categories:example headlines by category}, \myfig{fig:Categories:margianl contribution to sharpe ratio categores cs strategy diff} and \myfig{fig:Categories:CS strategy sharpe ratio all categories vs top three} about here}  
		\end{center}
		\subsection{Decomposing Predictability: Positive vs. Negative News} \label{sec:Mechanism:StrengthVsWeakness}
		So far, I have examined the predictive power of the \textit{AIFX index}. By construction, this variable is composed of two components: (i) the \textit{Strength ratio} (\myeq{eq:VariableConstruction:definition of pos ratio}), which measures the proportion of positive news, and (ii) the \textit{Weakness ratio} (\myeq{eq:VariableConstruction:definition of neg ratio}), which captures the proportion of negative news. In this section, I investigate the predictive power of \textit{Strength ratio} and \textit{Weakness ratio} separately. First, I construct cross-sectional trading strategies using each variable as the trading signal. Second, I estimate panel regressions with future currency returns as the dependent variable and the \textit{Strength ratio} and \textit{Weakness ratio} as explanatory variables. Third, I explore the reason behind the asymmetric predictive power.
		
		\subsubsection{Cross-sectional Trading Strategies}			\label{sec:CS Trading Strategy: Positive and Negative Ratio as Signal}
		I construct two sets of cross-sectional trading strategies, following the methodology of \mysec{sec:Cross Sectional Trading Strategy: Diff Ratio as the signal}. In the first set, I use the \textit{Strength ratio} as the trading signal. For each choice of lookback period, at the end of each month, currencies are sorted based on their \textit{Strength ratio}, and I take long positions in the top two currencies with the highest values and short positions in the bottom two with the lowest. Portfolios are rebalanced at the end of each calendar month. In the second set of strategies, I use the \textit{Weakness ratio} as the signal and follow the same procedure as before with the only difference that currencies are sorted in reverse order. In other words, I go long the two currencies with the lowest \textit{Weakness ratio} and short the two with the highest. \myfig{fig:DifferentialImpact:sharpe ratio cs strategy positive negative ratio together} presents the annualized Sharpe ratios of the two strategies across different lookback periods. The strategy based on the \textit{Strength ratio} consistently delivers economically meaningful performance, with Sharpe ratios exceeding 0.5 in most cases. In contrast, the strategy based on the \textit{Weakness ratio} exhibits weak performance. Notably, for lookback periods greater than 16 months, the strategy tends to misdirect predictions and generates slightly negative Sharpe ratios. These findings suggest that the predictive power of the \textit{AIFX index} is primarily driven by the \textit{Strength ratio}. In other words, positive news appear to possess more predictive signal than negative news in forecasting exchange rate movements.
		\begin{center}
			\textsc{\myfig{fig:DifferentialImpact:sharpe ratio cs strategy positive negative ratio together} about here} \vspace{0cm}
		\end{center} 
		\subsubsection{Panel Regressions}		\label{sec:Panel Regressions: Pos_neg_ratio}
		Following a similar methodology to \mysec{sec:Panel Regression}, I estimate the following panel regression model for each choice of lookback period:
		\begin{equation}
			R_{c,t+1} = \alpha_{t,\tau} +  \beta_{1,\tau} \text{Strength}_{c,t,\tau} + \beta_{2,\tau} \text{Weakness}_{c,t,\tau} + \epsilon_{c,t,\tau}
			\label{eq:DifferentialImpact:regression return on positive negative ratio cs}
		\end{equation}
		in \myeq{eq:DifferentialImpact:regression return on positive negative ratio cs}, \(R_{c,t+1}\) denotes the monthly excess return of currency \(c\) at time \(t+1\). Accordingly, this regression framework examines whether the \textit{Strength ratio} and \textit{Weakness ratio} of a currency at the end of a given month can predict the currency's return in the subsequent month. Therefore, we are interested in testing the statistical significance of \(\beta_1\) and \(\beta_2\). \myfig{fig:DifferentialImpact:t-stat for regression of return on positive negative ratio CS} reports the \(t\)-statistics for \(\beta_1\) and \(\beta_2\) across different lookback periods. The results reveal that \(\beta_1\) is statistically significant across nearly all specifications, whereas \(\beta_2\) is consistently insignificant. In other words, the \textit{Strength ratio} reliably forecasts next-month currency returns, while the \textit{Weakness ratio} does not. These findings reinforce the earlier conclusion that the predictive power of the \textit{AIFX index} is primarily driven by positive news.
		\begin{center}
			\textsc{\myfig{fig:DifferentialImpact:t-stat for regression of return on positive negative ratio CS} about here} \vspace{0cm}
		\end{center} 
		\subsubsection{Further Investigation of the Asymmetric Predictive Power} \label{sec:Mechanism:Further Investigating of the Asymmetric Predictive Power}
		In this part, I explore the answer to the question of why positive news, as captured by the \textit{Strength ratio}, carries more forward-looking information than negative news, as captured by the \textit{Weakness ratio}. A straightforward hypothesis is that the dataset may contain more positive than negative news. A higher number of data points can increase the signal-to-noise ratio, leading to more predictive power. \mytab{tab:data:headline frequency table} reports the distribution of positive, negative, and neutral news in the dataset. The table clearly shows that the news classifications are fairly balanced, with no substantial skew toward positive or negative observations, thereby refuting this hypothesis. A second hypothesis relates to the potential synchronization of business cycles among G10 economies. If most countries in the sample experience downturns or negative shocks simultaneously, then the incidence of negative news would be relatively synchronized across currencies. As a result, the \textit{Weakness ratio} would fail to generate sufficient cross-sectional variation to differentiate between currencies, limiting its usefulness in forecasting returns. \myfigIA{fig:DifferentialImpact:cs_std_per_period_48_months} presents the cross-sectional standard deviations for the \textit{AIFX index}, \textit{Strength ratio}, and \textit{Weakness ratio} over time. The figure reveals that all three series exhibit fairly similar levels of dispersion over time, suggesting that the lack of cross-sectional variation in the \textit{Weakness ratio} is not a plausible explanation. In the third exercise, I compare the average realized returns on the release day\footnote{This analysis differs from the rest of the paper, which has focused exclusively on monthly data. Moreover, precise within-month release dates are only reliably available in the dataset from 2008 onward. As a result, the daily analysis presented here is restricted to the post-2008 period.} across positive, negative, and neutral news, as shown in \myfig{fig:DifferentialImpact:pos neg differential impact smaller window}. On average, currencies appreciate by 1.92 basis points on days with positive news, depreciate by 0.28 basis points on days with neutral news, and experience a considerably larger depreciation of 4.99 basis points on days with negative news. Additionally, \myfigIA{fig:DifferentialImpact:pos neg differential impact expanded window} displays the average return in a three-day event window that includes the day before, the day of, and the day after the release. The results reveal a stronger immediate market reaction to negative news relative to positive news. This asymmetry is consistent with the findings of \citet{ANDERSEN/ETAL:2003}, who document stronger market reactions to negative economic surprises in the FX market during the 1992–1998 period. The results from these daily return analyses suggest a plausible explanation for the asymmetry in predictive power: negative news induces a stronger and more immediate exchange rate adjustment, thereby exhausting its informational content in the short run. Consequently, the \textit{Weakness ratio}, which aggregates negative news, carries limited predictive power at the monthly horizon. On the other hand, positive news prompts a more gradual adjustment, leaving room for further price (exchange rate) discovery. As a result, the \textit{Strength ratio} retains a greater degree of predictive content, helping explain why it is the primary driver of the overall performance of the \textit{AIFX index}.
		\begin{center}
			\textsc{\myfig{fig:DifferentialImpact:pos neg differential impact smaller window} about here} \vspace{0cm}
		\end{center} 
		\subsection{A Potential Mechanism}
		In this section, I propose and discuss a potential mechanism through which economic fundamentals influence exchange rates. The three categories found in \mysec{sec:Impact of Different Categories of Fundamentals} to have the highest predictive power are \textit{Inflation data}, \textit{Employment data}, and \textit{Broad economic activity indicators}, all of which enter the monetary policy framework proposed by \citet{TAYLOR:1993}. The Taylor rule posits that central banks adjust short-term interest rates in response to deviations in inflation from its target and to output gaps, measured by unemployment and broad economic activity indicators. Under this framework, a higher inflation or a positive output gap typically prompts monetary tightening, whereas lower inflation or a negative output gap leads to easing. Such monetary policy adjustments affect interest rate differentials, thereby influencing the attractiveness of holding a currency. Consistent with this, in \myexmpl{examples:VariableConstruction:second example} and \myexmpl{examples:VariableConstruction:third example}, the AI model explicitly evaluates economic news in terms of its likely effect on monetary policy under a Taylor-type framework, and then based on that, infers the impact on exchange rates. These observations suggest that the AI model incorporates reasoning consistent with economic theory when interpreting fundamental news. In addition, empirical evidence from previous literature supports the predictive power of Taylor-rule fundamentals for exchange rates \citep{CLARIDA/WALDMAN:2007, MOLODTSOVA/PAPELL:2009}. More recently, \citet{ENGLE/WU:2024} argue that as central banks became more independent and credible in adhering to Taylor-type policies, exchange rates became more systematically tied to economic fundamentals. They find that the explanatory power of exchange rate models has improved over time, particularly as inflation targeting and policy credibility have strengthened. Taken together, these findings suggest that the Taylor rule framework is a key mechanism through which economic fundamentals affect exchange rates, and that this mechanism is reflected both in the AI model’s reasoning and in the empirical results presented in this paper. Furthermore, the asymmetric way central banks implement the Taylor rule could explain the stronger market reaction to negative news, compared to positive news, documented in \mysec{sec:Mechanism:Further Investigating of the Asymmetric Predictive Power}. The original Taylor rule prescribes a linear and symmetric policy response; however, various scholars have identified asymmetric tendencies in actual monetary policy decisions \citep{DOLADO/ETAL:2004, BRUGGEMANN/RIEDEL:2011, KOMLAN:2013}. Notably, \citet{HOFMANN/BOGDANOVA:2012} show that actual policy interest rates in both advanced and emerging economies have often been lower than those prescribed by standard Taylor rules. They attribute these deviations to the asymmetric tendencies of central banks. Specifically, central banks cut rates quickly in downturns but raise them slowly or not at all in booms. The political economy of monetary policy also reinforces this asymmetry. Rate cuts are generally popular because they support growth and reduce unemployment, while rate hikes can be politically unpopular as they may slow the economy or increase borrowing costs. Even independent central banks are not immune to political pressures. For instance, on several occasions\footnote{Example of such political pressures reported by Bloomberg: www.bloomberg.com/news/articles/2025-06-06/trump-pressures-fed-s-powell-to-cut-rates-a-full-point}, U.S. President Donald Trump publicly criticized Federal Reserve Chair Jerome Powell for not lowering interest rates, despite inflation remaining above the policy target and no clear evidence of a negative output gap. If we accept this asymmetric tendency in central banks' behaviour, investors' immediate and more pronounced reaction to negative news, compared to positive news, could be rationalized. Specifically, investors rationally anticipate that central banks will respond aggressively to negative news (implying an easing of monetary policy and possible currency depreciation), prompting them to react strongly. Conversely, they respond less strongly to positive news (implying a tightening of monetary policy and possible currency appreciation), as they anticipate a more cautious reaction from the central bank. To conclude, the empirical findings of this paper highlight the importance of the Taylor rule and monetary policy in exchange rate determination. That said, I do not rule out alternative explanations and leave a more formal investigation of competing mechanisms to future research.
		\section{Conclusions}	\label{sec:Conclusions}
		This paper uses generative artificial intelligence (AI) to forecast currency returns from economic fundamentals, offering fresh insights into the long-standing exchange rate disconnect puzzle, originally documented by \citet{Meese/Rogoff:1983}. Leveraging models like GPT-4o and DeepSeek-V3, I analyze a large dataset of economic data releases for major currency pairs and construct AI-based variables that capture each currency’s underlying macroeconomic strength. These AI-derived signals exhibit strong predictive power in both cross-sectional and time-series settings, with performance robust to different lookback windows and alternative variable specifications. To ensure the reliability of these findings, I conduct a series exercises that rule out look-ahead bias and demonstrate that the AI model's performance stems from genuine reasoning rather than memorized information. I also explore the mechanisms behind this predictability and provide evidence that monetary policy, particularly the Taylor rule framework used by central banks to set interest rates, plays a central role in linking economic fundamentals to exchange rate movements. Further investigating the mechanisms driving this predictive power represents a promising avenue for future research. In addition, this paper offers preliminary evidence that AI can act as a financial analyst. Future work could explore this more deeply, in a spirit similar to \citet{CAO/ETAL:2024}, who show that an AI analyst, leveraging large-scale data and machine learning, can outperform human analysts in forecasting stock returns.

\newpage
\begin{spacing}{1}
\setlength{\bibsep}{8pt}
\raggedright
\phantomsection
\addcontentsline{toc}{section}{References}
\bibliographystyle{apalike}
\bibliography{xtra/biblio/biblio}
\end{spacing}

\newpage
\clearpage

%
\begin{landscape}
	
	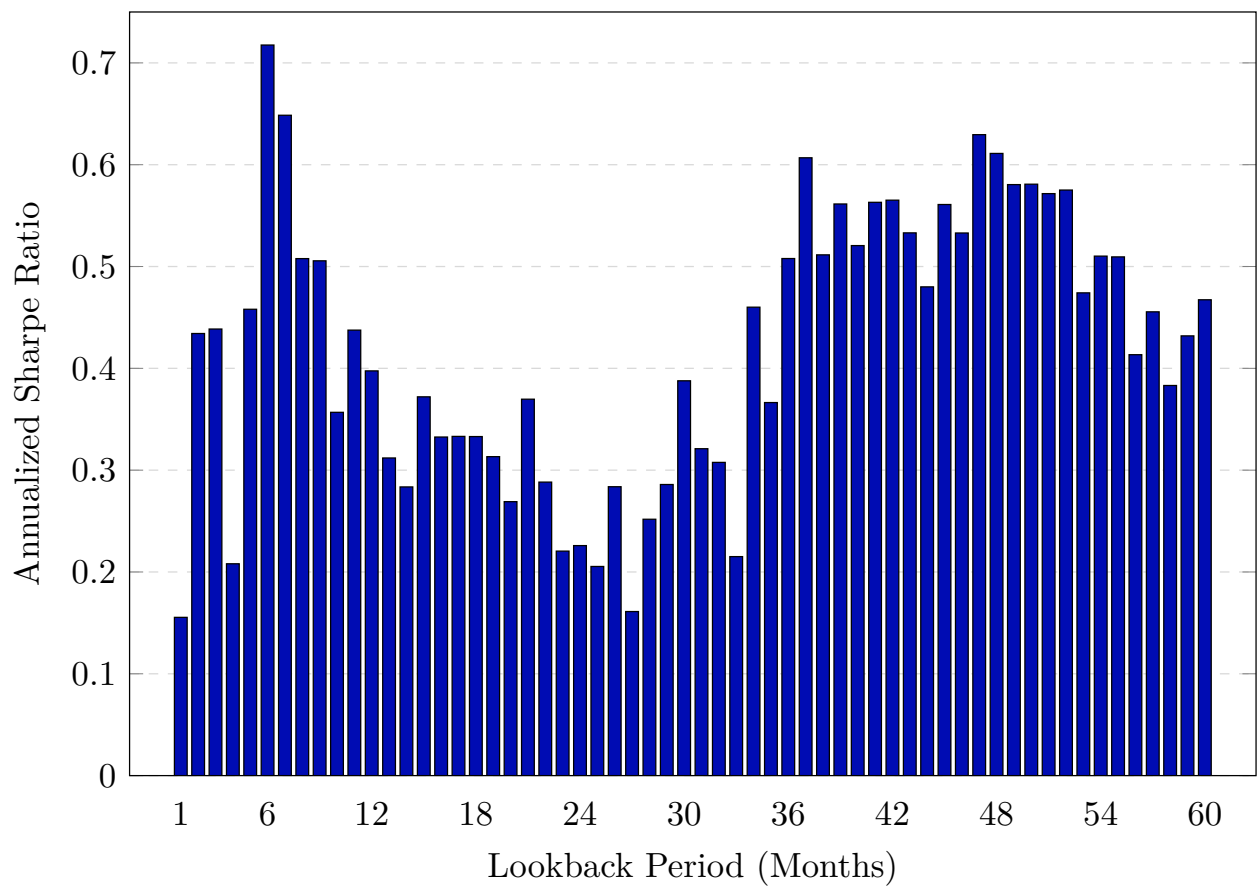
\begin{figure}

    \begin{center}
    \caption{\bf Sharpe Ratios}
    \scalebox{1.2}{
\pgfplotsset{compat=newest}
	\begin{tikzpicture}
		\begin{axis}[
			width=14cm,   
			height=10cm,
			ybar,
			bar width=4pt,
			enlarge x limits=0.05,
			xlabel={Lookback Period (Months)},
			ylabel={Annualized Sharpe Ratio},
			ymin=0,
			ymax=0.75,
			ytick={0,0.1,0.2,0.3,0.4,0.5,0.6,0.7,0.8},
			xtick={1,6,12,18,24,30,36,42,48,54,60},
			xtick style={draw=none},
			ymajorgrids=true,
			grid style={dashed,gray!30},
			ticklabel style={font=\small},
			label style={font=\small},
			]
			\addplot [
			fill=blue!70!black,
			draw=black
			] 
			table[
			x=Lookback,
			y=DiffStrategy,
			col sep=comma
			]{xtra/figures/Files/data/sharpe_ratio_cs_strategy_diff.csv};
		\end{axis}
	\end{tikzpicture}}
    \medskip
     \label{fig:Performance:sharpe ratio cs strategy diff}
    \end{center}

    \begin{small}
    This figure reports the annualized Sharpe ratios of cross-sectional strategies that use the \textit{AIFX index} (defined in \myeq{eq:VariableConstruction:definition of diff ratio}) as the trading signal. I refer to this strategy as the \textit{AIFX strategy}. For each choice of lookback period, currencies are sorted by their \textit{AIFX index} at the end of each month and I take long positions in the top two currencies with the highest \textit{AIFX index} and short positions in the bottom two with the lowest. Portfolios are rebalanced at the end of each calendar month. Lookback periods of 1 to 60 months are considered. The sample period spans from January 1996 to October 2024, but returns start at different dates due to differences in lookback periods.
    \end{small}

	\end{figure}

\end{landscape}

\begin{landscape}
	
	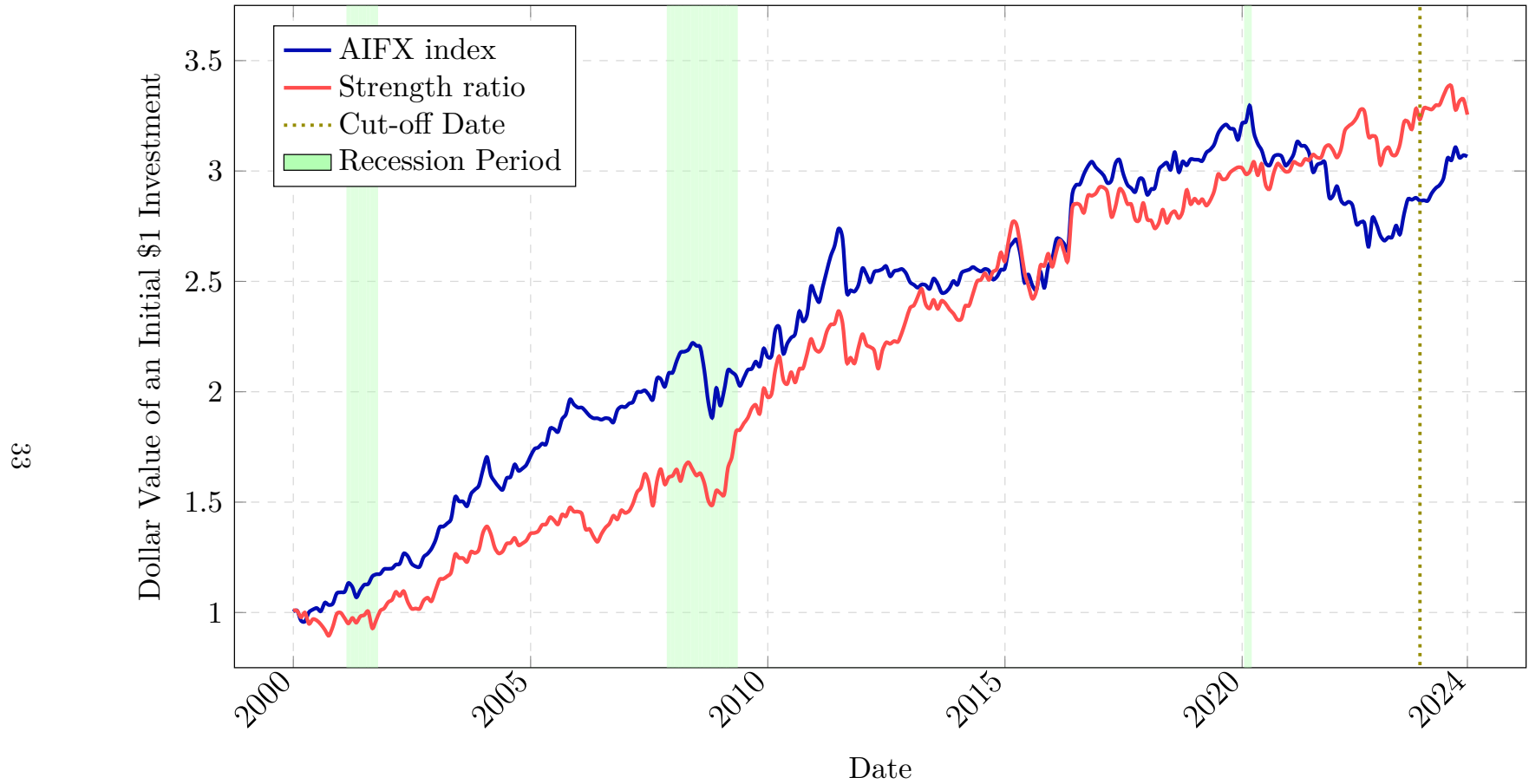
\begin{figure}

	    \begin{center}
	    	\caption{\bf Cumulative Returns}
		    \scalebox{1.2}{
\pgfplotsset{compat=1.9}
	\begin{tikzpicture}
		\begin{axis}[
			width=18cm,
			height=10cm,
			xlabel={Date},
			ylabel={Dollar Value of an Initial \$1 Investment},
			xticklabel style={rotate=45, anchor=east},
			xtick scale label code/.code={},
			grid=major,
			grid style={dashed,gray!30},
			ymajorgrids=true,
			y tick label style={font=\small},
			label style={font=\small},
			clip mode=individual,
			date coordinates in=x,        
			date ZERO={1999-12-31},       
			xmin={1999-12-31},            
			xmax={2024-09-30},            
			xtick={2000-01-01,2005-01-01,2010-01-01,2015-01-01,2020-01-01,2024-09-30},
			xticklabels={2000,2005,2010,2015,2020,2024},
			legend pos=north west,
			legend cell align={left},
			legend style={font=\small, /tikz/every even column/.append style={column sep=2pt}}, 
			enlarge x limits=0.05,
			ymax=3.75,
			ymin=0.75  
			]
			
			\addplot [
			ybar,
			bar width=1.4pt,
			fill=green!30,  
			draw=none,  
			opacity=0.4,  
			forget plot 
			] table [
			col sep=comma,
			x=date,
			y expr={\thisrow{NBER_Recession_indicator}==1 ? \pgfkeysvalueof{/pgfplots/ymax} : NaN}
			] {xtra/figures/Files/data/cs_strategy_performance_over_time_48_months_diff_positive.csv} \closedcycle;
			
			\addplot[mark=none, smooth, color=blue!70!black, line width=1.35pt] 
			table[col sep=comma, x=date, y=Diff] 
			{xtra/figures/Files/data/cs_strategy_performance_over_time_48_months_diff_positive.csv};
			
			\addplot[mark=none, smooth, color=red!70!white, line width=1.35pt] 
			table[col sep=comma, x=date, y=Pos] 
			{xtra/figures/Files/data/cs_strategy_performance_over_time_48_months_diff_positive.csv};
			
			\addplot [
			color=olive,
			very thick,
			dotted
			] coordinates {(2023-10-01, 0) (2023-10-01, \pgfkeysvalueof{/pgfplots/ymax})};
			
			\addlegendimage{area legend, fill=green!30, draw=none}
			\addlegendentry{Recession Period}
			
			\legend{AIFX index, Strength ratio, Cut-off Date, Recession Period}
		\end{axis}
	\end{tikzpicture}}
		    \medskip
		     \label{fig:Performance:cs strategy performance over time 48 months diff positive}
	    \end{center}

	    \begin{small}
	    The graph displays the Dollar value of an initial \$1 investment in two cross-sectional strategies from January 2000 to October 2024. The blue line uses the \textit{AIFX index} (defined in \myeq{eq:VariableConstruction:definition of diff ratio}) and the red line uses the \textit{Strength ratio} (defined in \myeq{eq:VariableConstruction:definition of pos ratio}) as the signal. Both trading strategies have a lookback period of 48 months. The green shaded regions denote NBER recession periods. The vertical dotted line marks GPT-4o’s knowledge cutoff date, October 2023, when its training data ends.
	    \end{small}

	\end{figure}

\end{landscape}

\begin{landscape}
	
	\begin{figure}

	    \begin{center}
	    	\caption{\bf Portfolio Composition}
		    \scalebox{1.2}{
\pgfplotsset{compat=1.9}
	\begin{tikzpicture}
		\begin{groupplot}[
			group style={
				group size=3 by 3,
				horizontal sep=1cm,
				vertical sep=1cm,
			},
			width=6cm,           
			height=4cm,          
			xticklabel style={font=\tiny, rotate=45, anchor=east},
			xtick scale label code/.code={},
			grid=major,
			grid style={dashed,gray!30},
			ymajorgrids=true,
			y tick label style={font=\small},
			label style={font=\small},
			clip mode=individual,
			date coordinates in=x,
			date ZERO={1999-12-31},
			ymin=-1.2,
			ymax=1.2,
			xmin={1999-12-31},
			xmax={2024-09-30},
			xtick={2000-01-01,2005-01-01,2010-01-01,2015-01-01,2020-01-01,2024-09-30},
			xticklabels={2000,2005,2010,2015,2020,2024},
			enlarge x limits=0.05,
			every axis title/.append style={yshift=-5pt},
			]
			
			\nextgroupplot[ylabel={Position}, title={AUD}]
			\addplot [ybar,bar width=0.5pt, fill=green!30, draw=none, opacity=0.4, forget plot]
			table [col sep=comma, x=date, y expr={\thisrow{NBER_Recession_indicator}==1 ? \pgfkeysvalueof{/pgfplots/ymin} : NaN}]
			{xtra/figures/Files/data/portfolio_composition_48_months_CS_diff_ratio.csv};
			\addplot [ybar,bar width=0.5pt, fill=green!30, draw=none, opacity=0.4, forget plot]
			table [col sep=comma, x=date, y expr={\thisrow{NBER_Recession_indicator}==1 ? \pgfkeysvalueof{/pgfplots/ymax} : NaN}]
			{xtra/figures/Files/data/portfolio_composition_48_months_CS_diff_ratio.csv};
			\addplot[mark=none, color=blue!70!black, line width=1.35pt] 
			table[col sep=comma, x=date, y=AUD] {xtra/figures/Files/data/portfolio_composition_48_months_CS_diff_ratio.csv};
			
			\nextgroupplot[title={CHF}]
			\addplot [ybar, bar width=0.5pt, fill=green!30, draw=none, opacity=0.4, forget plot]
			table [col sep=comma, x=date, y expr={\thisrow{NBER_Recession_indicator}==1 ? \pgfkeysvalueof{/pgfplots/ymin} : NaN}]
			{xtra/figures/Files/data/portfolio_composition_48_months_CS_diff_ratio.csv};
			\addplot [ybar, bar width=0.5pt, fill=green!30, draw=none, opacity=0.4, forget plot]
			table [col sep=comma, x=date, y expr={\thisrow{NBER_Recession_indicator}==1 ? \pgfkeysvalueof{/pgfplots/ymax} : NaN}]
			{xtra/figures/Files/data/portfolio_composition_48_months_CS_diff_ratio.csv};
			\addplot[mark=none, color=blue!70!black, line width=1.35pt] 
			table[col sep=comma, x=date, y=CHF] {xtra/figures/Files/data/portfolio_composition_48_months_CS_diff_ratio.csv};
			
			\nextgroupplot[title={EUR}]
			\addplot [ybar, bar width=0.5pt, fill=green!30, draw=none, opacity=0.4, forget plot]
			table [col sep=comma, x=date, y expr={\thisrow{NBER_Recession_indicator}==1 ? \pgfkeysvalueof{/pgfplots/ymin} : NaN}]
			{xtra/figures/Files/data/portfolio_composition_48_months_CS_diff_ratio.csv};
			\addplot [ybar, bar width=0.5pt, fill=green!30, draw=none, opacity=0.4, forget plot]
			table [col sep=comma, x=date, y expr={\thisrow{NBER_Recession_indicator}==1 ? \pgfkeysvalueof{/pgfplots/ymax} : NaN}]
			{xtra/figures/Files/data/portfolio_composition_48_months_CS_diff_ratio.csv};
			\addplot[mark=none, color=blue!70!black, line width=1.35pt] 
			table[col sep=comma, x=date, y=EUR] {xtra/figures/Files/data/portfolio_composition_48_months_CS_diff_ratio.csv};
			
			\nextgroupplot[ylabel={Position}, title={GBP}]
			\addplot [ybar, bar width=0.5pt, fill=green!30, draw=none, opacity=0.4, forget plot]
			table [col sep=comma, x=date, y expr={\thisrow{NBER_Recession_indicator}==1 ? \pgfkeysvalueof{/pgfplots/ymin} : NaN}]
			{xtra/figures/Files/data/portfolio_composition_48_months_CS_diff_ratio.csv};
			\addplot [ybar, bar width=0.5pt, fill=green!30, draw=none, opacity=0.4, forget plot]
			table [col sep=comma, x=date, y expr={\thisrow{NBER_Recession_indicator}==1 ? \pgfkeysvalueof{/pgfplots/ymax} : NaN}]
			{xtra/figures/Files/data/portfolio_composition_48_months_CS_diff_ratio.csv};
			\addplot[mark=none, color=blue!70!black, line width=1.35pt] 
			table[col sep=comma, x=date, y=GBP] {xtra/figures/Files/data/portfolio_composition_48_months_CS_diff_ratio.csv};
			
			\nextgroupplot[title={JPY}]
			\addplot [ybar, bar width=0.5pt, fill=green!30, draw=none, opacity=0.4, forget plot]
			table [col sep=comma, x=date, y expr={\thisrow{NBER_Recession_indicator}==1 ? \pgfkeysvalueof{/pgfplots/ymin} : NaN}]
			{xtra/figures/Files/data/portfolio_composition_48_months_CS_diff_ratio.csv};
			\addplot [ybar, bar width=0.5pt, fill=green!30, draw=none, opacity=0.4, forget plot]
			table [col sep=comma, x=date, y expr={\thisrow{NBER_Recession_indicator}==1 ? \pgfkeysvalueof{/pgfplots/ymax} : NaN}]
			{xtra/figures/Files/data/portfolio_composition_48_months_CS_diff_ratio.csv};
			\addplot[mark=none, color=blue!70!black, line width=1.35pt] 
			table[col sep=comma, x=date, y=JPY] {xtra/figures/Files/data/portfolio_composition_48_months_CS_diff_ratio.csv};
			
			\nextgroupplot[title={NZD}]
			\addplot [ybar, bar width=0.5pt, fill=green!30, draw=none, opacity=0.4, forget plot]
			table [col sep=comma, x=date, y expr={\thisrow{NBER_Recession_indicator}==1 ? \pgfkeysvalueof{/pgfplots/ymin} : NaN}]
			{xtra/figures/Files/data/portfolio_composition_48_months_CS_diff_ratio.csv};
			\addplot [ybar, bar width=0.5pt, fill=green!30, draw=none, opacity=0.4, forget plot]
			table [col sep=comma, x=date, y expr={\thisrow{NBER_Recession_indicator}==1 ? \pgfkeysvalueof{/pgfplots/ymax} : NaN}]
			{xtra/figures/Files/data/portfolio_composition_48_months_CS_diff_ratio.csv};
			\addplot[mark=none, color=blue!70!black, line width=1.35pt] 
			table[col sep=comma, x=date, y=NZD] {xtra/figures/Files/data/portfolio_composition_48_months_CS_diff_ratio.csv};
			
			\nextgroupplot[ylabel={Position}, xlabel={Date}, title={SEK}]
			\addplot [ybar, bar width=0.5pt, fill=green!30, draw=none, opacity=0.4, forget plot]
			table [col sep=comma, x=date, y expr={\thisrow{NBER_Recession_indicator}==1 ? \pgfkeysvalueof{/pgfplots/ymin} : NaN}]
			{xtra/figures/Files/data/portfolio_composition_48_months_CS_diff_ratio.csv};
			\addplot [ybar, bar width=0.5pt, fill=green!30, draw=none, opacity=0.4, forget plot]
			table [col sep=comma, x=date, y expr={\thisrow{NBER_Recession_indicator}==1 ? \pgfkeysvalueof{/pgfplots/ymax} : NaN}]
			{xtra/figures/Files/data/portfolio_composition_48_months_CS_diff_ratio.csv};
			\addplot[mark=none, color=blue!70!black, line width=1.35pt]
			table[col sep=comma, x=date, y=SEK] {xtra/figures/Files/data/portfolio_composition_48_months_CS_diff_ratio.csv};
			
			\nextgroupplot[xlabel={Date}, title={CAD}]
			\addplot [ybar, bar width=0.5pt, fill=green!30, draw=none, opacity=0.4, forget plot]
			table [col sep=comma, x=date, y expr={\thisrow{NBER_Recession_indicator}==1 ? \pgfkeysvalueof{/pgfplots/ymin} : NaN}]
			{xtra/figures/Files/data/portfolio_composition_48_months_CS_diff_ratio.csv};
			\addplot [ybar, bar width=0.5pt, fill=green!30, draw=none, opacity=0.4, forget plot]
			table [col sep=comma, x=date, y expr={\thisrow{NBER_Recession_indicator}==1 ? \pgfkeysvalueof{/pgfplots/ymax} : NaN}]
			{xtra/figures/Files/data/portfolio_composition_48_months_CS_diff_ratio.csv};
			\addplot[mark=none, color=blue!70!black, line width=1.35pt]
			table[col sep=comma, x=date, y=CAD] {xtra/figures/Files/data/portfolio_composition_48_months_CS_diff_ratio.csv};
			
			\nextgroupplot[xlabel={Date}, title={NOK}]
			\addplot [ybar, bar width=0.5pt, fill=green!30, draw=none, opacity=0.4, forget plot]
			table [col sep=comma, x=date, y expr={\thisrow{NBER_Recession_indicator}==1 ? \pgfkeysvalueof{/pgfplots/ymin} : NaN}]
			{xtra/figures/Files/data/portfolio_composition_48_months_CS_diff_ratio.csv};
			\addplot [ybar, bar width=0.5pt, fill=green!30, draw=none, opacity=0.4, forget plot]
			table [col sep=comma, x=date, y expr={\thisrow{NBER_Recession_indicator}==1 ? \pgfkeysvalueof{/pgfplots/ymax} : NaN}]
			{xtra/figures/Files/data/portfolio_composition_48_months_CS_diff_ratio.csv};
			\addplot[mark=none, color=blue!70!black, line width=1.35pt]
			table[col sep=comma, x=date, y=NOK] {xtra/figures/Files/data/portfolio_composition_48_months_CS_diff_ratio.csv};
			
		\end{groupplot}
	\end{tikzpicture}}
		    \medskip
		     \label{fig:portfolio composition 4 months CS diff ratio}
	    \end{center}

	    \begin{small}
	   	This figure shows the positions taken in each currency over time for the \textit{AIFX strategy} with a 48-month lookback period. A value of 1 indicates a long position, 0 indicates no position, and $-1$ indicates a short position. The performance is reported for the period from January 2000 to October 2024. The green shaded regions denote NBER recession periods.
	    \end{small}

	\end{figure}
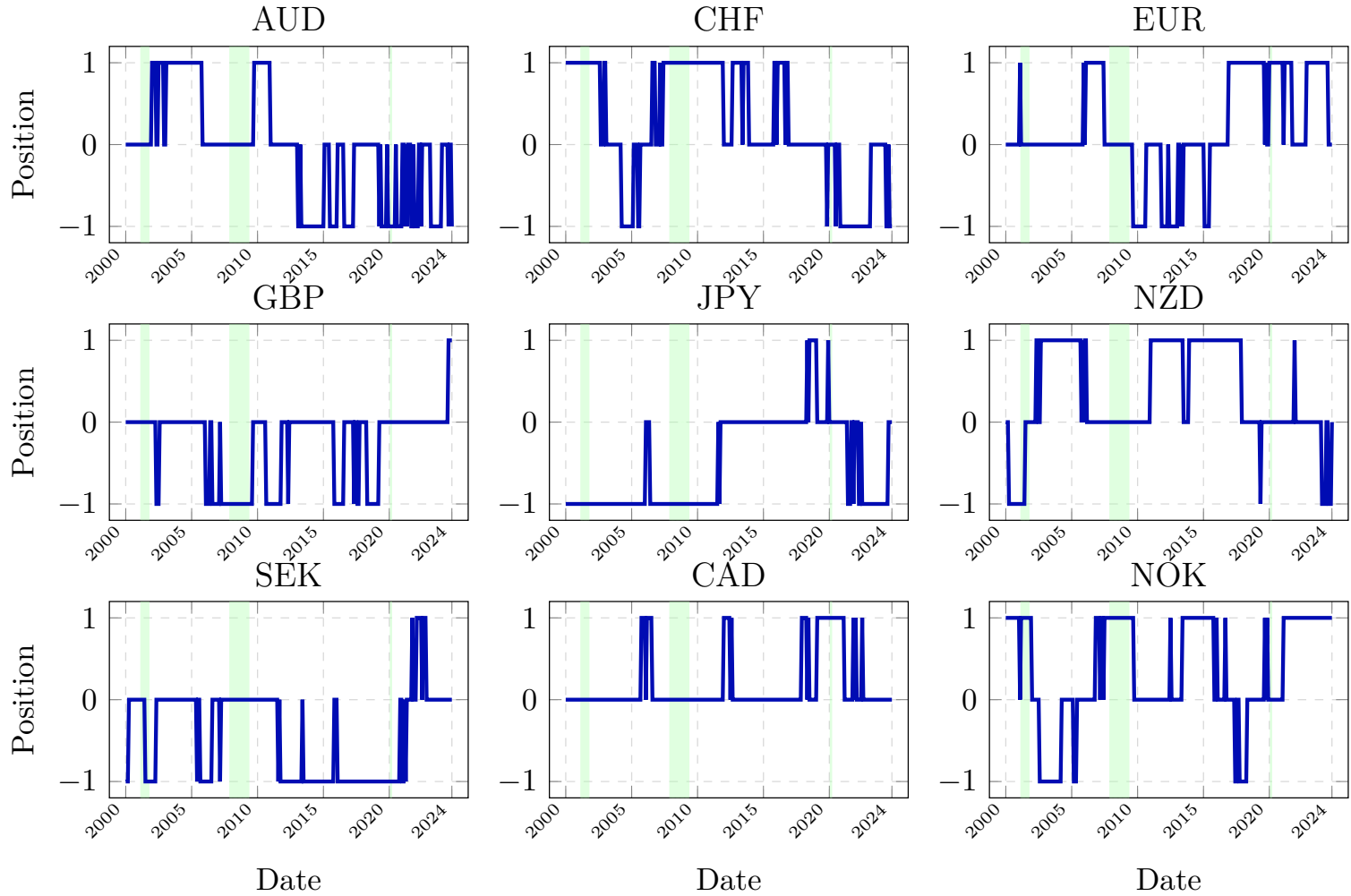

\end{landscape}

\begin{landscape}
	
	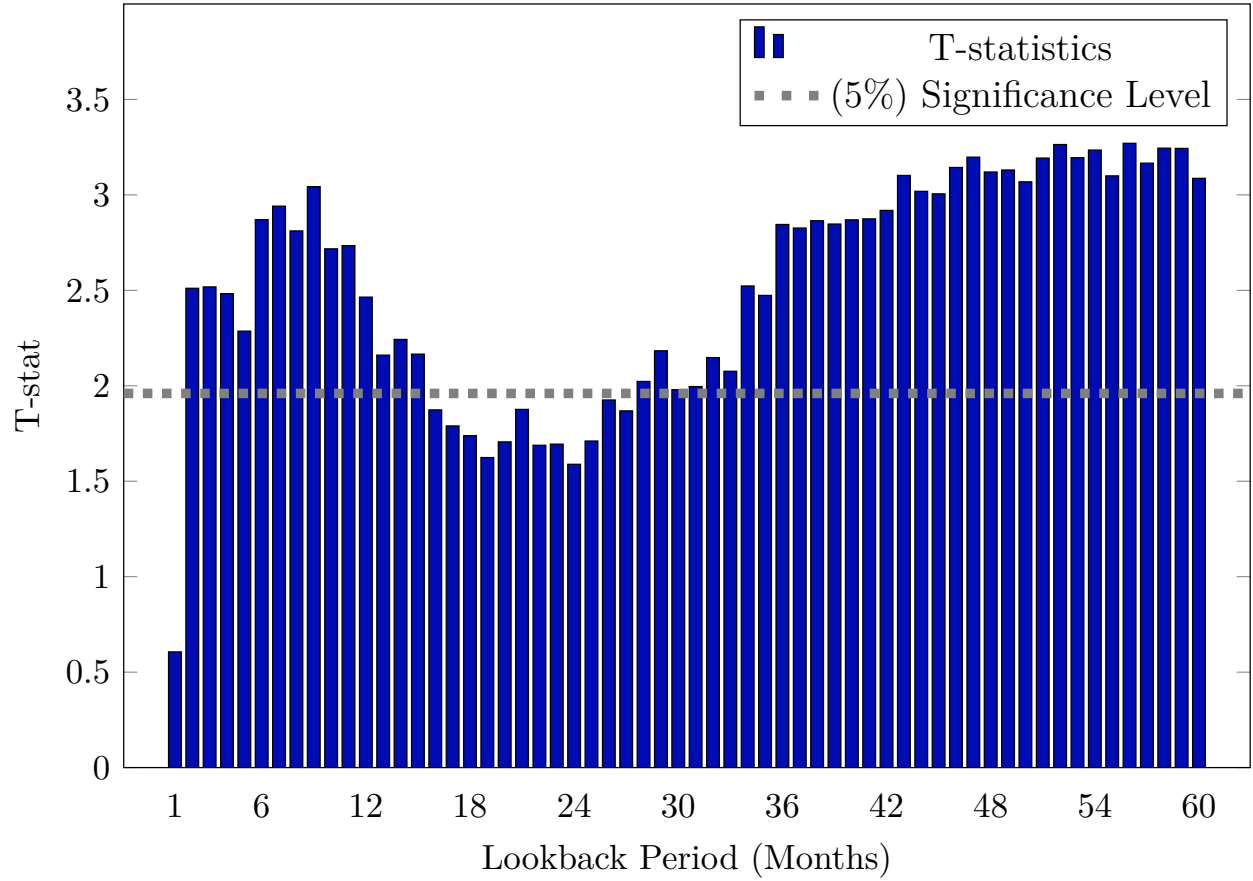
\begin{figure}

	    \begin{center}
	    	\caption{\bf $t$-statistics}
		    \scalebox{1.2}{
\pgfplotsset{compat=newest}
	\begin{tikzpicture}
		\begin{axis}[
			width=14cm,   
			height=10cm,
			ybar,
			bar width=4pt,
			enlarge x limits=0.05,
			xlabel={Lookback Period (Months)},
			ylabel={T-stat},
			ymin=0,
			ymax=4,
			ytick={0,0.5,1,1.5,2,2.5,3,3.5},
			xtick={1,6,12,18,24,30,36,42,48,54,60},
			xtick style={draw=none},
			ticklabel style={font=\small},
			label style={font=\small},
			]
			\addplot [
			fill=blue!70!black,
			draw=black
			] 
			table[
			x=number_of_lookback_months,
			y=ratio_diff_t_stat,
			col sep=comma
			]{xtra/figures/Files/data/t-stat_for_regression_of_return_on_diff_ratio_CS.csv};
			\addlegendentry{T-statistics}
			\draw [gray, dashed, line width=3pt] (rel axis cs:0,1.96/4) -- (rel axis cs:1,1.96/4);
			\addlegendimage{line legend, gray, dashed, line width=3pt}
			\addlegendentry{(5\%) Significance Level}
		\end{axis}
	\end{tikzpicture}}
		    \medskip
		     \label{fig:Performance:t-stat for regression of return on diff ratio CS}
	    \end{center}

	    \begin{small}
	    This figure displays the $t$-statistics of $\beta$ in \myeq{eq:Performance:regression return on diff ratio cs}. The regression is repeated for different choices of lookback period. This regression examines whether the \textit{AIFX index} of a currency at the end of a given month can predict the currency's excess return in the subsequent month. Standard errors are clustered at the currency and time level. The regression includes time fixed effects, although not reported here. The sample period spans from January 1996 to October 2024, but returns start at different dates due to differences in lookback periods. The dashed horizontal line marks the 5\% statistical significance threshold.

	    \end{small}

	\end{figure}

\end{landscape}

\begin{landscape}
	
	\begin{figure}

	    \begin{center}
	    	\caption{\bf Sharpe Ratios (GPT-4o vs DeepSeek-V3)}
		    \scalebox{1.2}{
\pgfplotsset{compat=newest}
	\begin{tikzpicture}
		\begin{axis}[
			width=14cm,   
			height=10cm,
			ybar,
			bar width=2pt,
			enlarge x limits=0.05,
			xlabel={Lookback Period (Months)},
			ylabel={Annualized Sharpe Ratio},
			ymax=0.85,
			ytick={-0.3,-0.2,-0.1,0.0,0.1,0.2,0.3,0.4,0.5,0.6,0.7},
			xtick={1,6,12,18,24,30,36,42,48,54,60},
			xtick style={draw=none},
			ymajorgrids=true,
			grid style={dashed,gray!30},
			ticklabel style={font=\small},
			label style={font=\small},
			legend pos=north west, 
			legend style={font=\small}, 
			legend cell align={left} 
			]
			
			\addplot [
			fill=blue!70!black,
			draw=black,
			bar shift=-1.5pt 
			] 
			table[
			x=lookback_period_number_of_months,
			y=deepseek_diff_strategy,
			col sep=comma
			]{xtra/figures/Files/data/deepseekv3_sharpe_ratio_cs_strategy_diff.csv};
			\addlegendentry{DeepSeek V3}
			
			\addplot [
			fill=red!70!white,
			draw=black,
			bar shift=1pt 
			] 
			table[
			x=lookback_period_number_of_months,
			y=gpt4o_DiffStrategy,
			col sep=comma
			]{xtra/figures/Files/data/deepseekv3_sharpe_ratio_cs_strategy_diff.csv};
			\addlegendentry{GPT 4o}

		\end{axis}
	\end{tikzpicture}}
		    \medskip
		     \label{fig:Performance:GPT-4o vs DeepSeek-V3}
	    \end{center}

	    \begin{small}
	    This figure compares the Sharpe ratios of the \textit{AIFX strategy} constructed using GPT-4o and DeepSeek-V3. In one strategy the \textit{AIFX index} (defined in \myeq{eq:VariableConstruction:definition of diff ratio}) was constructed using GPT-4o’s outputs and in the other strategy, the \textit{AIFX index} was constructed using DeepSeek V3’s outputs. Both strategies are cross-sectional ones that use the \textit{AIFX index} as the trading signal. For each choice of lookback period, currencies are sorted by their \textit{AIFX index} at the end of each month and I take long positions in the top two currencies with the highest \textit{AIFX index} and short positions in the bottom two with the lowest. Portfolios are rebalanced at the end of each calendar month. Lookback periods of 1 to 60 months are considered. The sample period spans from January 1996 to October 2024, but returns start at different dates due to differences in lookback periods.
	    \end{small}

	\end{figure}
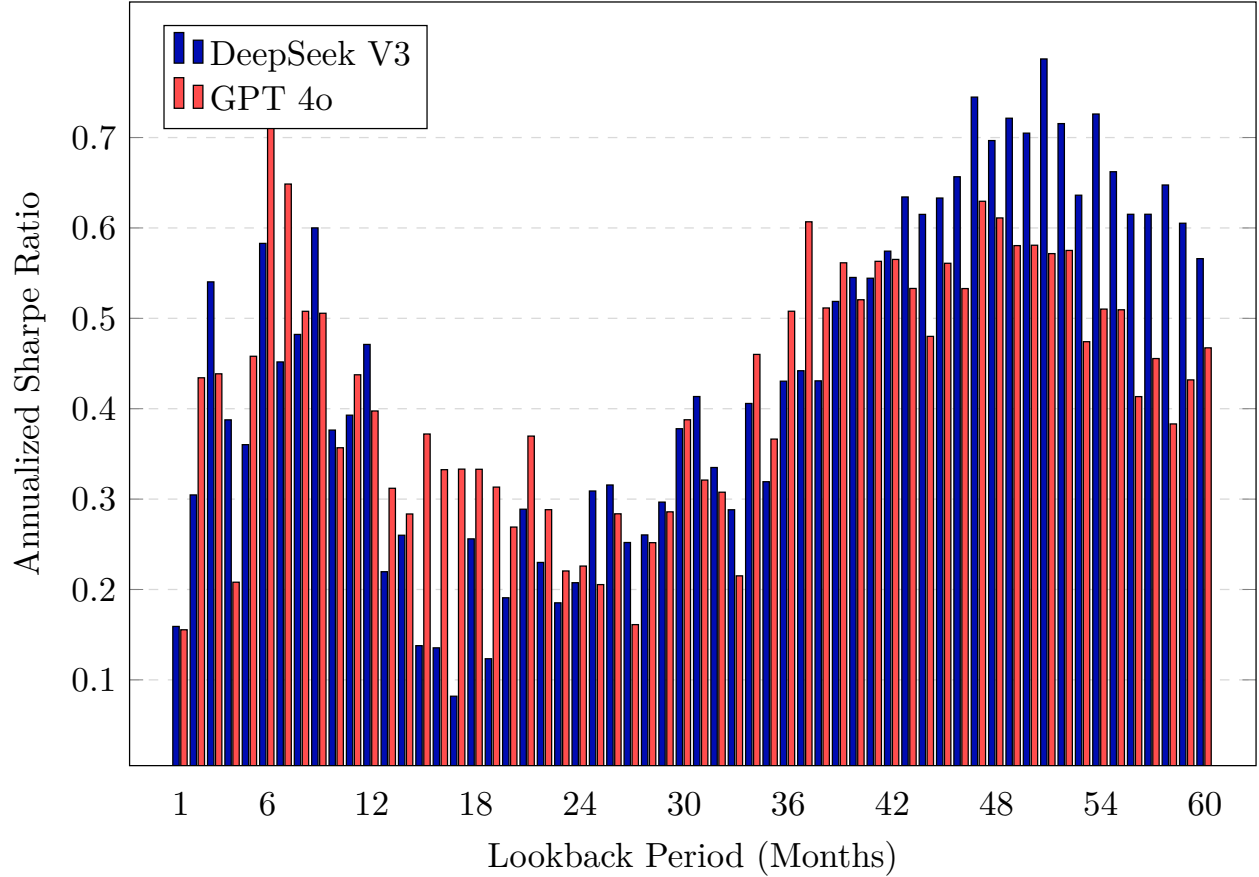

\end{landscape}

\begin{landscape}
	
	\begin{figure}

	    \begin{center}
	    	\caption{\bf Alternative Specification of the AIFX index}
		    \scalebox{1.2}{
\pgfplotsset{compat=newest}
	\begin{tikzpicture}
		\begin{axis}[
			width=14cm,   
			height=10cm,
			ybar,
			bar width=1.75pt,
			enlarge x limits=0.05,
			xlabel={Lookback Period (Months)},
			ylabel={Annualized Sharpe Ratio},
			ymax=0.75,
			ytick={-0.3,-0.2,-0.1,0.0,0.1,0.2,0.3,0.4,0.5,0.6,0.7},
			xtick={1,6,12,18,24,30,36,42,48,54,60},
			xtick style={draw=none},
			ymajorgrids=true,
			grid style={dashed,gray!30},
			ticklabel style={font=\small},
			label style={font=\small},
			legend pos= south east, 
			legend style={font=\small}, 
			legend cell align={left} 
			]
			
			\addplot [
			fill=red!70!white,
			draw=black,
			bar shift=-1.5pt 
			] 
			table[
			x=lookback_period_number_of_months,
			y=diff_strategy_weighted,
			col sep=comma
			]{xtra/figures/Files/data/sharpe_ratio_cs_strategy_diff_ratio_weighted_squared_and_unweighted.csv};
			\addlegendentry{Weighted}
			
			\addplot [
			fill=blue!70!black,
			draw=black,
			bar shift=1pt 
			] 
			table[
			x=lookback_period_number_of_months,
			y=diff_strategy_unweighted,
			col sep=comma
			]{xtra/figures/Files/data/sharpe_ratio_cs_strategy_diff_ratio_weighted_squared_and_unweighted.csv};
			\addlegendentry{Unweighted}

		\end{axis}
	\end{tikzpicture}}
		    \medskip
		     \label{fig:Performance:diff ratio weighted squared and unweighted}
	    \end{center}

	    \begin{small}
	    This figure reports the annualized Sharpe ratios of cross-sectional strategies that use the \textit{AIFX index} (defined in \myeq{eq:VariableConstruction:definition of diff ratio}) and \textit{Weighted AIFX index} (defined in \myeq{eq:Performance:definition of weighted diff ratio}) as the trading signal. For each choice of lookback period, currencies are sorted based on the signal at the end of each month and I take long positions in the top two currencies with the highest signal value and short positions in the bottom two with the lowest. Portfolios are rebalanced at the end of each calendar month. Lookback periods of 1 to 60 months are considered. The sample period spans from January 1996 to October 2024, but returns start at different dates due to differences in lookback periods.
	    \end{small}

	\end{figure}
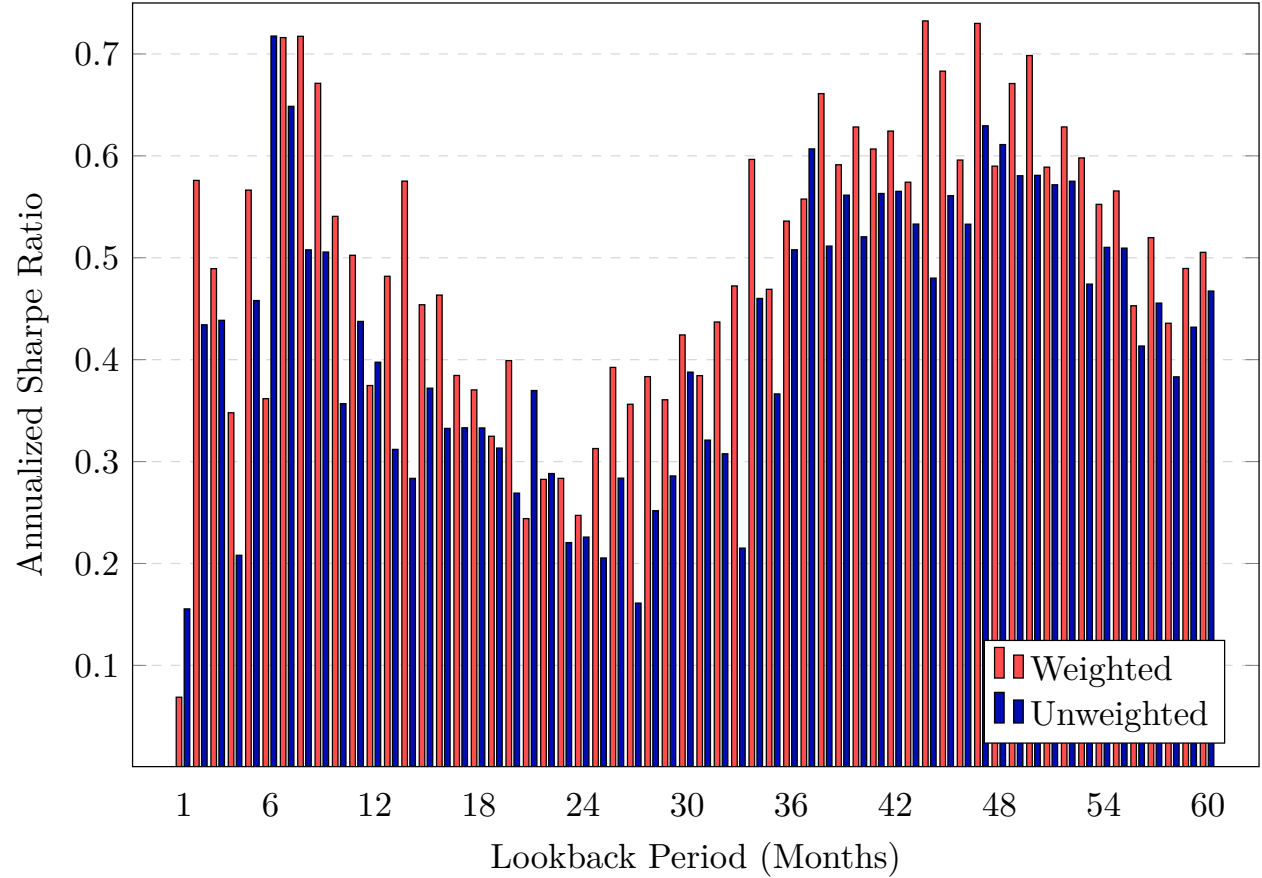

\end{landscape}

\begin{landscape}
	
	\begin{figure}

	    \begin{center}
	    	\caption{\bf Distribution of Data Releases}
		    \scalebox{1.2}{
\pgfplotsset{compat=newest}
	\begin{tikzpicture}
		\begin{axis}[
			width=14cm,   
			height=10cm,
			ybar,
			bar width=8pt,
			enlarge x limits=0.05,
			xtick=data, 
			xlabel={Year},
			ylabel={Number of Economic Data Releases},
			ymin=0,
			ymax=5500,
			xtick style={draw=none},
			ymajorgrids=true,
			grid style={dashed,gray!30},
			ticklabel style={font=\small},
			label style={font=\small},
			xticklabel style={
				rotate=90,
				anchor=east,
				/pgf/number format/1000 sep={}
			},
			]
			\addplot [
			fill=blue!70!black,
			draw=black
			] 
			table[
			x=year,
			y=economic_events_per_year,
			col sep=comma
			]{xtra/figures/Files/data/economic_events_per_year.csv};
		\end{axis}
	\end{tikzpicture}}
		    \medskip
		     \label{fig:LookAheadBias:economic events per year}
	    \end{center}

	    \begin{small}
	    This figure shows the distribution of data releases in the dataset across years. The sample period spans from January 1996 to October 2024. The dataset is collected from Investing.com’s economic calendar data. The economic calendar for each currency lists the dates and times of key economic releases, such as GDP reports, employment statistics, inflation readings, central bank decisions, and other economic indicators.
	    \end{small}

	\end{figure}
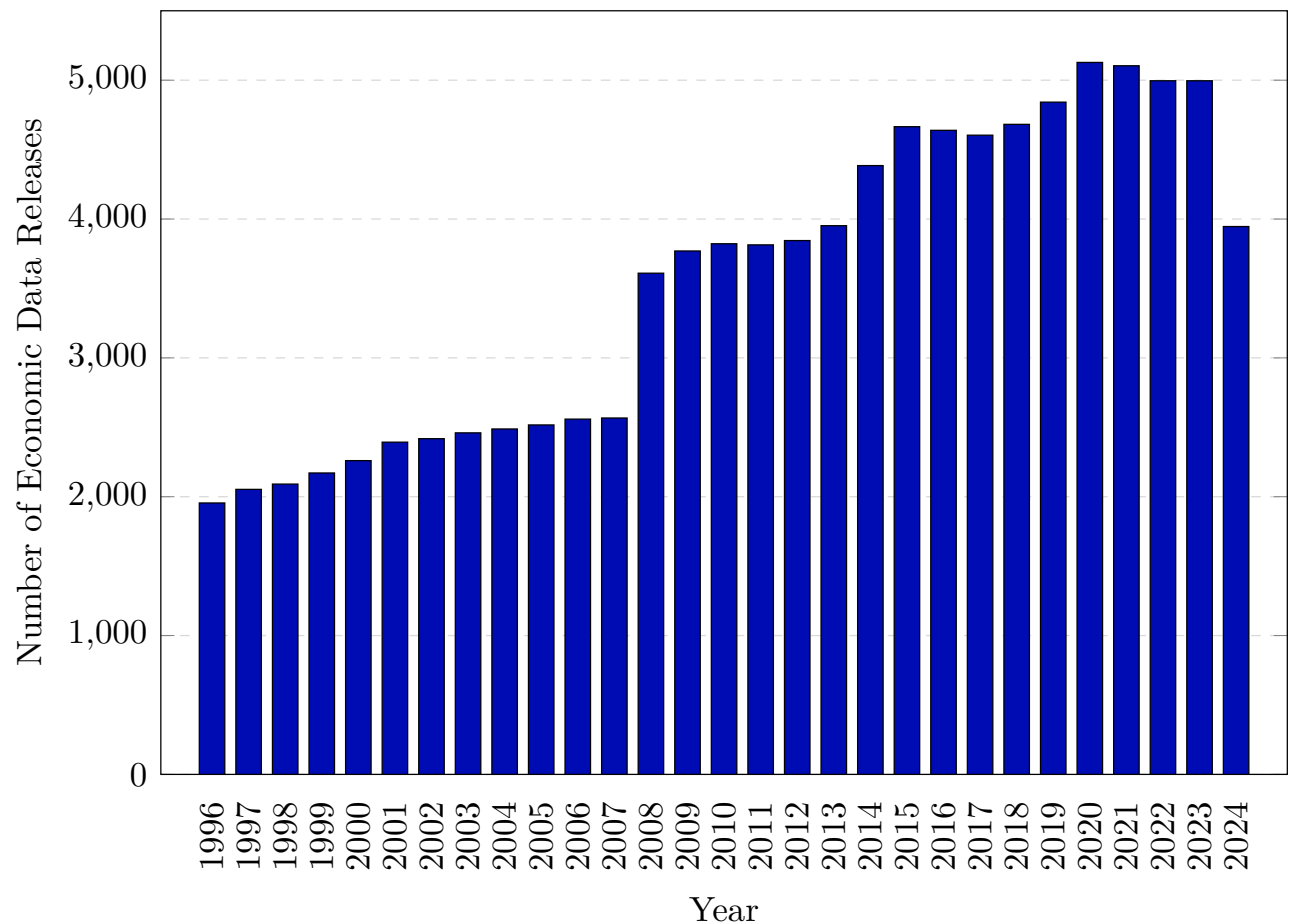

\end{landscape}

\begin{landscape}
	
	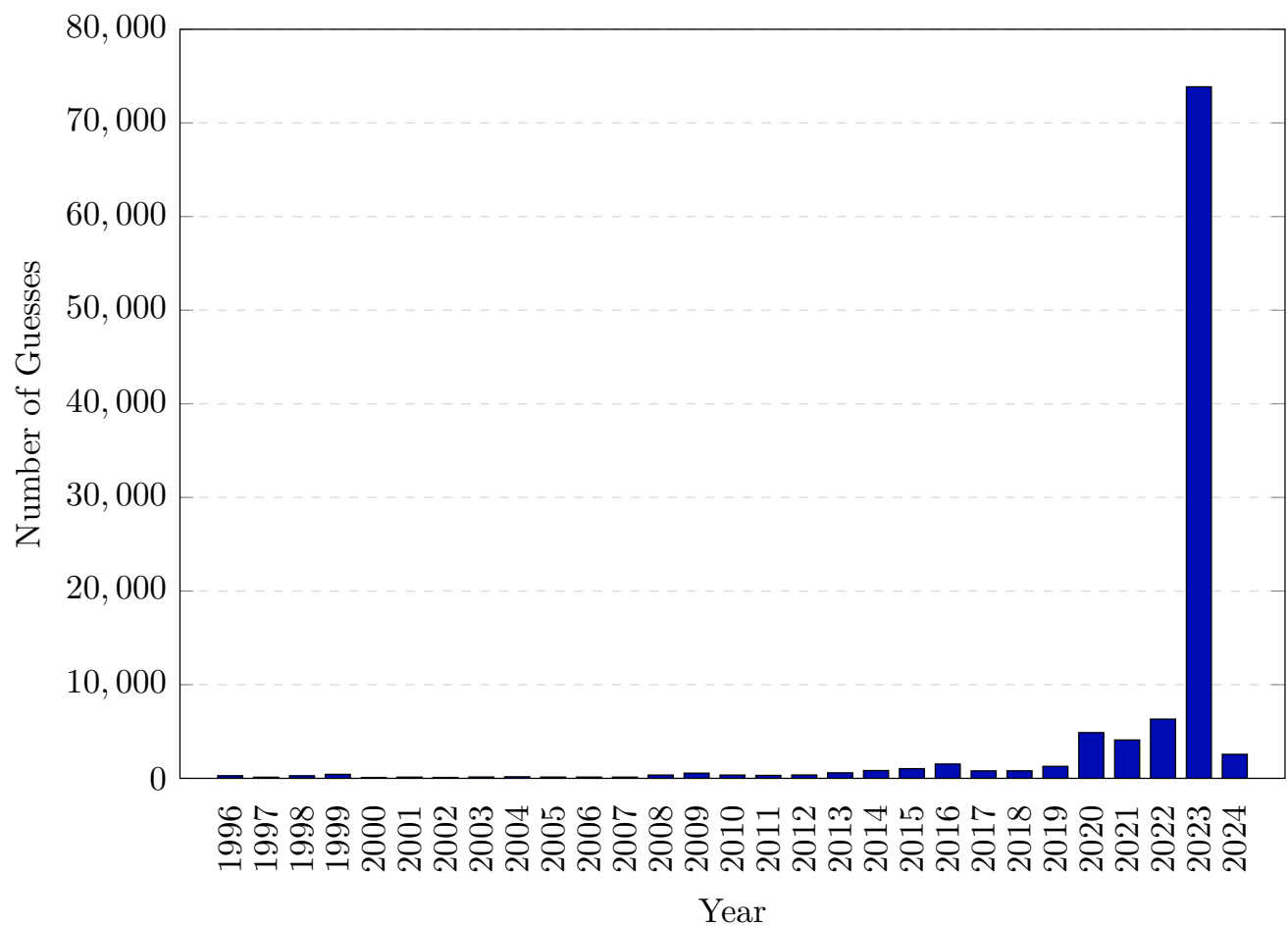
\begin{figure}

	    \begin{center}
	    	\caption{\bf Distribution of AI's Guesses}
		    \scalebox{1.2}{
\pgfplotsset{compat=newest}
	\begin{tikzpicture}
		\begin{axis}[
			width=14cm,   
			height=10cm,
			ybar,
			bar width=8pt,
			enlarge x limits=0.05,
			xtick=data, 
			xlabel={Year},
			ylabel={Number of Guesses},
			ymin=0,
			ymax=80000,
			scaled y ticks=false,  
			xtick style={draw=none},
			ymajorgrids=true,
			grid style={dashed,gray!30},
			ticklabel style={font=\small},
			label style={font=\small},
			xticklabel style={
				rotate=90,
				anchor=east,
				/pgf/number format/1000 sep={}
			},
			y tick label style={
				/pgf/number format/1000 sep={,}
			},
			]
			\addplot [
			fill=blue!70!black,
			draw=black
			] 
			table[
			x=year,
			y=how_many_guess_each_year,
			col sep=comma
			]{xtra/figures/Files/data/how_many_guess_each_year.csv};
		\end{axis}
	\end{tikzpicture}}
		    \medskip
		     \label{fig:LookAheadBias:how many guess each year}
	    \end{center}

	    \begin{small}
	    This figure displays the distribution of GPT-4o’s guesses when tasked with identifying the year (not the exact date) in which each data release occurred, using \myprp{prompt:LookAheadBias:guess the year for lookahead bias analysis}. This exercise serves as a test for potential look-ahead bias. In the experiment, I use the same inputs originally provided to the AI model via \myprp{prompt:VariableConstruction:financial analyst prompt}, but instead of asking for an economic analysis, the model is asked to guess the year of the data release. The sample period spans from January 1996 to October 2024.
	    \end{small}

	\end{figure}

\end{landscape}

\begin{landscape}
	
	\begin{figure}

	    \begin{center}
	    	\caption{\bf Distribution of AI's Correlation Estimates}
		    \scalebox{1.2}{
\pgfplotsset{compat=newest}
	\begin{tikzpicture}
		\begin{axis}[
			width=14cm,   
			height=10cm,
			grid=major,
			xlabel={Correlation},
			ylabel={Frequency},
			legend pos=north west,
			legend style={
				font=\small,
				scale=0.2,         
				transform shape,   
			},
			ymin=0, ymax=600,            
			enlarge x limits=0.05,
			scaled y ticks=false,
			scaled x ticks=false,
			yticklabel style={
				/pgf/number format/fixed,
				/pgf/number format/precision=3
			},
			xticklabel style={
				/pgf/number format/fixed,
				/pgf/number format/precision=3
			}
			]
			
			\addplot+ [
			ybar,
			hist={ bins=10 },
			fill=blue,
			draw=black,   
			no marks
			] 
			table[
			x=Correlation, 
			col sep=comma
			]{xtra/figures/Files/data/GBP_CPI_MOM.csv};
			\addlegendentry{GPT-4o Estimates}
			
			\draw[color=red!80!white, line width=2pt, dashed] 
			(axis cs:-0.018,0) -- (axis cs:-0.018,700);
			
			\addlegendimage{line legend, color=red!80!white, dashed, line width=2pt}
			\addlegendentry{True Realized Value}
			
		\end{axis}
	\end{tikzpicture}}
		    \medskip
		      \label{fig:LookAheadBias:GBP CPI(MOM)}
	    \end{center}

	    \begin{small}
	    Distribution of GPT-4o’s correlation estimates between the United Kingdom’s monthly CPI and next-month GBPUSD returns over the period from January 1996 to October 2023, compared with the true realized correlation over the same period. The correlation estimates were generated using \myprp{prompt:LookAheadBias:remember the correlaton macro var currency returns}, and the exercise was repeated 1{,}000 times to obtain the distribution.
	    \end{small}

	\end{figure}
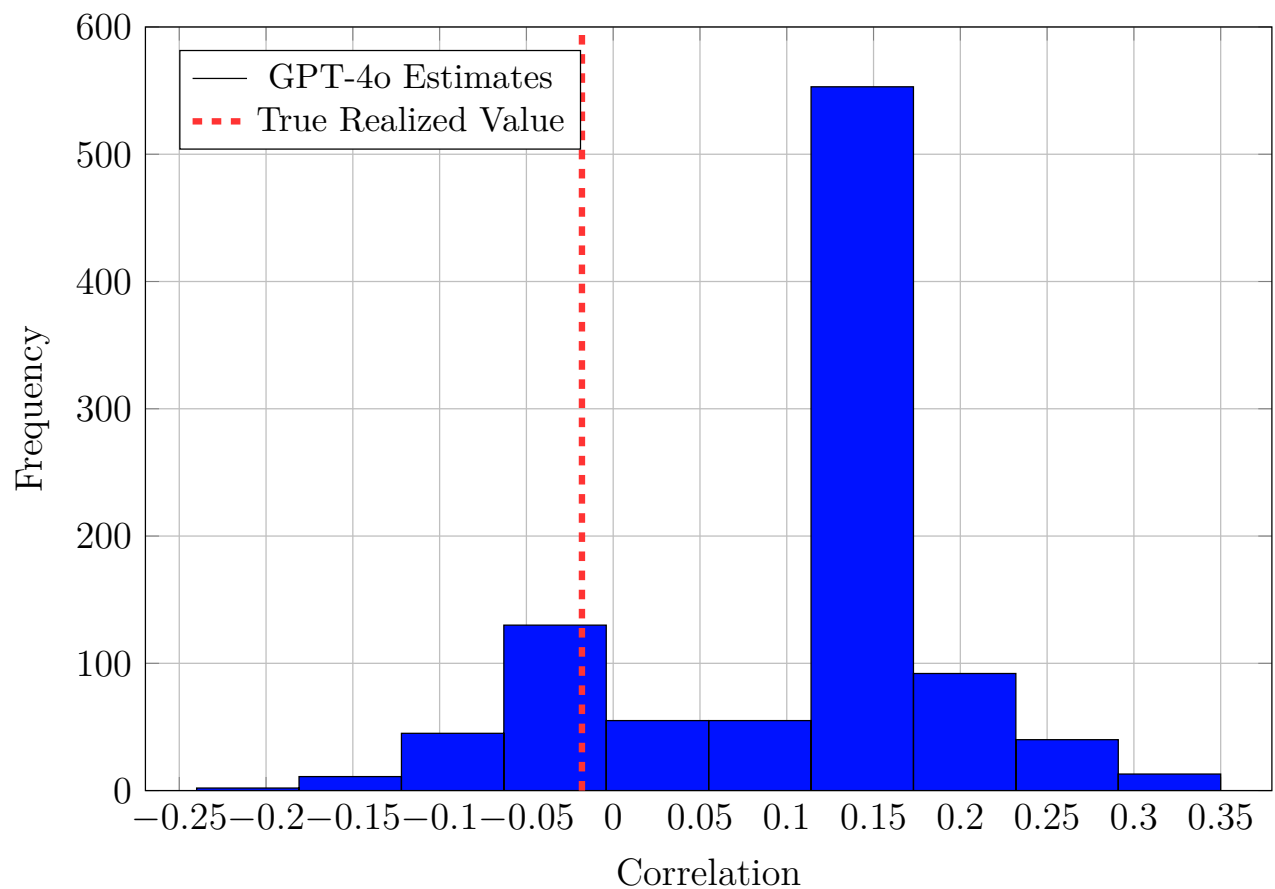

\end{landscape}

\begin{landscape}
	
	\begin{figure}

	    \begin{center}
	    	\caption{\bf Distribution of AI's Correlation Estimates}   
		    \scalebox{1.2}{
\pgfplotsset{compat=newest}
	\begin{tikzpicture}
		\begin{axis}[
			width=14cm,   
			height=10cm,
			grid=major,
			xlabel={Correlation},
			ylabel={Frequency},
			legend pos=north east,
			legend style={
				font=\small,
				scale=0.2,         
				transform shape,   
			},
			ymin=0, ymax=650,            
			enlarge x limits=0.05,
			scaled y ticks=false,
			scaled x ticks=false,
			yticklabel style={
				/pgf/number format/fixed,
				/pgf/number format/precision=3
			},
			xticklabel style={
				/pgf/number format/fixed,
				/pgf/number format/precision=3
			}
			]
			
			\addplot+ [
			ybar,
			hist={ bins=10 },
			fill=blue,
			draw=black,   
			no marks
			] 
			table[
			x=Correlation, 
			col sep=comma
			]{xtra/figures/Files/data/GBP_monthly_unemployment_rate.csv};
			\addlegendentry{GPT-4o Estimates}
			
			\draw[color=red!80!white, line width=2pt, dashed] 
			(axis cs:0.043,0) -- (axis cs:0.043,700);
			
			\addlegendimage{line legend, color=red!80!white, dashed, line width=2pt}
			\addlegendentry{True Realized Value}
			
		\end{axis}
	\end{tikzpicture}}
		    \medskip
		    \label{fig:LookAheadBias:GBP monthly unemployment rate}
	    \end{center}

	    \begin{small}
	    Distribution of GPT-4o’s correlation estimates between the United Kingdom’s monthly unemployment rate and next-month GBPUSD returns over the period from January 1996 to October 2023, compared with the true realized correlation over the same period. The correlation estimates were generated using \myprp{prompt:LookAheadBias:remember the correlaton macro var currency returns}, and the exercise was repeated 1{,}000 times to obtain the distribution.
	    \end{small}

	\end{figure}
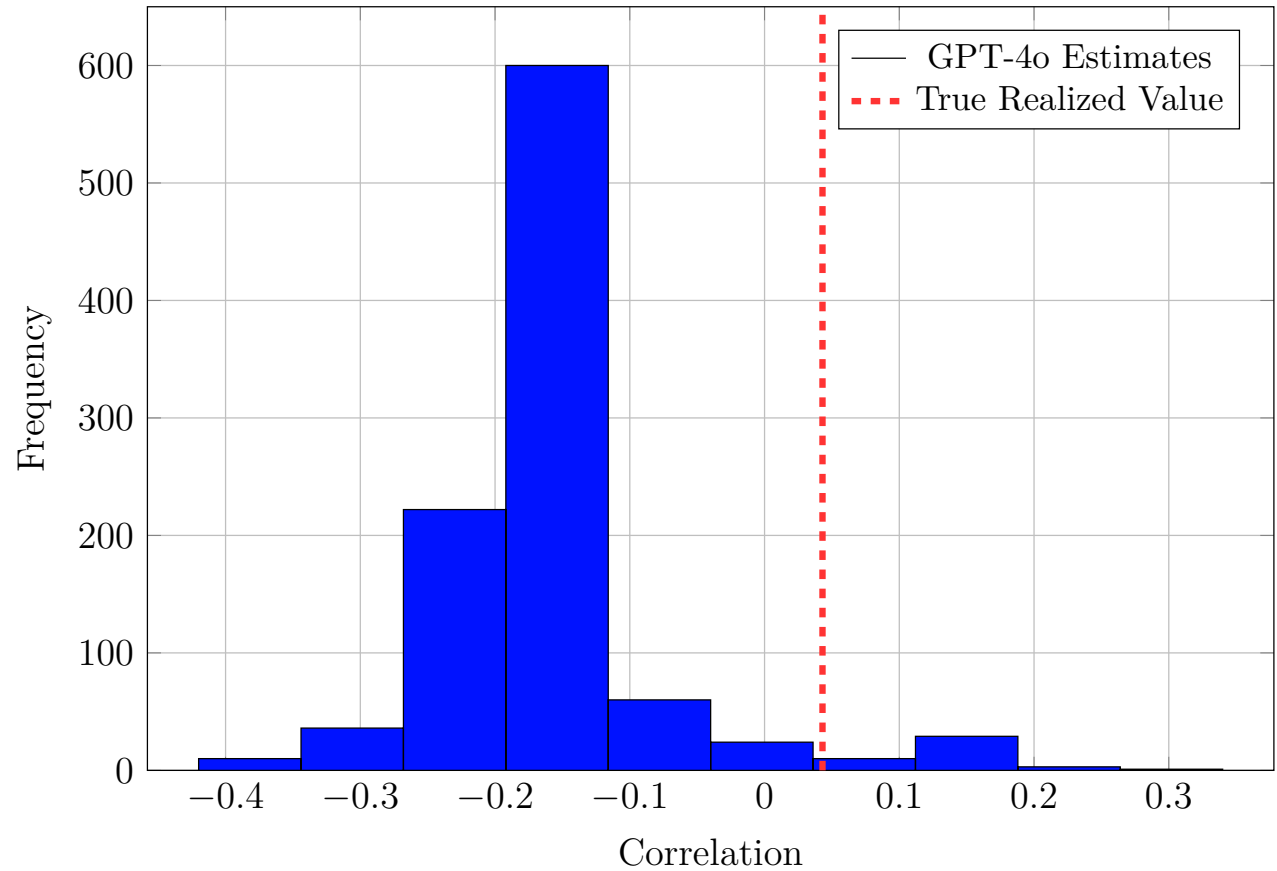

\end{landscape}

\begin{landscape}
	
	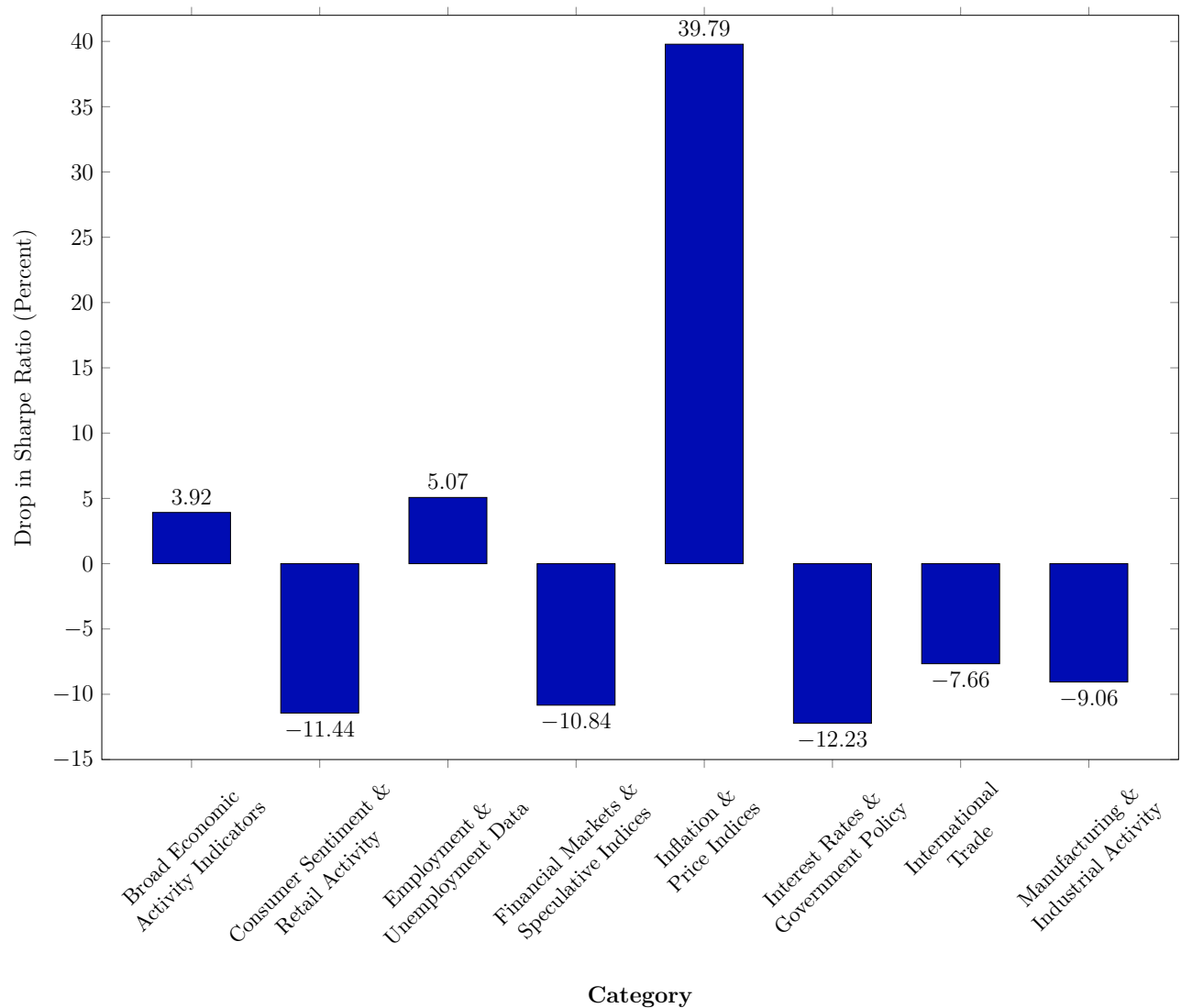
\begin{figure}

	    \begin{center}
	    	\caption{\bf Percentage Drop in Sharpe Ratio}
		    \scalebox{0.8}{
\pgfplotsset{compat=1.17}
	\begin{tikzpicture}
		\begin{axis}[
			width=21cm,   
			height=15cm,
			ybar,
			bar width=40pt,
			symbolic x coords={
				{Broad Economic \\ Activity Indicators},
				{Consumer Sentiment \& \\ Retail Activity},
				{Employment \& \\ Unemployment Data},
				{Financial Markets \& \\ Speculative Indices},
				{Inflation \& \\ Price Indices},
				{Interest Rates \& \\ Government Policy},
				{International \\ Trade},
				{Manufacturing \& \\ Industrial Activity}
			},
			xtick=data,
			x tick label style={rotate=45, align=center, font=\small},
			xlabel={\textbf{Category}},
			ylabel={Drop in Sharpe Ratio (Percent)},
			ymin=-15, 
			ymax=42,
			nodes near coords,
			enlarge x limits=0.1
			]
			\addplot[
			fill=blue!70!black,
			draw=black
			] 
			coordinates {
				({Broad Economic \\ Activity Indicators}, 3.923)
				({Consumer Sentiment \& \\ Retail Activity}, -11.441)
				({Employment \& \\ Unemployment Data}, 5.073)
				({Financial Markets \& \\ Speculative Indices}, -10.837)
				({Inflation \& \\ Price Indices}, 39.786)
				({Interest Rates \& \\ Government Policy}, -12.226)
				({International \\ Trade}, -7.662)
				({Manufacturing \& \\ Industrial Activity}, -9.062)
			};
		\end{axis}
	\end{tikzpicture}}
		    \medskip
		     \label{fig:Categories:margianl contribution to sharpe ratio categores cs strategy diff}
	    \end{center}

	    \begin{small}
	    This figure reports the percentage drop in Sharpe ratio of the \textit{AIFX strategy}, following a leave-one-out approach. Specifically, the baseline strategy is re-estimated multiple times, each time excluding one of the eight categories from the construction of the \textit{AIFX index} (defined in \myeq{eq:VariableConstruction:definition of diff ratio}). The reported values represent the average percentage drop in Sharpe ratio across lookback periods ranging from 36 to 60 months. The sample period spans from January 1996 to October 2024.
	    \end{small}

	\end{figure}

\end{landscape}

\begin{landscape}
	
	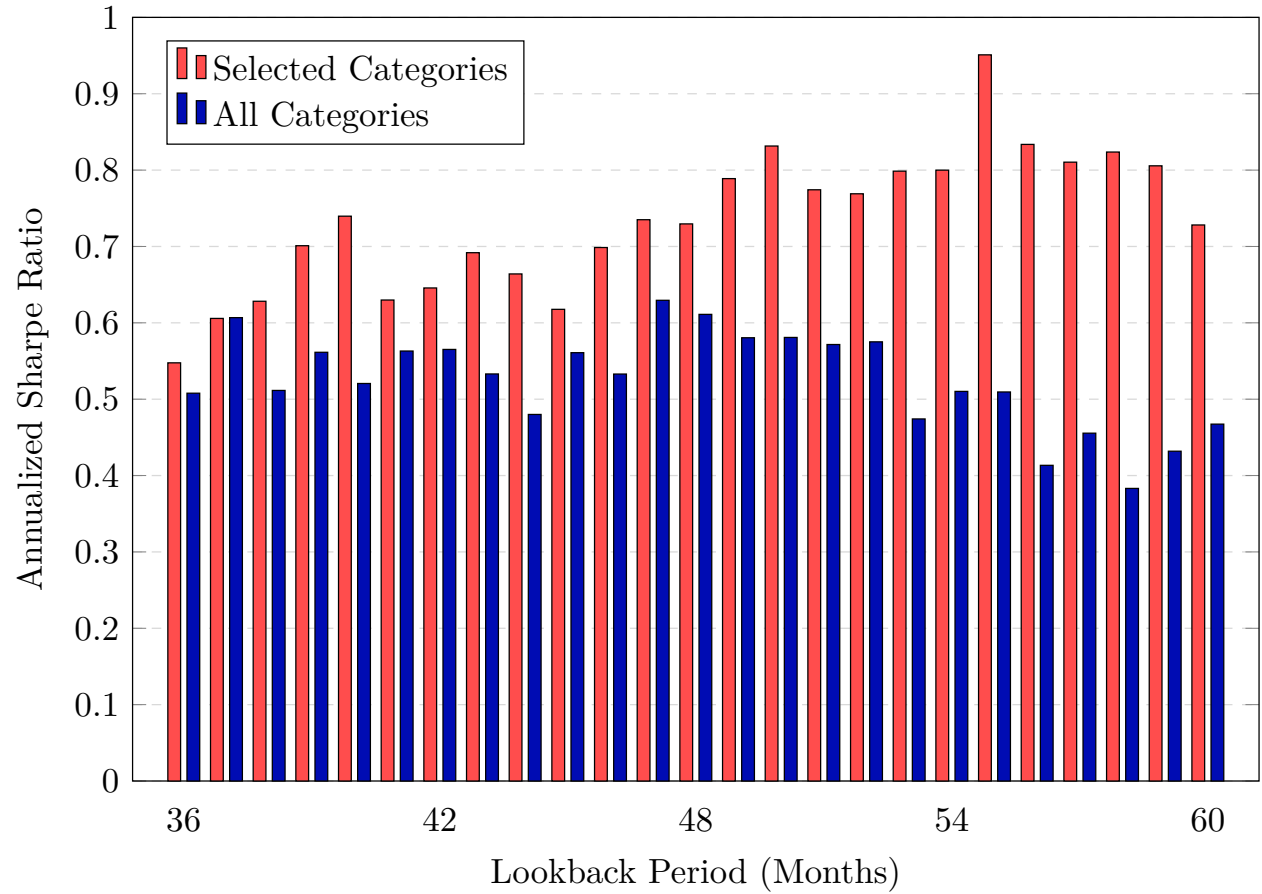
\begin{figure}

	    \begin{center}
	    	\caption{\bf Performance of Selected Categories}
		    \scalebox{1.2}{
\pgfplotsset{compat=newest}
	\begin{tikzpicture}
		\begin{axis}[
			width=14cm,   
			height=10cm,
			ybar,
			bar width=4pt,
			enlarge x limits=0.05,
			xlabel={Lookback Period (Months)},
			ylabel={Annualized Sharpe Ratio},
			ymin=0,
			ymax=1,
			ytick={0,0.1,0.2,0.3,0.4,0.5,0.6,0.7,0.8,0.9,1},
			xtick={36,42,48,54,60},
			xtick style={draw=none},
			ymajorgrids=true,
			grid style={dashed,gray!30},
			ticklabel style={font=\small},
			label style={font=\small},
			legend pos=north west, 
			legend style={font=\small}, 
			legend cell align={left} 
			]
			\addplot [
			fill=red!70!white,
			draw=black
			] 
			table[
			x=lookback_period_number_of_months,
			y=selected_categories_diff_strategy,
			col sep=comma
			]{xtra/figures/Files/data/CS_strategy_sharpe_ratio_all_categories_vs_top_three.csv};
			\addlegendentry{Selected Categories} 
			
			\addplot [
			fill=blue!70!black,
			draw=black
			] 
			table[
			x=lookback_period_number_of_months,
			y=all_categories_diff_strategy,
			col sep=comma
			]{xtra/figures/Files/data/CS_strategy_sharpe_ratio_all_categories_vs_top_three.csv};
			\addlegendentry{All Categories} 
		\end{axis}
	\end{tikzpicture}}
		    \medskip
		     \label{fig:Categories:CS strategy sharpe ratio all categories vs top three}
	    \end{center}

	    \begin{small}
	    The figures compares the performance of the \textit{AIFX strategy}, constructed using all eight categories of data releases, with an alternative strategy that uses only the top three categories: ‘Inflation data’, ‘Employment data’, and ‘Broad economic activity indicators’. The cross-sectional strategy uses the \textit{AIFX index} (defined in \myeq{eq:VariableConstruction:definition of diff ratio}) as the trading signal. For each choice of lookback period, currencies are sorted by their \textit{AIFX index} at the end of each month and I take long positions in the top two currencies with the highest \textit{AIFX index} and short positions in the bottom two with the lowest. Portfolios are rebalanced at the end of each calendar month. Lookback periods of 36 to 60 months are considered. The sample period spans from January 1996 to October 2024, but returns start at different dates due to differences in lookback periods.
	    \end{small}

	\end{figure}

\end{landscape}

\begin{landscape}
	
	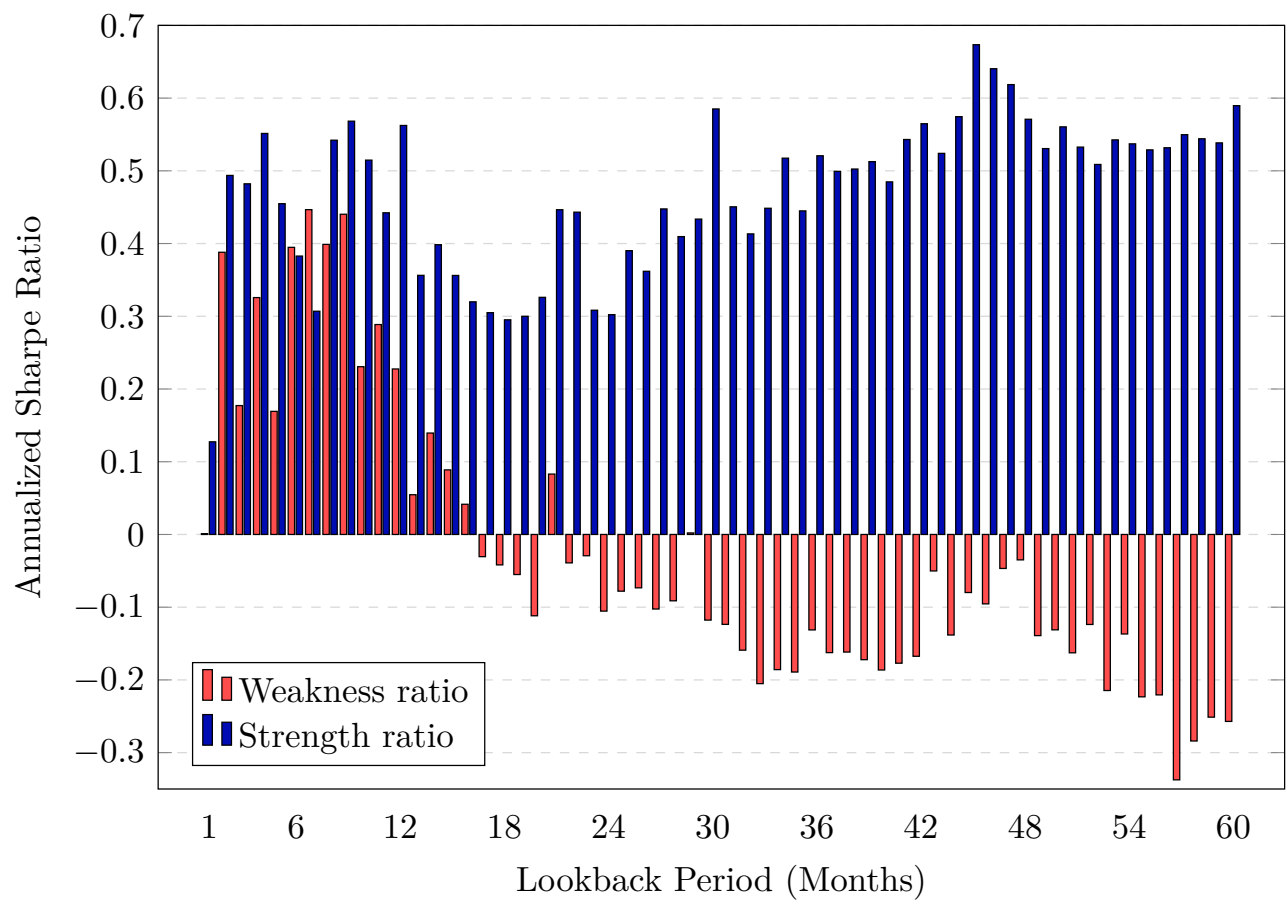
\begin{figure}

	    \begin{center}
	    	\caption{\bf Sharpe Ratios (\textit{Strength ratio} vs \textit{Weakness ratio})}
		    \scalebox{1.2}{
\pgfplotsset{compat=newest}
	\begin{tikzpicture}
		\begin{axis}[
			width=14cm,   
			height=10cm,
			ybar,
			bar width=2pt,
			enlarge x limits=0.05,
			xlabel={Lookback Period (Months)},
			ylabel={Annualized Sharpe Ratio},
			ymin=-0.35,
			ymax=0.7,
			ytick={-0.3,-0.2,-0.1,0.0,0.1,0.2,0.3,0.4,0.5,0.6,0.7},
			xtick={1,6,12,18,24,30,36,42,48,54,60},
			xtick style={draw=none},
			ymajorgrids=true,
			grid style={dashed,gray!30},
			ticklabel style={font=\small},
			label style={font=\small},
			legend pos=south west, 
			legend style={font=\small}, 
			legend cell align={left} 
			]
			
			\addplot [
			fill=red!70!white,
			draw=black,
			bar shift=-1.5pt 
			] 
			table[
			x=lookback_period_number_of_months,
			y=negative_ratio_strategy_sharpe,
			col sep=comma
			]{xtra/figures/Files/data/sharpe_ratio_cs_strategy_positive_negative_ratio_together.csv};
			\addlegendentry{Weakness ratio}
			
			\addplot [
			fill=blue!70!black,
			draw=black,
			bar shift=1pt 
			] 
			table[
			x=lookback_period_number_of_months,
			y=positive_ratio_strategy_sharpe,
			col sep=comma
			]{xtra/figures/Files/data/sharpe_ratio_cs_strategy_positive_negative_ratio_together.csv};
			\addlegendentry{Strength ratio}

		\end{axis}
	\end{tikzpicture}}
		    \medskip
		     \label{fig:DifferentialImpact:sharpe ratio cs strategy positive negative ratio together}
	    \end{center}

	    \begin{small}
	    This figure reports the annualized Sharpe ratios of cross-sectional strategies that use the \textit{Strength ratio} (defined in \myeq{eq:VariableConstruction:definition of pos ratio}) and \textit{Weakness ratio} (defined in \myeq{eq:VariableConstruction:definition of neg ratio}) as the trading signal. For the first strategy, for each choice of lookback period, currencies are sorted based on their \textit{Strength ratio} at the end of each month, and I take long positions in the top two currencies with the highest values and short positions in the bottom two with the lowest. Portfolios are rebalanced at the end of each calendar month. In the second strategy, I use the \textit{Weakness ratio} as the signal but sort currencies in reverse order. Specifically, at the end of each month, I go long the two currencies with the lowest \textit{Weakness ratio} and short the two with the highest. The rebalancing schedule and lookback period specifications remain the same as those in the strategy based on the \textit{Strength ratio}. Lookback periods of 1 to 60 months are considered. The sample period spans from January 1996 to October 2024, but returns start at different dates due to differences in lookback periods.
	    \end{small}
	    
	\end{figure}

\end{landscape}

\begin{landscape}
	
	\begin{figure}

	    \begin{center}
	    	\caption{\bf $t$-statistics (\textit{Strength ratio} vs \textit{Weakness ratio})}
		    \scalebox{1.2}{
\pgfplotsset{compat=newest}
	\begin{tikzpicture}
		\begin{axis}[
			width=14cm,   
			height=10cm,
			ybar,
			bar width=2pt,
			enlarge x limits=0.05,
			xlabel={Lookback Period (Months)},
			ylabel={T-stat},
			ymin=-2.2,
			ymax=4.5,
			xtick={1,6,12,18,24,30,36,42,48,54,60},
			xtick style={draw=none},
			ticklabel style={font=\small},
			label style={font=\small},
			legend pos=north west, 
			legend style={font=\small}, 
			legend cell align={left} 
			]
			
			\addplot [
			fill=red!70!white,
			draw=black,
			bar shift=-1.5pt 
			] 
			table[
			x=number_of_lookback_months,
			y=ratio_negative_t_stat,
			col sep=comma
			]{xtra/figures/Files/data/t-stat_for_regression_of_return_on_positive_negative_ratio_CS.csv};
			\addlegendentry{T-statistics (Weakness ratio)}
			
			\addplot [
			fill=blue!70!black,
			draw=black,
			bar shift=1pt 
			] 
			table[
			x=number_of_lookback_months,
			y=ratio_positive_t_stat,
			col sep=comma
			]{xtra/figures/Files/data/t-stat_for_regression_of_return_on_positive_negative_ratio_CS.csv};
			\addlegendentry{T-statistics (Strength ratio)}
			
			\pgfplotsextra{
				\draw[dashed,gray,line width=1.5pt] 
				(axis cs:\pgfkeysvalueof{/pgfplots/xmin},1.96) -- 
				(axis cs:\pgfkeysvalueof{/pgfplots/xmax},1.96);
				\draw[dashed,gray,line width=1.5pt] 
				(axis cs:\pgfkeysvalueof{/pgfplots/xmin},-1.96) -- 
				(axis cs:\pgfkeysvalueof{/pgfplots/xmax},-1.96);
			}
			\addlegendimage{line legend, gray, dashed, line width=3pt}
			\addlegendentry{(5\%) Significance Level}
		\end{axis}
	\end{tikzpicture}}
		    \medskip
		     \label{fig:DifferentialImpact:t-stat for regression of return on positive negative ratio CS}
	    \end{center}

	    \begin{small}
	    This figure displays the t-statistics of \(\beta_1\) and \(\beta_2\) in \myeq{eq:DifferentialImpact:regression return on positive negative ratio cs}. The regression is repeated for different choices of lookback period. This regression examines whether the \textit{Strength ratio} and \textit{Weakness ratio} of a currency at the end of a given month can predict the currency's excess return in the subsequent month. Standard errors are clustered at the currency and time level. The regression includes time fixed effects, although not reported here. The sample period spans from January 1996 to October 2024, but returns start at different dates due to differences in lookback periods. The dashed horizontal lines mark the 5\% statistical significance threshold. 
	    \end{small}

	\end{figure}
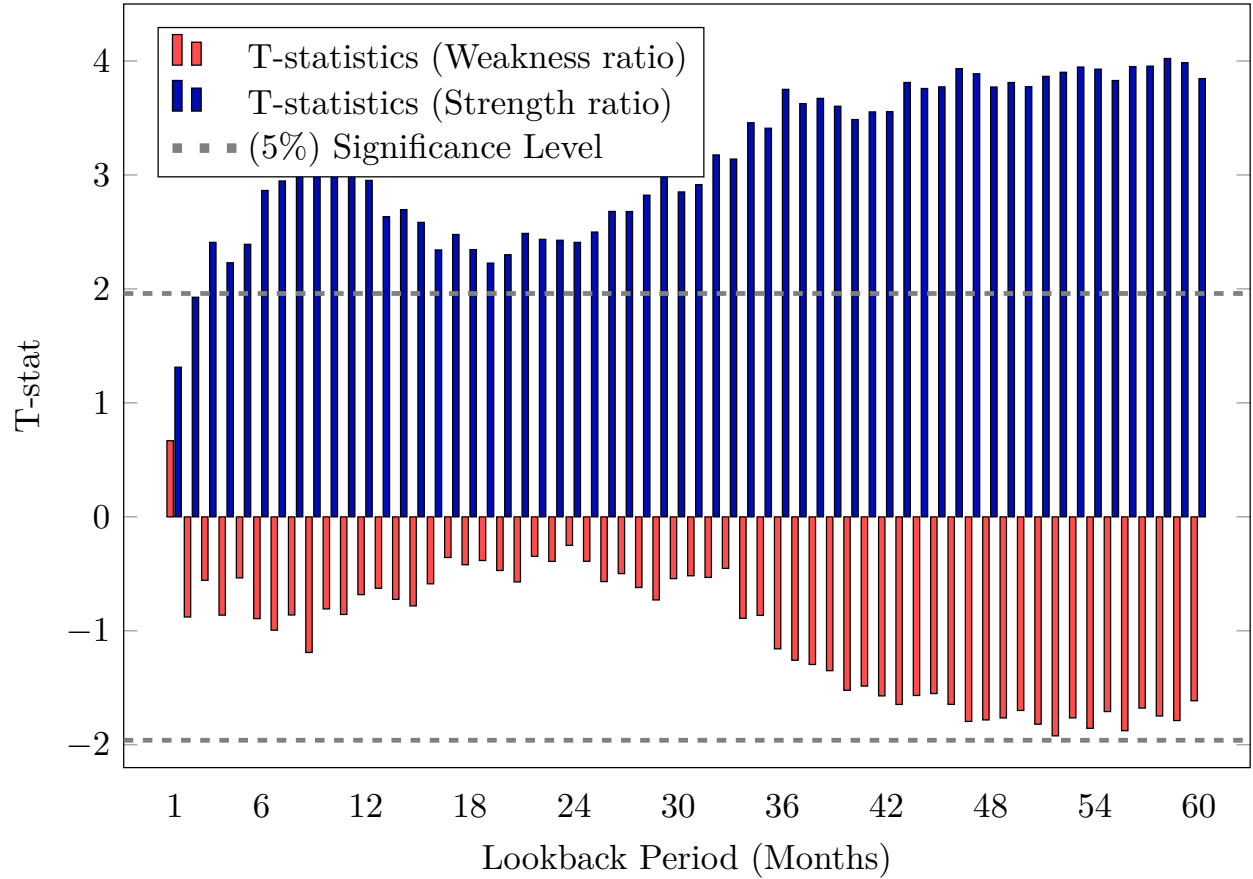

\end{landscape}

\begin{landscape}
	
	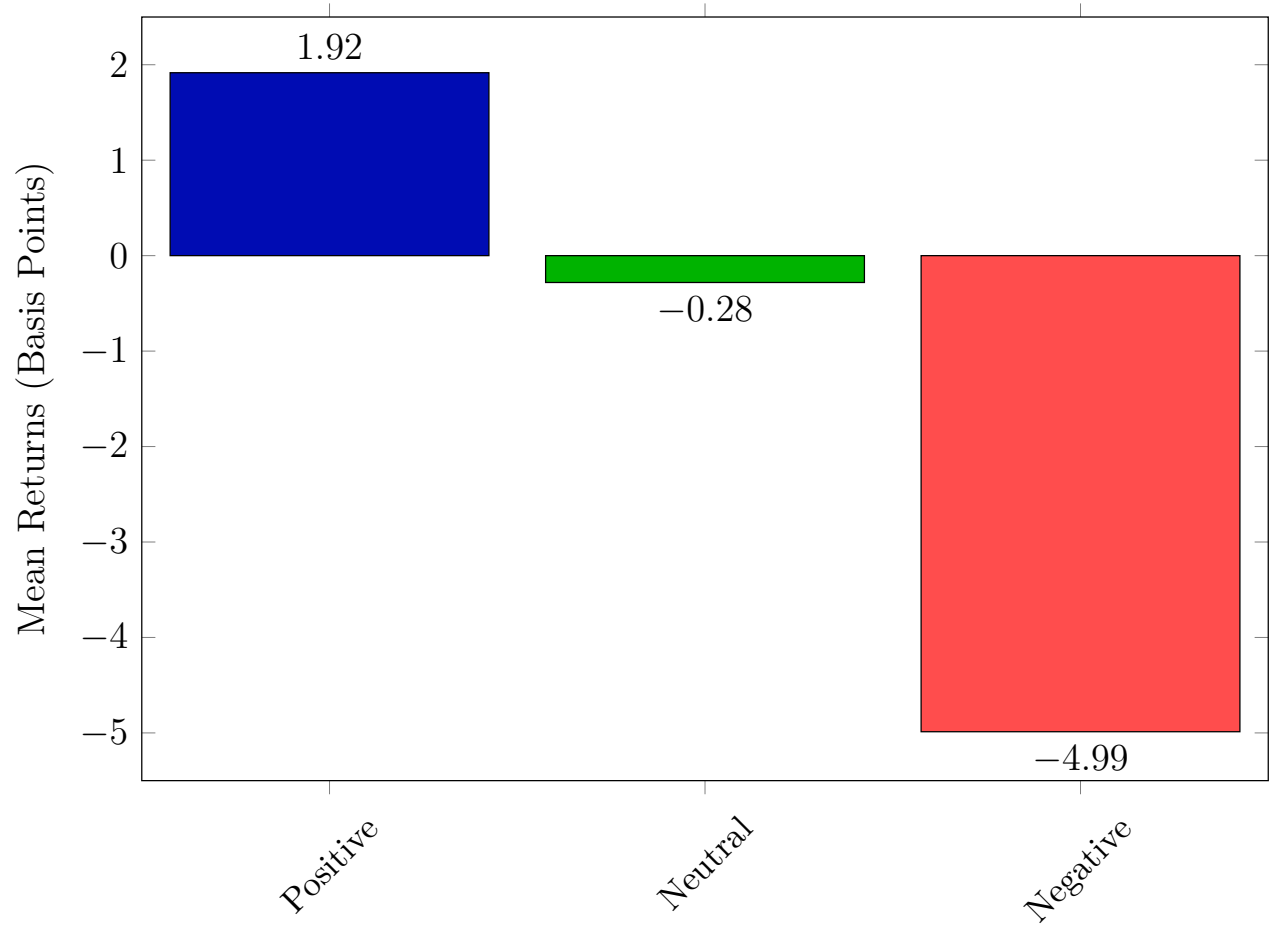
\begin{figure}

	    \begin{center}
	    	\caption{\bf Average Daily Returns}
		    \scalebox{1.2}{
\pgfplotsset{compat=1.17}
	\begin{tikzpicture}
		\begin{axis}[
			width=14cm,
			height=10cm,
			ybar,
			bar width=100pt,
			symbolic x coords={Positive, Neutral, Negative},
			xtick={Positive, Neutral, Negative},
			x tick label style={rotate=45, align=center, font=\small},
			ylabel={Mean Returns (Basis Points)},
			ymin=-5.5, 
			ymax=2.5,
			bar shift=0pt,
			nodes near coords,
			nodes near coords style={
				text=black,
				/pgf/number format/.cd,
				fixed,
				fixed zerofill,
				precision=2
			},
			enlarge x limits=0.25,
			scaled ticks=false,
			]
			
			\addplot+[fill=blue!70!black,draw=black] 
			coordinates {(Positive, 1.9164)};
			
			\addplot+[fill=green!70!black,draw=black] 
			coordinates {(Neutral, -0.2813)};
			
			\addplot+[fill=red!70!white,draw=black] 
			coordinates {(Negative, -4.9876)};
		\end{axis}
	\end{tikzpicture}}
		    \medskip
		     \label{fig:DifferentialImpact:pos neg differential impact smaller window}
	    \end{center}

	    \begin{small}
	    This figure compares the average realized returns on days associated with each type of news. Each of the data releases in the dataset is labeled as positive, neutral or negative using the output of \myprp{prompt:VariableConstruction:financial analyst prompt}. Specifically, I refer to data releases with the direction labeled as STRENGTHEN in the output as positive news, those labeled as WEAKEN as negative news, and those labeled as INSIGNIFICANT OR UNCERTAIN as neutral news. The sample period for this exercise is from 2008 to 2024.
	    \end{small}

	\end{figure}

\end{landscape}

 \begin{landscape}
	
	\begin{table}[ht]
		\caption{\bf Frequency Table of Data Releases} \label{tab:data:headline frequency table}
		
		\begin{footnotesize}
			This table displays the distribution of data releases for each of the ten major currencies included in the dataset, which was sourced from Investing.com's economic calendar data spanning January 1996 to October 2024. 'Total' indicates the overall number of observations for each currency, while 'Positive,' 'Negative,' and 'Neutral' denote the classification of data releases by GPT-4o using \myprp{prompt:VariableConstruction:financial analyst prompt}, based on their impact on the currency's returns. The percentages in parentheses represent the proportion of each category relative to the total observations for that currency.	
		\end{footnotesize}

		\bigskip
		
		\newcolumntype{L}[1]{>{\raggedright\let\newline\\\arraybackslash\hspace{0pt}}m{#1}}
		\newcolumntype{C}[1]{>{\centering\let\newline\\\arraybackslash\hspace{0pt}}m{#1}}
		\newcolumntype{R}[1]{>{\raggedleft\let\newline\\\arraybackslash\hspace{0pt}}m{#1}}
		\setlength\extrarowheight{1pt}

		\sisetup{ 
			detect-all,
			table-number-alignment = center, 
			output-decimal-marker = {,}, 
			table-align-text-pre = false,
			table-align-text-post = false,
			group-digits = integer,
			table-format = 3.3
		}

	\centering
	\scalebox{1.00}{					
		\begin{tabular}{*{8}{c}}
	
	\toprule
	\multicolumn{1}{c}{\textbf{Currency}} & 
	\multicolumn{1}{c}{\textbf{Total}} & 
	\multicolumn{1}{c}{\textbf{Positive}} & 
	\multicolumn{1}{c}{\textbf{(\%)}} & 
	\multicolumn{1}{c}{\textbf{Negative}} & 
	\multicolumn{1}{c}{\textbf{(\%)}} & 
	\multicolumn{1}{c}{\textbf{Neutral}} & 
	\multicolumn{1}{c}{\textbf{(\%)}} \\ 
	\midrule
	AUD & 12,134 & 4,347 & 36 & 4,885 & 40 & 2,902 & 24 \\ 
	CAD & 12,652 & 4,462 & 35 & 4,837 & 38 & 3,353 & 27 \\ 
	CHF & 5,554  & 1,825 & 33 & 1,952 & 35 & 1,777 & 32 \\ 
	EUR & 13,698 & 4,097 & 30 & 4,550 & 33 & 5,051 & 37 \\ 
	GBP & 20,261 & 6,266 & 31 & 7,246 & 36 & 6,749 & 33 \\ 
	JPY & 17,846 & 5,736 & 32 & 6,760 & 38 & 5,350 & 30 \\ 
	NZD & 9,261  & 3,589 & 39 & 3,792 & 41 & 1,880 & 20 \\ 
	NOK & 4,864  & 1,509 & 31 & 1,547 & 32 & 1,808 & 37 \\ 
	SEK & 6,490  & 2,075 & 32 & 2,498 & 38 & 1,917 & 30 \\ 
	USD & 72,060 & 21,223 & 29 & 20,508 & 28 & 30,329 & 42 \\ 
	ALL & 174,820 & 55,129 & 32 & 58,575 & 33 & 61,116 & 35 \\ 
	\bottomrule	    
\end{tabular}
	}
		
	\end{table}
\end{landscape}

\begin{landscape}
	\begin{table}[h]
		\caption{\bf Summary Statistics}   \label{tab:fx returns summary stats}
	
		\begin{footnotesize}
			This table reports summary statistics for exchange rate returns. All rates are defined as the number of U.S. dollars (USD) required to purchase one unit of foreign currency (FCU). Panel A presents statistics for excess returns, while Panel B shows statistics for spot returns. The reported statistics include mean returns, standard deviation, skewness, excess kurtosis, first-order autocorrelation (AR(1)), and Sharpe ratios. All measures are based on monthly returns, though means, standard deviations, and Sharpe ratios are annualized. The sample period spans January 1996 to October 2024, with the exception of the EUR, for which excess return data begin in January 1999.
		\end{footnotesize}

		\bigskip
		
		\newcolumntype{L}[1]{>{\raggedright\let\newline\\\arraybackslash\hspace{0pt}}m{#1}}
		\newcolumntype{C}[1]{>{\centering\let\newline\\\arraybackslash\hspace{0pt}}m{#1}}
		\newcolumntype{R}[1]{>{\raggedleft\let\newline\\\arraybackslash\hspace{0pt}}m{#1}}
		\setlength\extrarowheight{1pt}

		\sisetup{ 
			detect-all,
			table-number-alignment = center, 
			output-decimal-marker = {.}, 
			group-digits = integer,
			table-format = +1.3,  
			parse-numbers = false  
		}
		
		\centering
		\scalebox{1.00}{
			\begin{tabular}{l | S[table-column-width=1.5cm]
		S[table-column-width=1.4cm]
		S[table-column-width=1.4cm]
		S[table-column-width=1.4cm]
		S[table-column-width=1.4cm]
		S[table-column-width=1.4cm]
		S[table-column-width=1.4cm]
		S[table-column-width=1.4cm]
		S[table-column-width=1.4cm]}
	\hline
	 & {\textbf{AUD}} & {\textbf{CAD}} & {\textbf{CHF}} & {\textbf{EUR}} & {\textbf{GBP}} & {\textbf{JPY}} & {\textbf{NZD}} & {\textbf{NOK}} & {\textbf{SEK}} \\ 
	 & \multicolumn{9}{c}{\textbf{Panel A. Excess Returns}} \\  \hline
	Mean               & 0.993 & -0.141 & -0.953 & -1.105 & -0.182 & -3.873 & 1.602 & -1.325 & -2.080 \\
	Standard deviation & 11.916 & 8.199 & 9.812 & 9.393 & 8.410 & 10.352 & 12.353 & 11.347 & 10.765 \\
	Skewness           & -0.379 & -0.486 & 0.186 & -0.073 & -0.337 & 0.446 & -0.265 & -0.196 & 0.024 \\
	Excess kurtosis    & 1.599 & 3.338 & 1.578 & 1.259 & 1.304 & 2.356 & 1.123 & 0.693 & 0.364 \\
	AR(1)              & 0.031 & -0.064 & -0.044 & 0.038 & 0.022 & 0.005 & -0.014 & 0.010 & 0.036 \\
	Sharpe Ratio       & 0.083 & -0.017 & -0.097 & -0.118 & -0.022 & -0.374 & 0.130 & -0.117 & -0.193 \\ 
	 & \multicolumn{9}{c}{\textbf{Panel B. Spot Returns}} \\ 		
	 \hline
	Mean               & -0.437 & -0.048 & 1.175 & -0.603 & -0.556 & -1.220 & -0.403 & -1.825 & -1.483 \\
	Standard deviation & 11.864 & 8.193 & 9.779 & 9.257 & 8.399 & 10.344 & 12.337 & 11.308 & 10.722 \\
	Skewness           & -0.421 & -0.501 & 0.164 & -0.126 & -0.343 & 0.492 & -0.282 & -0.216 & 0.001 \\
	Excess kurtosis    & 1.710 & 3.360 & 1.608 & 1.237 & 1.373 & 2.600 & 1.209 & 0.765 & 0.403 \\
	AR(1)              & 0.022 & -0.065 & -0.052 & 0.030 & 0.019 & 0.001 & -0.017 & 0.003 & 0.027 \\
	Sharpe Ratio       & -0.037 & -0.006 & 0.120 & -0.065 & -0.066 & -0.118 & -0.033 & -0.161 & -0.138 \\ \hline
\end{tabular}
						}
	\end{table}
\end{landscape}

\begin{landscape}
	\begin{table}[h]
		\caption{\bf Performance Statistics}   \label{tab:Performance:cs AI strategy diff performance statistics}
	
		\begin{footnotesize}
			This table reports the performance statistics for the cross-sectional strategies that use the \textit{AIFX index} (defined in \myeq{eq:VariableConstruction:definition of diff ratio}) as the trading signal. I refer to this strategy as the \textit{AIFX strategy}. For each choice of lookback period, currencies are sorted by their \textit{AIFX index} at the end of each month and I take long positions in the top two currencies with the highest \textit{AIFX index} and short positions in the bottom two with the lowest. Portfolios are rebalanced at the end of each calendar month. Results are shown for different lookback periods. The statistics include mean returns, standard deviation, skewness, excess kurtosis, first-order autocorrelation (AR(1)) and Sharpe Ratios. The measures are based on monthly returns, but means, standard deviations and Sharpe ratios are annualized. The performance reported is from January 2001 to October 2024.
		\end{footnotesize}

		\bigskip
		
		\newcolumntype{L}[1]{>{\raggedright\let\newline\\\arraybackslash\hspace{0pt}}m{#1}}
		\newcolumntype{C}[1]{>{\centering\let\newline\\\arraybackslash\hspace{0pt}}m{#1}}
		\newcolumntype{R}[1]{>{\raggedleft\let\newline\\\arraybackslash\hspace{0pt}}m{#1}}
		\setlength\extrarowheight{1pt}

		\sisetup{ 
			detect-all,
			table-number-alignment = center, 
			output-decimal-marker = {.}, 
			group-digits = integer,
			table-format = +1.3,  
			parse-numbers = false  
		}
		
		\centering
		\scalebox{1.00}{
			\begin{tabular}{l|S[table-column-width=2.5cm]S[table-column-width=2.5cm]S[table-column-width=2.5cm]S[table-column-width=2.5cm]S[table-column-width=2.5cm]}
	\hline
	& {\textbf{36 months}}
	& {\textbf{42 months}}
	& {\textbf{48 months}}
	& {\textbf{54 months}}
	& {\textbf{60 months}} \\ \hline
	Mean & 4.060 & 4.261 & 4.354 & 3.603 & 3.197 \\
	Standard deviation & 7.031 & 7.056 & 7.335 & 7.341 & 6.840 \\
	Skewness & -0.086 & 0.058 & -0.152 & -0.137 & -0.480 \\
	Excess kurtosis &  1.777 & 1.902 & 2.634 & 2.566 &1.960 \\
	AR(1)  & -0.015 & -0.001 & 0.008 & -0.006 & -0.010 \\ 
	Sharpe Ratio  &  0.577 &  0.604 & 0.594 &  0.491 & 0.467 \\\hline
\end{tabular}

						}
	\end{table}
\end{landscape}

 \begin{landscape}
 	
	\begin{table}[ht]
		\caption{\bf Performance Over Benchmark Factors} \label{tab:Performance:cs performance over fx strategies}
		
		\begin{footnotesize}
			This table presents the results from the contemporaneous regression specified in \myeq{eq:Performance:regression for performance over fx strategies CS Strategy Diff}, which examines whether the monthly returns of the \textit{AIFX strategy} can be explained by common currency factors. The results are reported for different lookback periods. Benchmark factors include Dollar, Dollar Carry, Cross-sectional Carry, Cross-sectional One-month Momentum, and Cross-sectional Value strategies, as described in \mysecIA{Appendix:Definition of FX factors}. The performance reported is from January 2001 to October 2024. \citet{NEWEY/WEST:1987} standard errors are reported in parentheses. ***, **, and * indicate statistical significance at the 1\%, 5\%, and 10\% levels, respectively.
		\end{footnotesize}

		\bigskip
		
		\newcolumntype{L}[1]{>{\raggedright\let\newline\\\arraybackslash\hspace{0pt}}m{#1}}
		\newcolumntype{C}[1]{>{\centering\let\newline\\\arraybackslash\hspace{0pt}}m{#1}}
		\newcolumntype{R}[1]{>{\raggedleft\let\newline\\\arraybackslash\hspace{0pt}}m{#1}}
		\setlength\extrarowheight{1pt}

		\sisetup{
			detect-all,
			table-format = 1.2,        
			group-digits = false,
			table-number-alignment = center
		}
	
		\centering
		\scalebox{1.00}{					
			\begin{tabular}
	{l
	S[table-format=1.2]
	S[table-format=1.2]
	S[table-format=1.2]
	S[table-format=1.2]
	S[table-format=1.2]}
	
	\toprule
	& \multicolumn{5}{c}{\textbf{Return of the \textit{AIFX strategy}}}\\
	\midrule
	& \textbf{36 months} & \textbf{42 months} & \textbf{48 months} & \textbf{54 months} & \textbf{60 months}\\
	\midrule
	Alpha & 2.88\textsuperscript{**} & 2.88\textsuperscript{**} & 3.24\textsuperscript{**} & 2.40\textsuperscript{*} & 2.04\textsuperscript{*}\\
	& {(1.32)} & {(1.32)} & {(1.32)} & {(1.32)} & {(1.20)}\\
	Dollar & 0.10\textsuperscript{*}  & 0.09\textsuperscript{*}  & 0.09 & 0.06 & 0.06\\
	& {(0.05)} & {(0.05)} & {(0.05)} & {(0.05)} & {(0.04)}\\
	Dollar Carry & 0.14\textsuperscript{***} & 0.12\textsuperscript{***} & 0.16\textsuperscript{***} & 0.15\textsuperscript{***} & 0.17\textsuperscript{***}\\
	& {(0.05)} & {(0.05)} & {(0.05)} & {(0.05)} & {(0.04)}\\
	Carry & 0.29\textsuperscript{***} & 0.35\textsuperscript{***} & 0.33\textsuperscript{***} & 0.41\textsuperscript{***} & 0.41\textsuperscript{***}\\
	& {(0.06)} & {(0.05)} & {(0.06)} & {(0.05)} & {(0.05)}\\
	Value & -0.03 & -0.02 & -0.13\textsuperscript{**} & -0.18\textsuperscript{***} & -0.18\textsuperscript{***}\\
	& {(0.06)} & {(0.05)} & {(0.06)} & {(0.05)} & {(0.05)} \\
	Momentum & -0.03 & -0.07 & -0.05 & -0.06 & -0.02\\
	& {(0.05)} & {(0.05)} & {(0.06)} & {(0.05)} & {(0.05)}\\
	\midrule
	\multicolumn{1}{l}{$R^2$(\%)} & \multicolumn{1}{c}{20.5} & \multicolumn{1}{c}{25.1} & \multicolumn{1}{c}{24.2} & \multicolumn{1}{c}{31.3} & \multicolumn{1}{c}{36.8}\\
	\multicolumn{1}{l}{N} & \multicolumn{1}{c}{286} & \multicolumn{1}{c}{286} & \multicolumn{1}{c}{286} & \multicolumn{1}{c}{286} & \multicolumn{1}{c}{286}\\
	\bottomrule
\end{tabular}

			}

	\end{table}
\end{landscape}

 \begin{landscape}
	
	\begin{table}[ht]
		\caption{\bf Correlations of the \textit{AIFX index}} \label{tab:LookAheadBias:correlation diff ratio before after chatgpt3.5vs4}
		
		\begin{footnotesize}
		This table reports the correlation between the \textit{AIFX index} (defined in \myeq{eq:VariableConstruction:definition of diff ratio}) constructed using outputs from GPT-3.5 and GPT-4o. For each currency, the table presents the correlation of the signals across two distinct periods: January 1996 to September 2021—when both models had access to training data—and September 2021 to October 2023—a period only included in GPT-4o’s training data. 
		\end{footnotesize}

		\bigskip
		
		\newcolumntype{L}[1]{>{\raggedright\let\newline\\\arraybackslash\hspace{0pt}}m{#1}}
		\newcolumntype{C}[1]{>{\centering\let\newline\\\arraybackslash\hspace{0pt}}m{#1}}
		\newcolumntype{R}[1]{>{\raggedleft\let\newline\\\arraybackslash\hspace{0pt}}m{#1}}
		\setlength\extrarowheight{1pt}


\centering
\scalebox{1.00}{					
	\begin{tabular}{@{}lcc@{}}
	\toprule
	\textbf{Currency} & \textbf{1996 - 2021} & \textbf{2021 - 2023} \\ 
	\midrule
	AUD & 0.86 & 0.83 \\
	CAD & 0.81 & 0.89 \\
	CHF & 0.76 & 0.72 \\
	EUR & 0.87 & 0.94 \\
	GBP & 0.86 & 0.89 \\
	JPY & 0.81 & 0.79 \\
	NZD & 0.84 & 0.83 \\
	NOK & 0.74 & 0.67 \\
	SEK & 0.84 & 0.86 \\
	\bottomrule
\end{tabular}
}

	\end{table}
\end{landscape}

 \begin{landscape}
	
	\begin{table}[ht]
		\caption{\bf Difference-in-Differences Exercise} \label{tab:LookAheadBias:diff in diff gpt3.5vs4}
		
		\begin{footnotesize}
			This table presents the results of the difference-in-differences analysis, described in \myeq{eq:LookAheadBias:diff-in-diff gpt3.5vs4}, investigating the differential output of the two AI models GPT-3.5 and GPT-4o across two distinct periods: from January 1996 to September 2021, which is within both models' training sets, against the period from September 2021 to October 2023, which only GPT-4o was trained on. \(T_i\) represents the treatment, which is a dummy variable assigned the value 1 for GPT-4o and 0 for GPT-3.5. Similarly, \(\text{After}_t\) is a dummy variable that is 1 for the period from September 2021 to October 2023 and 0 from January 1996 to September 2021. \(T_i* \text{After}_i\) is the interaction term. The analysis uses two model specifications to check robustness. Column 1 does not cluster standard errors, while column 2 clusters standard errors on the currency level.
		\end{footnotesize}

		\bigskip
		
		\newcolumntype{L}[1]{>{\raggedright\let\newline\\\arraybackslash\hspace{0pt}}m{#1}}
		\newcolumntype{C}[1]{>{\centering\let\newline\\\arraybackslash\hspace{0pt}}m{#1}}
		\newcolumntype{R}[1]{>{\raggedleft\let\newline\\\arraybackslash\hspace{0pt}}m{#1}}
		\setlength\extrarowheight{1pt}


\centering
\scalebox{1.00}{
	
\begin{tabular}{l
		S[table-format=1.2]
		S[table-format=1.2]
		S[table-format=1.2]
		S[table-format=1.2]
		S[table-format=1.2]}
	\toprule
	\multicolumn{1}{c}{} & \textbf{(1)} & \textbf{(2)} \\ 
	\midrule
	\(T_i\) & 0.03\textsuperscript{***} & 0.03\textsuperscript{***}  \\ 
	& {(0.01)} & {(0.01)} \\ 
	
	\(\text{After}_i\) &  -0.01 & -0.01  \\ 
	& {(0.02)} & {(0.02)} \\ 
	
	\(T_i \times \text{After}_i\) & 0.01 & 0.01 \\ 
	& {(0.02)} & {(0.01)} \\ 
	
	\midrule
	\multicolumn{1}{l}{Currency Fixed Effect} & \multicolumn{1}{c}{Y} & \multicolumn{1}{c}{Y}   \\ 
	\multicolumn{1}{l}{Clustered Standard Errors} & \multicolumn{1}{c}{N} & \multicolumn{1}{c}{Y}   \\ 
	\multicolumn{1}{l}{\(R^2\) (\%)} & \multicolumn{1}{c}{2.80} & \multicolumn{1}{c}{2.80}   \\ 
	\multicolumn{1}{l}{\# of Obs.} & \multicolumn{1}{c}{5958} & \multicolumn{1}{c}{5958}  \\ 
	\bottomrule
\end{tabular}

}

	\end{table}
\end{landscape}

\begin{landscape}
	\begin{table}[h]
		\caption{\bf Pure Hindsight Portfolio as Control Factor}   \label{tab:LookaheadBias:pure hindsight portfolio regression}
	
		\begin{footnotesize}
			This table reports the results for the contemporaneous regression specified in \myeq{eq:LookAheadBias:regression ai strategy on pure hindsight strategy} which examines whether the monthly returns of the \textit{AIFX strategy} can be explained by the \textit{pure hindsight portfolio}. The  \textit{pure hindsight portfolio} is solely based on what the AI model may remember about historical currency returns and uses the outputs of \myprp{prompt:LookAheadBias:remember the direction for lookahead bias analysis}. The portfolio is rebalanced at the end of each calendar month, same as the \textit{AIFX strategy}. The results are reported for different lookback periods. The performance reported is from January 2001 to October 2024. \citet{NEWEY/WEST:1987} standard errors are reported in parentheses. ***, **, and * indicate statistical significance at the 1\%, 5\%, and 10\% levels, respectively.
		\end{footnotesize}

		\bigskip
		
		\newcolumntype{L}[1]{>{\raggedright\let\newline\\\arraybackslash\hspace{0pt}}m{#1}}
		\newcolumntype{C}[1]{>{\centering\let\newline\\\arraybackslash\hspace{0pt}}m{#1}}
		\newcolumntype{R}[1]{>{\raggedleft\let\newline\\\arraybackslash\hspace{0pt}}m{#1}}
		\setlength\extrarowheight{1pt}

		\centering
		\scalebox{1.00}{
			\begin{tabular}
	{l
		S[table-format=1.2]
		S[table-format=1.2]
		S[table-format=1.2]
		S[table-format=1.2]
		S[table-format=1.2]}
	
	\toprule
	& \multicolumn{5}{c}{\textbf{Return of the \textit{AIFX strategy}}}\\
	\midrule
	& \textbf{36 months} & \textbf{42 months} & \textbf{48 months} & \textbf{54 months} & \textbf{60 months}\\
	\midrule
	Alpha & 5.52\textsuperscript{***} & 5.99\textsuperscript{***} & 6.06\textsuperscript{***} & 5.45\textsuperscript{***} & 5.09\textsuperscript{***}\\
	& {(1.47)} & {(1.48)} & {(1.56)} & {(1.57)} & {(1.46)}\\
	Beta & -0.36\textsuperscript{***} & -0.44\textsuperscript{***} & -0.38\textsuperscript{***} & -0.38\textsuperscript{***} & -0.40\textsuperscript{***}\\
	& {(0.09)} & {(0.09)} & {(0.10)} & {(0.10)} & {(0.09)}\\
	\midrule
	\multicolumn{1}{l}{Information Ratio} & \multicolumn{1}{c}{0.78} & \multicolumn{1}{c}{0.85} & \multicolumn{1}{c}{0.83} & \multicolumn{1}{c}{0.75} & \multicolumn{1}{c}{0.76}\\
	\multicolumn{1}{l}{$R^2$(\%)} & \multicolumn{1}{c}{4.80} & \multicolumn{1}{c}{7.15} & \multicolumn{1}{c}{5.10} & \multicolumn{1}{c}{5.00} & \multicolumn{1}{c}{6.46}\\
	\multicolumn{1}{l}{N} & \multicolumn{1}{c}{298} & \multicolumn{1}{c}{292} & \multicolumn{1}{c}{286} & \multicolumn{1}{c}{280} & \multicolumn{1}{c}{274}\\
	\bottomrule
\end{tabular}

						}
	\end{table}
\end{landscape}

 \begin{landscape}
	
	\begin{table}[ht]
		\caption{\bf Categories Descriptive Table} \label{tab:Categories:example headlines by category}
		
		\begin{footnotesize}
			The table presents the eight categories, along with representative examples of data releases included in each category and the number of observations in the dataset for each group. The methodology used to assign data releases to categories is detailed in \mysecIA{Appendix:Categories prodecure}. The data releases are sourced from Investing.com's economic calendar data spanning January 1996 to October 2024.
		\end{footnotesize}

		\bigskip
		
		\newcolumntype{L}[1]{>{\raggedright\let\newline\\\arraybackslash\hspace{0pt}}m{#1}}
		\newcolumntype{C}[1]{>{\centering\let\newline\\\arraybackslash\hspace{0pt}}m{#1}}
		\newcolumntype{R}[1]{>{\raggedleft\let\newline\\\arraybackslash\hspace{0pt}}m{#1}}
		\setlength\extrarowheight{1pt}

		\sisetup{ 
			detect-all,
			table-number-alignment = center, 
			output-decimal-marker = {,}, 
			table-align-text-pre = false,
			table-align-text-post = false,
			group-digits = integer,
			table-format = 2.3
		}

		\centering
		\scalebox{1}{					
				\begin{tabular}{|>{\centering\arraybackslash}m{6.25cm}|>{\centering\arraybackslash}m{10.25cm}|>{\centering\arraybackslash}m{2cm}|}
	\hline
	\textbf{\parbox{5cm}{\centering Category}} & \textbf{\parbox{7cm}{\centering Example Headlines}} & \textbf{\parbox{2cm}{\centering Frequency}} \\
	\hline
	Employment and Unemployment Data & Unemployment Rate, Employment Change, Participation Rate,  Overtime Pay, Continuing Jobless Claims & 17,717 \\
	\hline
	Consumer Sentiment, Retail or Services Activity & Retail Sales, Westpac Consumer Sentiment, NAB Business Confidence, Michigan Consumer Sentiment, Chain Store Sales & 31,129 \\
	\hline
	International Trade & Exports, Imports, Trade Balance, Current Account, Terms of Trade Index & 14,685 \\
	\hline
	Manufacturing, Industrial or Construction Activity & Building Approvals, HIA New Home Sales, AIG Manufacturing Index, Capacity Utilization Rate, Chicago PMI & 37,945 \\
	\hline
	Inflation and Price Indices & PPI, CPI, Wage Price Index, Import Price Index, ISM Manufacturing Prices & 36,753 \\
	\hline
	Interest Rates, Monetary Policy, Government Budget, or Bond Issuance & Bank Lending, M2 Money Stock, Government Operating Balance, Fed's Balance Sheet, Deposit Facility Rate & 20,587 \\
	\hline
	Financial Markets and Speculative Indices & Net Investment Flow, Mortgage Refinance Index, CFTC S\&P 500 speculative net positions, Foreign Investments in Japanese Stocks, CFTC Gold speculative net positions & 21,227 \\
	\hline
	Broad Economic Activity Indicators & GDP, Labour Productivity, Corporate Profits, Non-farm Productivity, Bankruptcy Filings & 22,473 \\
	\hline
\end{tabular}

			}
		
	\end{table}
\end{landscape}

%
%
%
%
%
%
%
%
%
%
%
%
%
%
%
%
%
%


\phantomsection
\addcontentsline{toc}{section}{Internet appendix}
\begin{appendices}  \label{appendix}
	\begin{center}
		
		{\Large Internet Appendix}
		
	\end{center}
	
	\pagenumbering{gobble}
	
	\setcounter{section}{0}
	\setcounter{subsection}{0}
	\renewcommand{\thesection}{\Alph{section}}
	\renewcommand{\thesubsection}{\thesection.\arabic{subsection}}
	
	\setcounter{equation}{0}
	\renewcommand{\theequation}{\thesection.\arabic{equation}}
	
	\setcounter{table}{0}
	\renewcommand\thetable{A.\arabic{table}}
	
	\setcounter{figure}{0}
	\renewcommand\thefigure{A.\arabic{figure}}
	
	\renewcommand{\thepage}{A-\arabic{page}}
	\setcounter{page}{1}
	
	\section{Definition of Currency Factors} \label{Appendix:Definition of FX factors}
		This section outlines the construction of benchmark FX strategies used throughout the paper. These strategies follow standard practices in the empirical FX literature \citep{CHERNOV/ETAL:2023} and serve as controls to evaluate the performance of the AI-powered signals.
		\begin{itemize}
			\item \textbf{Dollar}: A long-only portfolio that takes equal-weighted long positions in all non-USD currencies. This strategy captures the average performance of foreign currencies against the U.S. dollar.
			\item \textbf{Dollar Carry}: This strategy uses the average forward discount across all currencies as the signal. It goes long (short) all currencies versus the USD when the average forward discount is positive (negative).
		\end{itemize}
		\subsection*{Cross-sectional Strategies}
		
		\begin{itemize}
			\item \textbf{Carry}: This strategy uses each currency’s forward discount as an individual signal. Currencies are ranked by their forward discount, and positions are assigned using rank-based weights defined by:
			\begin{equation}
				w_{pt}^i = \kappa \left( \text{rank}(z_{pt}^i) - \frac{1}{N} \sum_{i=1}^N \text{rank}(z_{pt}^i) \right),
				\label{eq:Performance:rank weight formula}
			\end{equation}
			where \( w_{pt}^i \) denotes the weight of currency \( i \), \( z_{pt}^i \) is the forward discount signal, and \( \kappa \) is a scaling constant that ensures the portfolio is USD-neutral. With \( N = 9 \), possible weights are: \( \pm 0.4, \pm 0.3, \pm 0.2, \pm 0.1, 0.0 \).
			\item \textbf{Momentum}: Constructed in the same way as Carry, but the signal \( z_{pt}^i \) is defined as the excess return in the most recent month:
			\[
			z_{pt}^i = R_{t-1}^i.
			\]
			\item \textbf{Value}: This strategy uses the five-year detrended FX rate as the signal:
			\begin{equation}
				\widetilde{Q}_t^i = Q_t^i \left( \frac{1}{13} \sum_{j=-6}^{6} Q_{t-60+j}^i \right)^{-1},
				\label{eq:Performance:value strategy definition}
			\end{equation}
			where \( Q_t^i \) denotes the FX rate. The signal \( \widetilde{Q}_t^i \) is then used with the rank-weighting scheme in \myeq{eq:Performance:rank weight formula}.
		\end{itemize}
		
		\subsection*{Time-Series Strategies}
		Time-series (TS) strategies rely on sign-based signals. At each rebalancing date, currencies with a positive signal are held long, and those with a negative signal are held short, using the following weight scheme:
		\begin{equation}
			w_{\text{TS},t}^{i} = \frac{1}{N} \operatorname{sign}(z_t^i),
			\label{eq:Performance:sign weight formula}
		\end{equation}
		where \( N \) is the number of currencies. This ensures that the volatility of time-series strategies are comparable to cross-sectional strategies. The net USD exposure in not zero and varies over time.
		\begin{itemize}
			\item \textbf{TS-Carry}: Uses the forward discount as the signal \( z_t^i \). Positions are taken according to \myeq{eq:Performance:sign weight formula}.
			\item \textbf{TS-Momentum}: Uses the most recent monthly excess return \( R_{t-1}^i \) as the signal:
			\[
			z_t^i = R_{t-1}^i.
			\]
			Positions are taken according to \myeq{eq:Performance:sign weight formula}.
			\item \textbf{TS-Value}: Uses the five-year detrended FX rate defined in \myeq{eq:Performance:value strategy definition} as the signal:
			\[
			z_t^i = \widetilde{Q}_t^i.
			\]
			Positions are taken according to \myeq{eq:Performance:sign weight formula}.
		\end{itemize}
		All strategies are rebalanced monthly. Cross-sectional strategies are USD-neutral by construction, while TS strategies may exhibit time-varying USD exposure depending on the aggregate direction of the signals.
		\section{Time-Series Strategy; Additional Results} 	\label{Appendix: Time-Series Strategy; More Results}
		This section provides further insights into the performance of the time-series strategy introduced in \mysec{Time-Series Strategies}. \myfigIA{fig:Performance:ts strategy performance over time 48 months} illustrates the cumulative returns of the strategy over time, while \mytabIA{tab:Performance:ts AI strategy diff performance statistics} reports key performance statistics, including the mean, standard deviation, skewness, excess kurtosis, and first-order autocorrelation of returns. To assess whether the strategy's returns can be explained by benchmark time-series FX strategies, I estimate contemporaneous regressions similar to \myeq{eq:Performance:regression for performance over fx strategies CS Strategy Diff}. The construction of the benchmark strategies is detailed in \mysecIA{Appendix:Definition of FX factors}. The regression results are presented in \mytabIA{tab:Performance:ts performance over fx strategies}. The findings indicate that the strategy’s returns are not fully explained by the benchmark currency factors, and this conclusion holds across various lookback periods. For instance, with a 54-month lookback window, the strategy generates a statistically significant alpha that accounts for approximately 69\% of the average return of the strategy, implying that only about one-third of the return is subsumed by the benchmark strategies. 

		\section{Categorization Procedure} \label{Appendix:Categories prodecure}
		This section outlines the methodology used to classify each data release in the dataset into one of eight categories. These categories are designed to represent distinct groups of economic fundamentals. The categories are: 1- Employment and Unemployment Data 2- Consumer Sentiment, Retail, or Services Activity 3- International Trade 4- Manufacturing, Industrial, or Construction Activity 5- Inflation and Price Indices 6- Interest Rates, Monetary Policy, Government Budget, or Bond Issuance 7- Financial Markets and Speculative Indices 8- Broad Economic Activity Indicators. 
		The selection of these eight categories was made with the assistance of GPT-4o and guided by two key criteria: (i) economic interpretability, ensuring that each category captures a distinct dimension of economic conditions, and (ii) data sufficiency, ensuring that each category contains a meaningful number of observations. To assign each data release to its appropriate category, I interact with GPT-4o using the prompt shown in \myprp{prompt:Categories:category analysis}. This prompt instructs the AI model to classify the headline of each economic data release into one of the predefined categories.
		\begin{prompt}[H]
			\centering
			\fbox{
				\parbox{0.97\textwidth}{
					\textit{\textbf{Prompt:} Forget all previous instructions. I will give you a headline. Please list the category it belongs to among the following categories: '1. Employment and Unemployment Data; 2. Consumer Sentiment, Retail or Services Activity; 3. International Trade; 4. Manufacturing, Industrial or Construction Activity; 5. Inflation and Price Indices; 6. Interest Rates, Monetary Policy, Government Budget or Bond Issuance; 7. Financial Markets and Speculative Indices; 8. Broad Economic Activity Indicators.' Generate the output in this format: \{(CATEGORY: CATEGORY NUMBER AND NAME)\}}
				}
			}
			\caption{}
			\label{prompt:Categories:category analysis}
		\end{prompt}
		\vspace{-20pt}This automated classification procedure ensures consistency across a large set of headlines and enables a systematic analysis of the heterogeneous predictive power of different types of economic fundamentals, as discussed in \mysec{sec:Impact of Different Categories of Fundamentals}. In addition, \mytab{tab:Categories:example headlines by category} presents the eight categories, along with representative examples of headlines and the number of observations within each category.

		\begin{landscape}
	
	\begin{figure}

	    \begin{center}
	    	\caption{\bf $t$-statistics (spot returns)}
		    \scalebox{1.2}{
\pgfplotsset{compat=newest}
	\begin{tikzpicture}
		\begin{axis}[
			width=14cm,
			height=10cm,
			ybar,
			bar width=4pt,
			enlarge x limits=0.05,
			xlabel={Lookback Period (Months)},
			ylabel={T-stat},
			ymin=0,
			ymax=3.5,
			ytick={0,0.5,1,1.5,2,2.5,3,3.5},
			xtick={1,6,12,18,24,30,36,42,48,54,60},
			xtick style={draw=none},
			ticklabel style={font=\small},
			label style={font=\small},
			legend style={
				at={(0.97,0.97)},
				anchor=north east,
				draw=black, 
				fill=white, 
				font=\small
			},
			legend cell align={left}
			]
			
			\addplot [
			fill=blue!70!black,
			draw=black
			] 
			table[
			x=number_of_lookback_months,
			y=ratio_diff_t_stat,
			col sep=comma
			]{xtra/figures/Files/data/t-stat_spot_for_regression_of_return_on_diff_ratio_CS.csv};
			\addlegendentry{T-statistics}
			
			\draw [gray, dashed, line width=3pt] (rel axis cs:0,1.96/3.5) -- (rel axis cs:1,1.96/3.5);
			\draw [red, dashed, line width=3pt] (rel axis cs:0,1.64/3.5) -- (rel axis cs:1,1.64/3.5);
			
			\addlegendimage{line legend, gray, dashed, line width=3pt}
			\addlegendentry{(5\%) Significance Level}
			
			\addlegendimage{line legend, red, dashed, line width=3pt}
			\addlegendentry{(10\%) Significance Level}
			
		\end{axis}
	\end{tikzpicture}}
		    \medskip
		    \label{fig:Performance:t-stat for regression of spot return on diff ratio CS}
		    \end{center}

		    \begin{small}
		    This figure displays the $t$-statistics of $\beta$ in \myeq{eq:Performance:regression return on diff ratio cs}, where the dependent variable is spot return. The regression is repeated for different choices of lookback period. This regression examines whether the \textit{AIFX index} of a currency at the end of a given month can predict the currency's spot return in the subsequent month. Standard errors are clustered at the currency and time level. The regression includes time fixed effects, although not reported here. The sample period spans from January 1996 to October 2024, but returns start at different dates due to differences in lookback periods. The dashed horizontal lines mark the 5\% and 10\% statistical significance thresholds.

		    \end{small}

	    \end{figure}
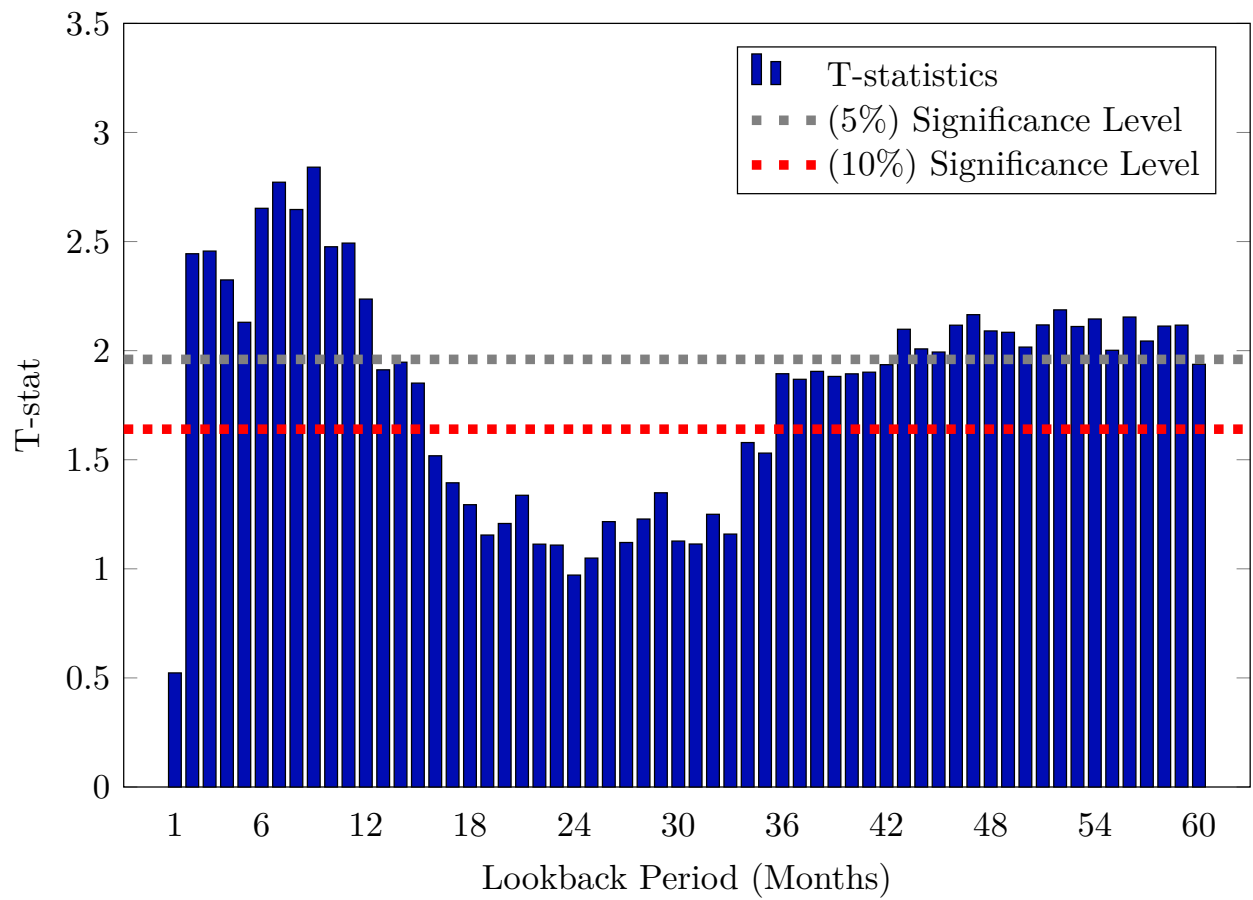
    
\end{landscape}

		\begin{landscape}
	
	\begin{figure}

    \begin{center}
    \caption{\bf Performance of AI-powered Time-series strategy}
    \scalebox{1.2}{
\pgfplotsset{compat=newest}
	\begin{tikzpicture}
		\begin{axis}[
			width=14cm,   
			height=10cm,
			ybar,
			bar width=4pt,
			enlarge x limits=0.05,
			xlabel={Lookback Period (Months)},
			ylabel={Annualized Sharpe Ratio},
			ymin=-0.1,
			ymax=0.5,
			ytick={-0.1,0,0.1,0.2,0.3,0.4,0.5},
			xtick={1,6,12,18,24,30,36,42,48,54,60},
			xtick style={draw=none},
			ymajorgrids=true,
			grid style={dashed,gray!30},
			ticklabel style={font=\small},
			label style={font=\small},
			]
			\addplot [
			fill=blue!70!black,
			draw=black
			] 
			table[
			x=lookback_period_number_of_months,
			y=currency_usd_diff_strategy_sharpe,
			col sep=comma
			]{xtra/figures/Files/data/ts_sharpe_diff_currency-usd.csv};
		\end{axis}
	\end{tikzpicture}}
    \medskip
     \label{fig:Performance:sharpe ratio TS strategy diff currency usd}
    \end{center}

    \begin{small}
     This figure reports the annualized Sharpe ratios of time-series strategies that use the \textit{Diff AIFX index} (defined in \myeq{eq:Performance:definition of diff currency usd ratio}) as the trading signal. For each choice of lookback period \(\tau\), at  the end of each month \(t\), I take a long position in currency \(c\) if \(\text{Diff AIFX}_{c,t,\tau}\) is positive and a short position if negative. Thus, portfolio weights are either \(+1\) or \(-1\), depending on the sign of the signal. Portfolios are rebalanced at the end of each calendar month. Lookback periods of 1 to 60 months are considered. The sample period spans from January 1996 to October 2024, but returns start at different dates due to differences in lookback periods.
    \end{small}

	\end{figure}
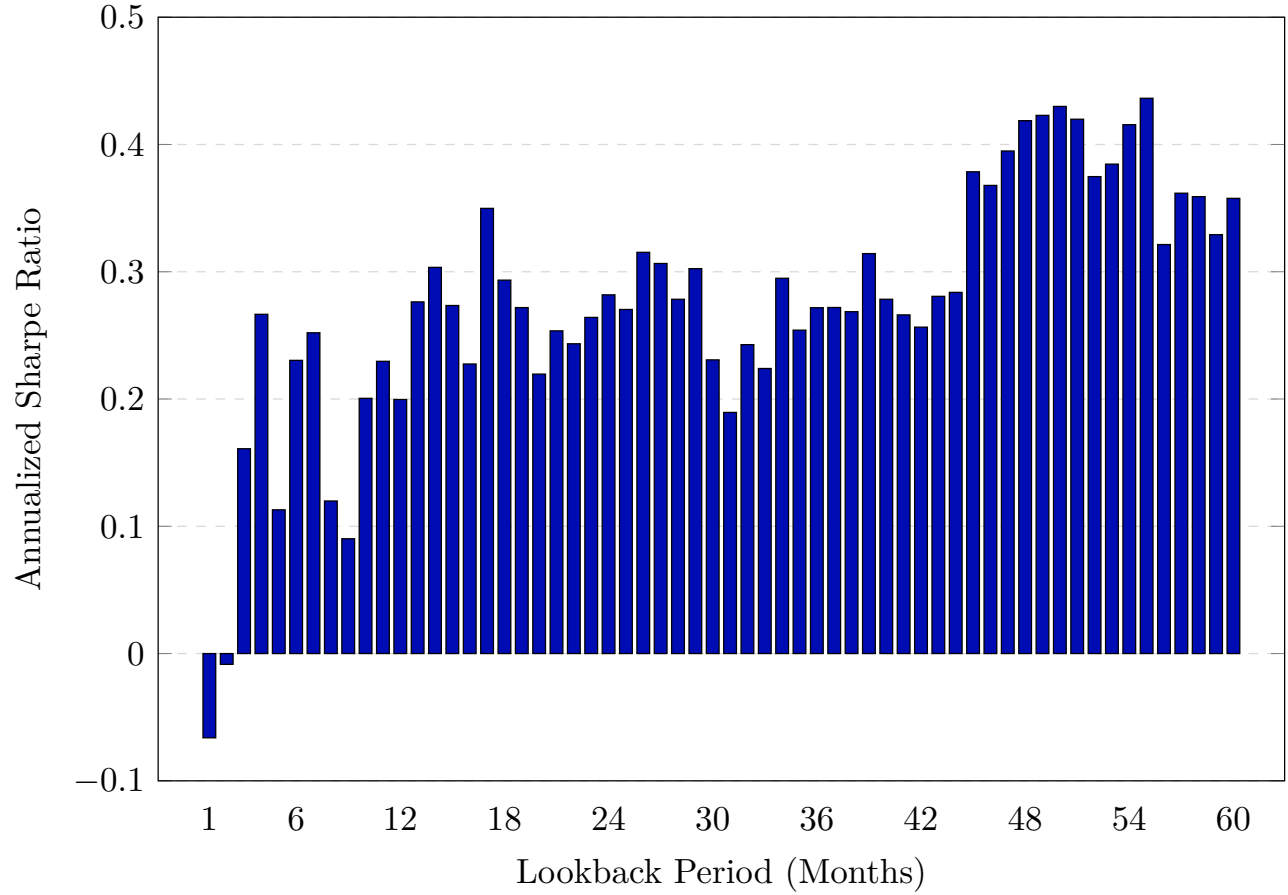

\end{landscape}

		\begin{landscape}
	
	\begin{figure}

    \begin{center}
    \caption{\bf Sharpe Ratios (\(\text{AIFX}_{c}\) vs \(\text{AIFX}_{USD}\))}
    \scalebox{1.2}{
\pgfplotsset{compat=newest}
	\begin{tikzpicture}
		\begin{axis}[
			width=14cm,   
			height=10cm,
			ybar,
			bar width=2pt,
			enlarge x limits=0.05,
			xlabel={Lookback Period (Months)},
			ylabel={Annualized Sharpe Ratio},
			ymin=-0.30,
			ymax=0.6,
			ytick={-0.3,-0.2,-0.1,0.0,0.1,0.2,0.3,0.4,0.5,0.6},
			xtick={1,6,12,18,24,30,36,42,48,54,60},
			xtick style={draw=none},
			ymajorgrids=true,
			grid style={dashed,gray!30},
			ticklabel style={font=\small},
			label style={font=\small},
			legend pos=north east, 
			legend style={font=\small}, 
			legend cell align={left} 
			]
			
			\addplot [
			fill=red!70!white,
			draw=black,
			bar shift=-1.5pt 
			] 
			table[
			x=lookback_period_number_of_months,
			y=currency_diff_strategy_sharpe,
			col sep=comma
			]{xtra/figures/Files/data/ts_sharpe_diff_currency_diff_usd.csv};
			\addlegendentry{\(\text{AIFX}_{c}\)}
			
			\addplot [
			fill=blue!70!black,
			draw=black,
			bar shift=1pt 
			] 
			table[
			x=lookback_period_number_of_months,
			y=usd_diff_strategy_sharpe,
			col sep=comma
			]{xtra/figures/Files/data/ts_sharpe_diff_currency_diff_usd.csv};
			\addlegendentry{\(\text{AIFX}_{USD}\)}

		\end{axis}
	\end{tikzpicture}}
    \medskip
     \label{fig:Performance:sharpe ratio TS strategy diff currency diff usd}
    \end{center}

    \begin{small}
     This figure reports the annualized Sharpe ratios of time-series strategies that use the \(\text{AIFX}_{c}\) and \(\text{AIFX}_{USD}\) as the trading signal. Portfolios are rebalanced at the end of each calendar month. Lookback periods of 1 to 60 months are considered. The sample period spans from January 1996 to October 2024, but returns start at different dates due to differences in lookback periods.
    \end{small}

	\end{figure}
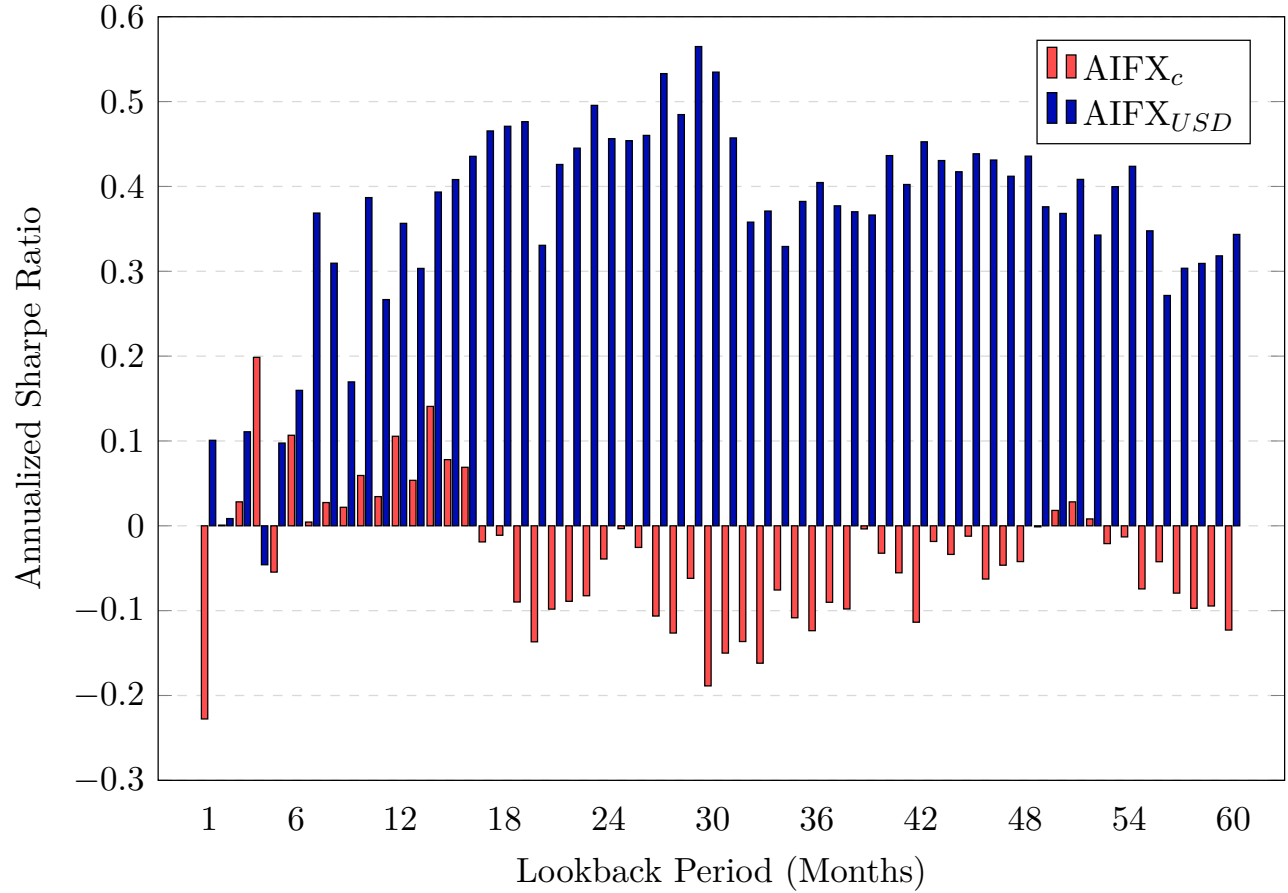

\end{landscape}

		\begin{landscape}
	
	\begin{figure}

	    \begin{center}
	    	\caption{\bf Cumulative Returns}
		    \scalebox{1.2}{
\pgfplotsset{compat=1.9}
	\begin{tikzpicture}
		\begin{axis}[
			width=18cm,
			height=10cm,
			xlabel={Date},
			ylabel={Dollar Value of an Initial \$1 Investment},
			xticklabel style={rotate=45, anchor=east},
			xtick scale label code/.code={},
			grid=major,
			grid style={dashed,gray!30},
			ymajorgrids=true,
			y tick label style={font=\small},
			label style={font=\small},
			clip mode=individual,
			date coordinates in=x,        
			date ZERO={1999-12-31},         
			xmin={1999-12-31},             
			xmax={2024-09-30},             
			xtick={2000-01-01,2005-01-01,2010-01-01,2015-01-01,2020-01-01,2024-09-30},
			xticklabels={2000,2005,2010,2015,2020,2024},
			legend pos=north west,
			legend cell align={left},
			enlarge x limits=0.05,
			ymax=2.6,
			ymin = 0.9  
			]
			
			\addplot [
			ybar,
			bar width=1.4pt, 
			fill=green!30,  
			draw=none,  
			opacity=0.4,  
			forget plot 
			] table [
			col sep=comma,
			x=date,
			y expr={\thisrow{NBER_Recession_indicator}==1 ? \pgfkeysvalueof{/pgfplots/ymax} : NaN}
			] {xtra/figures/Files/data/ts_performance_overtime_currency_usd_and_usd_diff.csv} \closedcycle;
			
			\addplot[mark=none, smooth, color=blue!70!black, line width=1.35pt] 
			table[col sep=comma, x=date, y=diff_currency_diff_usd] 
			{xtra/figures/Files/data/ts_performance_overtime_currency_usd_and_usd_diff.csv};
			
			\addplot[mark=none, smooth, color=red!70!white, line width=1.35pt] 
			table[col sep=comma, x=date, y=diff_usd] 
			{xtra/figures/Files/data/ts_performance_overtime_currency_usd_and_usd_diff.csv};
			
			\addplot [
			color=olive,
			very thick,
			dotted
			] coordinates {(2023-10-01, 0) (2023-10-01, \pgfkeysvalueof{/pgfplots/ymax})};
			
			\addlegendimage{area legend, fill=green!30, draw=none}
			\addlegendentry{Recession Period}
			
			\legend{Diff AIFX, \(\text{AIFX}_{USD}\), Cut-off Date, Recession Period}
		\end{axis}
	\end{tikzpicture}}
		    \medskip
		     \label{fig:Performance:ts strategy performance over time 48 months}
	    \end{center}

	    \begin{small}
	    The graph displays the Dollar value of an initial \$1 investment in two time-series strategies from January 2000 to October 2024. The blue line uses the \textit{Diff AIFX index} (defined in \myeq{eq:Performance:definition of diff currency usd ratio}) and the red line uses the \(\text{AIFX}_{USD}\) as the signal. Both trading strategies have a lookback period of 48 months. The green shaded regions denote NBER recession periods. The vertical dotted line marks the knowledge cutoff date of GPT-4o (October 2023).
	    \end{small}

	\end{figure}

\end{landscape}

		\begin{landscape}
	
	\begin{figure}

	    \begin{center}
	    	\caption{\bf Comparison of F1 Score and Strategy's Annual Return}
		    \scalebox{1.2}{
\pgfplotsset{compat=newest}
	\begin{tikzpicture}
		\begin{axis}[
			width=14cm,   
			height=10cm,
			enlarge x limits=0.05,
			xtick=data, 
			xlabel={Year},
			ylabel={Performance (\%)},
			xtick style={draw=none},
			ymajorgrids=true,
			grid style={dashed,gray!30},
			ticklabel style={font=\small},
			label style={font=\small},
			xticklabel style={
				rotate=90,
				anchor=east,
				/pgf/number format/1000 sep={}
			},
			]
			\addplot[
			color=blue!70!black,
			line width=2pt,
			mark=*,
			] 
			table[
			x=year,
			y=F1-score,
			col sep=comma
			]{xtra/figures/Files/data/f1_score_annual_return.csv};
				\addlegendentry{F1 Score}
			
			\addplot[
			color=red!70!white,
			line width=2pt,
			mark=*,
			] 
			table[
			x=year,
			y=annualized_return_percentage_48_months,
			col sep=comma
			]{xtra/figures/Files/data/f1_score_annual_return.csv};
				\addlegendentry{Annual Return}
				
			\node[anchor=north, font=\small] at (rel axis cs:0.5,1) {Correlation = $-0.36$};
		\end{axis}
	\end{tikzpicture}}
		    \medskip
		     \label{fig:LookAheadBias:f1 score annual return}
	    \end{center}

	    \begin{small}
	    The red line displays the annual return of the \textit{AIFX strategy} with a lookback period of 48 months. The blue line shows the F1 score (defined in \myeq{eq:Lookahead:definition of f1 score}) for each year. The performance is reported from January 2000 to October 2024.
	    \end{small}

	\end{figure}
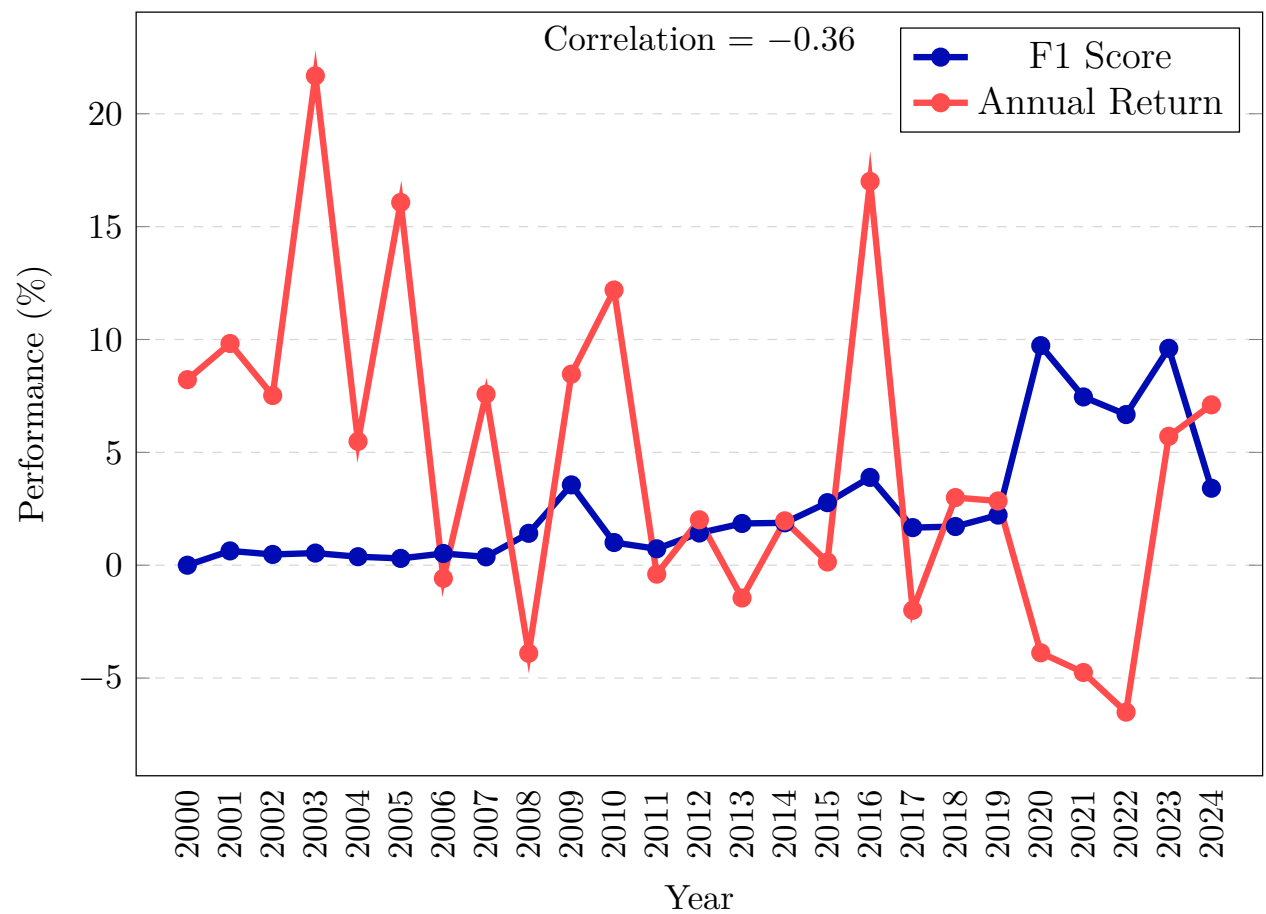

\end{landscape}

		\begin{landscape}
	
	\begin{figure}

	    \begin{center}
	    	\caption{\bf Pure Hindsight Portfolio's Performance Over Time}
		    \scalebox{1.2}{
\pgfplotsset{compat=1.9}
	\begin{tikzpicture}
		\begin{axis}[
			width=18cm,
			height=10cm,
			xlabel={Date},
			ylabel={Dollar Value of an Initial \$1 Investment},
			xticklabel style={rotate=45, anchor=east},
			xtick scale label code/.code={},
			grid=major,
			grid style={dashed,gray!30},
			ymajorgrids=true,
			y tick label style={font=\small},
			label style={font=\small},
			clip mode=individual,
			date coordinates in=x,        
			date ZERO={1996-02-29},         
			xmin={1996-02-29},             
			xmax={2024-10-31},             
			xtick={1996-01-01,2000-01-01,2005-01-01,2010-01-01,2015-01-01,2020-01-01,2024-10-31},
			xticklabels={1996,2000,2005,2010,2015,2020,2024},
			enlarge x limits=0.05,
			ymax=3,
			ymin=0.9,  
			legend pos=north west, 
			legend cell align={left}, 
			legend style={font=\small, /tikz/every even column/.append style={column sep=2pt}} 
			]
			
			\addplot [
			ybar,
			bar width=1.4pt, 
			fill=green!30,  
			draw=none,  
			opacity=0.4,  
			forget plot 
			] table [
			col sep=comma,
			x=date,
			y expr={\thisrow{recession_indicator}==1 ? \pgfkeysvalueof{/pgfplots/ymax} : NaN}
			] {xtra/figures/Files/data/pure_hindsight_performance_over_time.csv} \closedcycle;
			
			\addplot[mark=none, smooth, color=blue!70!black, line width=1.35pt] 
			table[col sep=comma, x=date, y=return] 
			{xtra/figures/Files/data/pure_hindsight_performance_over_time.csv};
			
			\addplot [
			color=olive,
			very thick,
			dotted
			] coordinates {(2023-10-01, 0) (2023-10-01, \pgfkeysvalueof{/pgfplots/ymax})};
			
			\addlegendimage{area legend, fill=green!30, draw=none}
			\addlegendentry{Recession Period}
			
			\legend{Pure-hindsight Portfolio, Cut-off Date, Recession Period}
		\end{axis}
	\end{tikzpicture}}
		    \medskip
		     \label{fig:LookAheadBias:pure hindsight performance over time}
	    \end{center}

	    \begin{small}
	    The graph displays the Dollar value of an initial \$1 investment in the \textit{pure hindsight portfolio} from January 1996 to October 2024. The trading strategy’s signal is solely based on what the AI model may remember about historical currency returns and uses the outputs of \myprp{prompt:LookAheadBias:remember the direction for lookahead bias analysis}. The portfolio is rebalanced at the end of each calendar month, same as the \textit{AIFX strategy}. The green shaded regions denote NBER recession periods. The vertical dotted line marks the knowledge cutoff date of GPT-4o (October 2023).
	    \end{small}

	\end{figure}
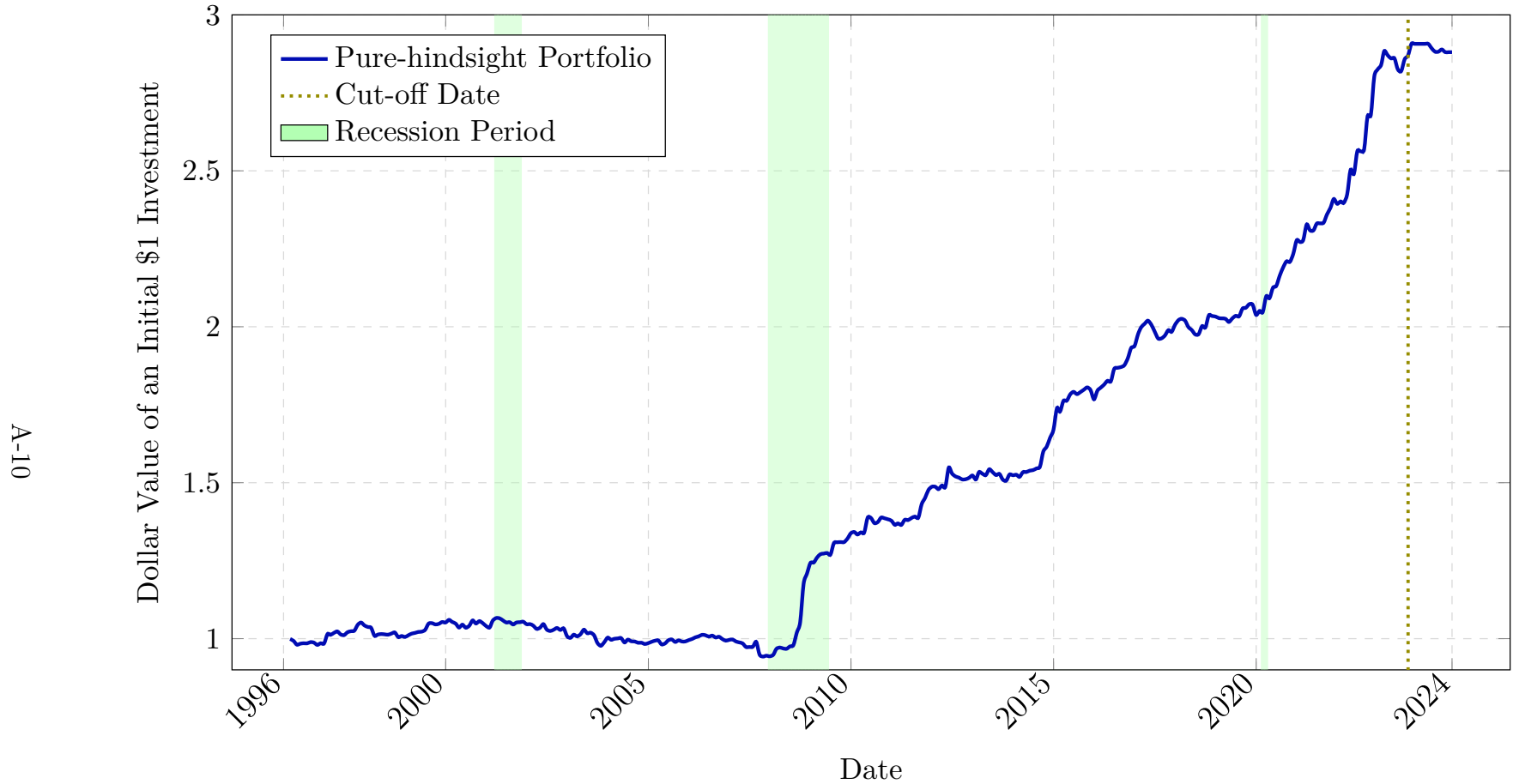

\end{landscape}

		\begin{landscape}
	
	\begin{figure}

	    \begin{center}
	    	\caption{\bf Cross-sectional Standard Deviation}
		    \scalebox{1.2}{
\pgfplotsset{compat=1.9}
	\begin{tikzpicture}
		\begin{axis}[
			width=14cm,
			height=10cm,
			xlabel={Date},
			ylabel={Standard Deviation},
			xticklabel style={rotate=45, anchor=east},
			xtick scale label code/.code={},
			grid=major,
			grid style={dashed,gray!30},
			ymajorgrids=true,
			y tick label style={/pgf/number format/fixed}, 
			scaled y ticks=false,                          
			label style={font=\small},
			clip mode=individual,
			date coordinates in=x,
			date ZERO={1999-12-31},
			xmin={1999-12-31},
			xmax={2024-09-30},
			xtick={2000-01-01,2005-01-01,2010-01-01,2015-01-01,2020-01-01,2024-09-30},
			xticklabels={2000,2005,2010,2015,2020,2024},
			legend pos=north east,
			legend cell align={left},
			enlarge x limits=0.05,
			]
			\addplot[mark=none, smooth, color=blue!70!black, line width=1.35pt] table[col sep=comma, x=date, y=positive_news_std_per_period] {xtra/figures/Files/data/cs_std_per_period_48_months.csv};
			\addplot[mark=none, smooth, color=red!70!white, line width=1.35pt] table[col sep=comma, x=date, y=negative_news_std_per_period] {xtra/figures/Files/data/cs_std_per_period_48_months.csv};
			\addplot[mark=none, smooth, color=gray, line width=1.35pt, dotted] table[col sep=comma, x=date, y=diff_std_per_period] {xtra/figures/Files/data/cs_std_per_period_48_months.csv};
			\legend{Strength ratio, Weakness ratio, AIFX index}
		\end{axis}
	\end{tikzpicture}}
		    \medskip
		     \label{fig:DifferentialImpact:cs_std_per_period_48_months}
	    \end{center}

	    \begin{small}
	    The figure displays the monthly cross-sectional standard deviation of three AI-powered variables from January 2000 to October 2024. The three variables include: \textit{AIFX index} (defined in \myeq{eq:VariableConstruction:definition of diff ratio}) , \textit{Strength ratio} (defined in \myeq{eq:VariableConstruction:definition of pos ratio}) and \textit{Weakness ratio} (defined in \myeq{eq:VariableConstruction:definition of neg ratio}). All the three variables have a lookback period of 48 months.
	    \end{small}

	\end{figure}
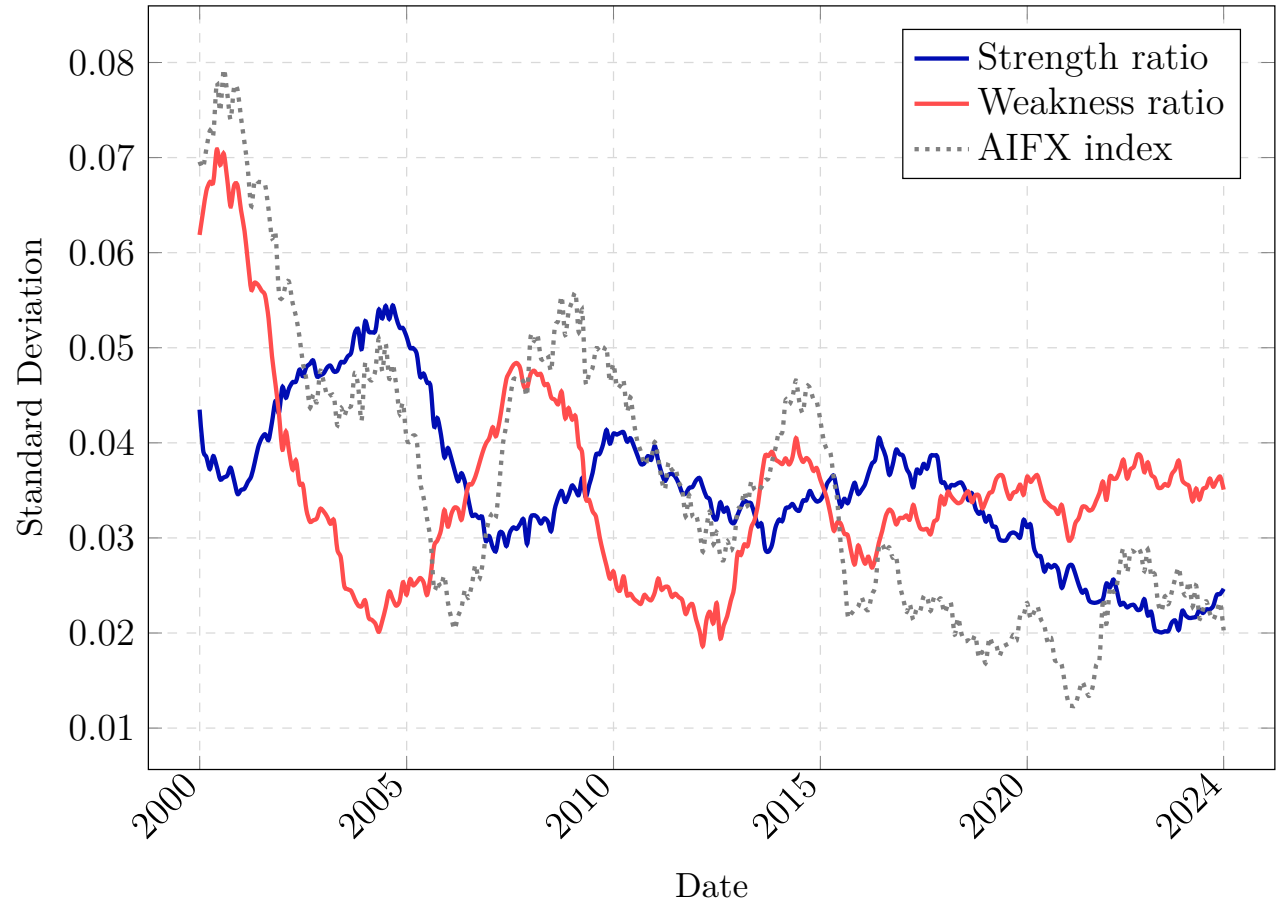

\end{landscape}

		\begin{landscape}
	
	\begin{figure}

	    \begin{center}
	    	\caption{\bf Average Daily Returns}
		    \scalebox{1.2}{
\pgfplotsset{compat=1.17}
	\begin{tikzpicture}
		\begin{axis}[
			width=14cm,
			height=10cm,
			ybar,
			bar width=50pt,
			symbolic x coords={
				{Day Before},
				{Same Day},
				{Day After}
			},
			xtick=data,
			x tick label style={rotate=45, align=center, font=\small},
			ylabel={Mean Returns (Basis Points)},
			ymin=-5.5, 
			ymax=2.5,
			nodes near coords,
			nodes near coords style={
				/pgf/number format/.cd,
				fixed,
				fixed zerofill,
				precision=2
			},
			enlarge x limits=0.25,
			scaled ticks=false,
			legend pos=south west,
			]
			\addplot[
			fill=blue!70!black,
			draw=black
			] 
			coordinates {
				({Day Before}, 0.0998)
				({Same Day}, 1.9164)
				({Day After}, 0.3048)
			};
			\addlegendentry{Positive News}
			\addplot[
			fill=red!70!white,
			draw=black
			] 
			coordinates {
				({Day Before}, -1.2444)
				({Same Day}, -4.9876)
				({Day After}, -0.7339)
			};
			\addlegendentry{Negative News}
		\end{axis}
	\end{tikzpicture}}
		    \medskip
		     \label{fig:DifferentialImpact:pos neg differential impact expanded window}
	    \end{center}

	    \begin{small}
	    This figure compares the average realized returns associated with positive and negative news on the day before, day of and the day after the release. Each of the data releases in the dataset is labeled as positive, neutral or negative using the output of \myprp{prompt:VariableConstruction:financial analyst prompt}. Specifically, I refer to data releases with the direction labeled as STRENGTHEN in the output as positive news, those labeled as WEAKEN as negative news, and those labeled as INSIGNIFICANT OR UNCERTAIN as neutral news. The sample period for this exercise is from 2008 to 2024.
	    \end{small}

	\end{figure}

\end{landscape}

		\begin{landscape}
	\begin{table}[h]
		\caption{\bf Performance Statistics}   \label{tab:Performance:fx strategy performance statistics}
	
		\begin{footnotesize}
			This table reports the performance statistics of the benchmark currency factors. The benchmarks include Dollar, Dollar Carry, Cross-sectional Carry, Cross-sectional One-month Momentum, and Cross-sectional Value strategies, constructed as described in \mysecIA{Appendix:Definition of FX factors}. The statistics include mean returns, standard deviation, skewness, excess kurtosis, first-order autocorrelation (AR(1)) and Sharpe Ratios. The measures are based on monthly returns, but means, standard deviations and Sharpe ratios are annualized. The performance reported is from January 2001 to October 2024.
		\end{footnotesize}

		\bigskip
		
		\newcolumntype{L}[1]{>{\raggedright\let\newline\\\arraybackslash\hspace{0pt}}m{#1}}
		\newcolumntype{C}[1]{>{\centering\let\newline\\\arraybackslash\hspace{0pt}}m{#1}}
		\newcolumntype{R}[1]{>{\raggedleft\let\newline\\\arraybackslash\hspace{0pt}}m{#1}}
		\setlength\extrarowheight{1pt}

		\sisetup{ 
			detect-all,
			table-number-alignment = center, 
			output-decimal-marker = {.}, 
			group-digits = integer,
			table-format = +1.3,  
			parse-numbers = false  
			}
		\centering
		\scalebox{1.00}{
		\begin{tabular}
	{l|S[table-column-width=2.5cm]S[table-column-width=2.5cm]S[table-column-width=2.5cm]S[table-column-width=2.5cm]S[table-column-width=2.5cm]}
	\hline
	& {\textbf{Dollar}}
	& {\textbf{Dollar Carry}}
	& {\textbf{Carry}}
	& {\textbf{Value}}
	& {\textbf{Momentum}}\\ \hline
	Mean & 0.249 & 2.111 & 3.201 & 2.282 & -0.374 \\
	Standard deviation & 8.296 & 8.273 & 7.653 &6.963 & 6.963 \\
	Skewness & -0.154 & -0.289 & -0.942 & 0.024 & 0.406 \\
	Excess kurtosis & 0.807 & 0.896 & 3.764 & 0.077 & 4.532 \\
	AR(1)  & 0.028 & 0.019 & 0.044 & 0.001 & -0.041 \\ 
	Sharpe Ratio  & 0.030 & 0.255 & 0.418 & 0.328 & -0.054 \\ \hline
\end{tabular}
						}

	\end{table}
\end{landscape}

		\begin{landscape}
	\begin{table}[h]
		\caption{\bf Performance Statistics}   \label{tab:Performance:ts AI strategy diff performance statistics}
	
		\begin{footnotesize}
			This table reports the performance statistics for the time-series strategies that use the \textit{Diff AIFX index} (defined in \myeq{eq:Performance:definition of diff currency usd ratio}) as the trading signal. For each choice of lookback period \(\tau\), at  the end of each month \(t\), I take a long position in currency \(c\) if \(\text{Diff AIFX}_{c,t,\tau}\) is positive and a short position if negative. Thus, portfolio weights are either \(+1\) or \(-1\), depending on the sign of the signal. Portfolios are rebalanced at the end of each calendar month. Results are shown for different lookback periods. The statistics include mean returns, standard deviation, skewness, excess kurtosis, first-order autocorrelation (AR(1)) and Sharpe Ratios. The measures are based on monthly returns, but means, standard deviations and Sharpe ratios are annualized. The performance reported is from January 2001 to October 2024.
		\end{footnotesize}

		\bigskip
		
		\newcolumntype{L}[1]{>{\raggedright\let\newline\\\arraybackslash\hspace{0pt}}m{#1}}
		\newcolumntype{C}[1]{>{\centering\let\newline\\\arraybackslash\hspace{0pt}}m{#1}}
		\newcolumntype{R}[1]{>{\raggedleft\let\newline\\\arraybackslash\hspace{0pt}}m{#1}}
		\setlength\extrarowheight{1pt}

		\sisetup{ 
			detect-all,
			table-number-alignment = center, 
			output-decimal-marker = {.}, 
			group-digits = integer,
			table-format = +1.3,  
			parse-numbers = false  
		}
		
		\centering
		\scalebox{1.00}{
			\begin{tabular}
	{l|S[table-column-width=2.5cm]S[table-column-width=2.5cm]S[table-column-width=2.5cm]S[table-column-width=2.5cm]S[table-column-width=2.5cm]}
	\hline
	& {\textbf{48 months}}
	& {\textbf{54 months}}
	& {\textbf{60 months}} \\ \hline
	Mean & 2.187 & 2.252 & 2.112 \\
	Standard deviation & 5.883 & 5.843 & 5.905 \\
	Skewness & -0.183 & -0.153 & -0.184 \\
	Excess kurtosis & 0.315 & 0.535 & 0.260 \\
	AR(1)  &  -0.043 & -0.057 & -0.039 \\ 
	Sharpe Ratio  & 0.372 & 0.385 & 0.358 \\ \hline
\end{tabular}
						}
	\end{table}
\end{landscape}

		 \begin{landscape}
 	
	\begin{table}[ht]
		\caption{\bf Performance Over Benchmark Factors} \label{tab:Performance:ts performance over fx strategies}
		
		\begin{footnotesize}
			This table presents the results from a contemporaneous regression similar to the one specified in \myeq{eq:Performance:regression for performance over fx strategies CS Strategy Diff}, which examines whether the monthly returns of the time-series strategy, using the \textit{Diff AIFX index} (defined in \myeq{eq:Performance:definition of diff currency usd ratio}) as the trading signal, can be explained by common currency factors. The time-series benchmark strategies are described in \mysecIA{Appendix:Definition of FX factors}. The performance reported is from January 2001 to October 2024. \citet{NEWEY/WEST:1987} standard errors are reported in parentheses. ***, **, and * indicate statistical significance at the 1\%, 5\%, and 10\% levels, respectively.
		\end{footnotesize}

		\bigskip
		
		\newcolumntype{L}[1]{>{\raggedright\let\newline\\\arraybackslash\hspace{0pt}}m{#1}}
		\newcolumntype{C}[1]{>{\centering\let\newline\\\arraybackslash\hspace{0pt}}m{#1}}
		\newcolumntype{R}[1]{>{\raggedleft\let\newline\\\arraybackslash\hspace{0pt}}m{#1}}
		\setlength\extrarowheight{1pt}

		\sisetup{
			detect-all,
			table-format = 1.2,        
			group-digits = false,
			table-number-alignment = center
		}
	
		\centering
		\scalebox{1.00}{					
			\begin{tabular}
	{l
		S[table-format=1.2]
		S[table-format=1.2]
		S[table-format=1.2]}
	
	\toprule
	& \multicolumn{3}{c}{\textbf{Return of the AI-powered Strategy}}\\
	\midrule
	& \textbf{48 months} & \textbf{54 months} & \textbf{60 months}\\
	\midrule
	Alpha & 1.42\textsuperscript{*} & 1.56\textsuperscript{**} & 1.44\textsuperscript{**}  \\
	& {(0.84)} & {(0.72)} & {(0.72)}\\
	Dollar &  -0.46\textsuperscript{***} & -0.52\textsuperscript{***} &  -0.56\textsuperscript{***}  \\
	& {(0.03)} & {(0.03)} & {(0.02)}\\
	Dollar Carry & 0.24\textsuperscript{***} & 0.28\textsuperscript{***} & 0.26\textsuperscript{***} \\
	& {(0.05)} & {(0.04)} & {(0.04)}\\
	Carry & 1.95\textsuperscript{***} & 0.89 & 0.99\textsuperscript{*}\\
	& {(0.71)} & {(0.63)} & {(0.59)}\\
	Value & 0.02 & 0.42 & 0.22\\
	& {(0.36)} & {(0.32)} & {(0.30)}\\
	Momentum & -0.10 & 0.23 & 0.33\\
	& {(0.34)} & {(0.30)} & {(0.28)}\\
	\midrule
	\multicolumn{1}{c}{$R^2$(\%)} & \multicolumn{1}{c}{54.9} & \multicolumn{1}{c}{64.7} & \multicolumn{1}{c}{69.5}\\
	\multicolumn{1}{c}{N} & \multicolumn{1}{c}{286} & \multicolumn{1}{c}{286} & \multicolumn{1}{c}{286}\\
	\bottomrule
\end{tabular}

			}

	\end{table}
\end{landscape}

		\begin{landscape}
	\begin{table}[h]
		\caption{\bf Evaluation of the Classification}   \label{tab:LookaheadBias:guess_year_detailed_table}
	
		\begin{footnotesize}
			This table provides a detailed description of the classification exercise of \mysec{sec:Guess the Year}. In the classification exercise, the AI model is tasked with identifying the year (not the exact date) in which each data release occurred, using \myprp{prompt:LookAheadBias:guess the year for lookahead bias analysis}. This exercise serves as a test for potential look-ahead bias. In the experiment, I use the same inputs originally provided to the AI model via \myprp{prompt:VariableConstruction:financial analyst prompt}, but instead of asking for an economic analysis, the model is asked to guess the year of the data release. The sample period spans from January 1996 to October 2024.
		\end{footnotesize}

		\bigskip
		
		\newcolumntype{L}[1]{>{\raggedright\let\newline\\\arraybackslash\hspace{0pt}}m{#1}}
		\newcolumntype{C}[1]{>{\centering\let\newline\\\arraybackslash\hspace{0pt}}m{#1}}
		\newcolumntype{R}[1]{>{\raggedleft\let\newline\\\arraybackslash\hspace{0pt}}m{#1}}
		\setlength\extrarowheight{1pt}

		\sisetup{ 
			detect-all,
			table-number-alignment = center, 
			output-decimal-marker = {.}, 
			group-digits = integer,
			table-format = +1.3,  
			parse-numbers = false  
		}
		
		\centering
		\scalebox{0.8}{
			\begin{tabular}{ccccccc}
	\toprule
	\textbf{Year} & \textbf{Number of Data Releases} & \textbf{All Guesses} & \textbf{Correct Guesses} & \textbf{Precision(\%)} & \textbf{Recall(\%)} & \textbf{F1-score(\%)} \\
	\midrule
	1996 & 1955 & 283 & 8 & 2.83 & 0.41 & 0.72 \\
	1997 & 2053 & 123 & 7 & 5.69 & 0.34 & 0.64 \\
	1998 & 2091 & 279 & 16 & 5.73 & 0.77 & 1.35 \\
	1999 & 2171 & 426 & 18 & 4.23 & 0.83 & 1.39 \\
	2000 & 2260 & 96 & 0 & 0.00 & 0.00 & 0.00 \\
	2001 & 2393 & 131 & 8 & 6.11 & 0.33 & 0.63 \\
	2002 & 2418 & 98 & 6 & 6.12 & 0.25 & 0.48 \\
	2003 & 2460 & 144 & 7 & 4.86 & 0.28 & 0.54 \\
	2004 & 2488 & 169 & 5 & 2.96 & 0.20 & 0.38 \\
	2005 & 2517 & 139 & 4 & 2.88 & 0.16 & 0.30 \\
	2006 & 2559 & 137 & 7 & 5.11 & 0.27 & 0.52 \\
	2007 & 2567 & 132 & 5 & 3.79 & 0.19 & 0.37 \\
	2008 & 3610 & 353 & 28 & 7.93 & 0.78 & 1.41 \\
	2009 & 3770 & 557 & 77 & 13.82 & 2.04 & 3.56 \\
	2010 & 3822 & 353 & 21 & 5.95 & 0.55 & 1.01 \\
	2011 & 3814 & 315 & 15 & 4.76 & 0.39 & 0.73 \\
	2012 & 3845 & 362 & 30 & 8.29 & 0.78 & 1.43 \\
	2013 & 3952 & 590 & 42 & 7.12 & 1.06 & 1.85 \\
	2014 & 4385 & 834 & 49 & 5.88 & 1.12 & 1.88 \\
	2015 & 4665 & 1036 & 79 & 7.63 & 1.69 & 2.77 \\
	2016 & 4639 & 1530 & 120 & 7.84 & 2.59 & 3.89 \\
	2017 & 4604 & 799 & 45 & 5.63 & 0.98 & 1.67 \\
	2018 & 4682 & 804 & 47 & 5.85 & 1.00 & 1.71 \\
	2019 & 4842 & 1282 & 68 & 5.30 & 1.40 & 2.22 \\
	2020 & 5128 & 4883 & 487 & 9.97 & 9.50 & 9.73 \\
	2021 & 5104 & 4099 & 343 & 8.37 & 6.72 & 7.45 \\
	2022 & 4996 & 6329 & 378 & 5.97 & 7.57 & 6.67 \\
	2023 & 4996 & 73860 & 3792 & 5.13 & 75.90 & 9.61 \\
	2024 & 3946 & 2566 & 111 & 4.33 & 2.81 & 3.41 \\
	\bottomrule
\end{tabular}
						}
	\end{table}
\end{landscape}

\end{appendices}

\end{document}